\documentclass[fleqn,usenatbib]{mnras}

\usepackage{newtxtext,newtxmath}

\usepackage[T1]{fontenc}
\usepackage{ae,aecompl}

\usepackage{epsfig,graphicx}	
\usepackage{amsmath}	
\usepackage{amssymb}	
\usepackage{natbib}
\usepackage{bm}
\usepackage{comment}

\title[GRS 1758$-$258 \& 1E 1740.7$-$2942]
{Jet--ISM Interactions near the Microquasars GRS 1758$-$258 and 1E 1740.7$-$2942}

\author[A.J. Tetarenko et al.]{A.J. Tetarenko$^{1}$\thanks{E-mail: a.tetarenko@eaobservatory.org},
E.W. Rosolowsky$^{2}$,
J.C.A Miller-Jones$^{3}$,
and G.R. Sivakoff$^{2}$
\\
$^1$East Asian Observatory, 660 N. A'oh\={o}k\={u} Place, University
Park, Hilo, Hawaii 96720, USA\\
$^2$Department of Physics, University of Alberta, CCIS 4-181, Edmonton, AB T6G 2E1, Canada\\
$^3$International Centre for Radio Astronomy Research - Curtin University, GPO Box U1987, Perth, WA 6845, Australia\\
}

\date{Accepted XXX. Received YYY; in original form ZZZ}

\pubyear{2020}

\defcitealias{kauff17}{K17}
\defcitealias{sol87}{S87}

\begin{document}
\label{firstpage}
\pagerange{\pageref{firstpage}--\pageref{lastpage}}
\maketitle

\begin{abstract}
We present Atacama Large Millimeter/Sub-millimeter Array observations of the candidate jet-ISM interaction zones near the black hole X-ray binaries GRS 1758$-$258 and 1E 1740.7$-$2942. Using these data, we map the molecular line emission in the regions, detecting emission from the HCN [$J=1-0$], HCO$^+$ [$J=1-0$], SiO [$J=2-1$], CS [$J=2-1$], $^{13}$CO [$J=1-0$], C$^{18}$O [$J=1-0$], HNCO [$J=4_{0,4}-3_{0,3}$], HNCO [$J=5_{0,5}-4_{0,4}$], and CH$_3$OH [$J=2_{1,1}-1_{1,0}$] molecular transitions.
Through examining the morphological, spectral, and kinematic properties of
this emission, we identify molecular structures that may trace jet-driven cavities in the gas surrounding these systems. Our results from the GRS 1758$-$258 region in particular, are consistent with recent work, which postulated the presence of a jet-blown cocoon structure in deep radio continuum maps of the region.
Using these newly discovered molecular structures as calorimeters, we estimate the time averaged jet power from these systems, finding $(1.1-5.7)\times10^{36}{\rm erg\,s}^{-1}$ over $0.12-0.31$ Myr for GRS 1758$-$258 and $(0.7-3.5)\times10^{37}{\rm erg\,s}^{-1}$ over $0.10-0.26$ Myr for 1E 1740.7$-$2942.
Additionally, the spectral line characteristics of the detected emission place these molecular structures in the central molecular zone of our Galaxy, thereby constraining the distances to the black hole X-ray binaries to be $8.0\pm1.0$ kpc.
Overall, our analysis solidifies the diagnostic capacity of molecular lines, and highlights how astro-chemistry can both identify jet-ISM interaction zones and probe jet feedback from Galactic X-ray binaries.

\end{abstract}



\section{Introduction}
\label{sec:intro}
Black holes regulate their local environments through the ejection of powerful relativistic jets. For example, the jets launched by super-massive black holes in Active Galactic Nuclei (AGN) are often observed to carve out huge cavities in the surrounding gas, depositing enough energy to affect large scale processes such as galaxy evolution, star formation, and even the distribution of matter in the early universe \citep{mag98,mac05,mac07,mir11}.
Additionally, stellar-mass black holes present in Galactic X-ray binaries (BHXBs) or ultra-luminous X-ray sources (ULXs; systems\footnote{We note that some ULXs are also known to contain neutron stars (e.g., \citealt{bach14}).} in nearby galaxies accreting near the Eddington limit, $L_X>10^{39}\,{\rm erg\,s}^{-1}$; e.g., \citealt{kaaret17}),
are known to release a significant portion of liberated accretion power into their local environments through jets, often driving jet-blown bubbles or ionized nebula structures surrounding the central compact objects \citep{heigrimm,gallo05,fen05,rus10,sor10,pak10,cseh14}.

\renewcommand\tabcolsep{1.5pt}
 \begin{table*}
\caption{Summary of molecular tracers sampled in this study}\quad
\centering
\begin{tabular}{ lcl}
 \hline\hline
 {\bf Molecule}&{\bf What it traces?}&{\bf Why?}\\[0.15cm]
  \hline
HCN&Density&High critical density ($10^4-10^{5} {\rm cm}^{-3}$) --- due to the high electric dipole moment of this molecule,\\ && gas needs to be in a high density environment, where many collisions will occur,  to become excited.\\[0.1cm]
HCO$^+$&Density&High critical density ($10^4-10^{5} {\rm cm}^{-3}$), and thus is preferentially excited in high density environments.\\[0.1cm]
&Ionization&HCO$^+$ abundance can be enhanced in regions with a higher fraction of ionization.\\[0.1cm]
SiO&Shocks&Silicon-bearing species --- silicon abundance is enhanced during the dust grain destruction process.\\[0.1cm]
HNCO&Density&High critical density ($10^5-10^{7} {\rm cm}^{-3}$), and thus can only be excited in high density environments.\\[0.1cm]
&Shocks&HNCO abundance has been found to increase in the presence of slower shocks, in contrast to SiO\\ && tracing faster shocks.\\[0.1cm]
CS&Density&High critical density ($10^4-10^{5} {\rm cm}^{-3}$), and thus can only be excited in high density environments. \\[0.1cm]
&Shocks&Sulfur-bearing species --- similar to silicon, sulfur abundance can be enhanced during the dust\\ && grain destruction process. \\[0.1cm]
CO&Density&Probes gas opacity ($^{12}$CO tends to be optically thick, while $^{13}$CO and C$^{18}$O tend to be optically thin),\\
&&and is a good tracer of where most of the gas mass is located.\\[0.1cm]
CH$_3$OH&Shocks&Traces the breakup of the icy mantles during the dust grain destruction process. Indicates the presence\\ && of slower shocks, in contrast to SiO, as the grains do not need to be completely destroyed to produce\\ && this molecule.\\[0.1cm]
 \hline
\end{tabular}\\
\label{table:trace}
\end{table*}
\renewcommand\tabcolsep{6pt}

Identifying and studying the physical conditions of the gas and dust in these jet interaction zones near black holes not only provides information about the impact black holes have on their environments, but also can be used to understand fundamental properties of the central engine and jet launching process. In particular, valuable information on unknown jet properties, most notably, the total jet power, radiative efficiency, jet speed, and the matter content are encoded within the interaction regions (e.g.\ \citealt{mac07,bur59,cas75,heinz06}). These jet properties, which are notoriously difficult to measure through other methods, quantify the matter and energy input into the jets, and thus are crucial in pinpointing how jets are launched and accelerated. For example, jet power estimates from broad-band spectral measurements or minimum energy calculations during time-resolved flaring events \citep{fen19} often rely on assumptions of spectral shape, radiative efficiency, composition,  and the ratio of energy in particles/magnetic field.  Alternatively, one can use jet-driven structures in {the interstellar medium (ISM)} as calorimeters to constrain the power that the jet needs to carry to produce and maintain such a structure in the surrounding medium, without the limitations of these assumptions \citep{gallo05,rus07,sell15}.

Galactic BHXBs offer excellent laboratories in which to study jet interaction zones, as their jets evolve on day to month timescales, they are located at nearby distances, and they are thought to be good analogues for AGN. Given their incredible diagnostic potential, over the last couple of decades there have been many observational campaigns searching for these highly sought after interaction sites near BHXBs. To date, several candidate jet-ISM interaction sites have been identified; SS 433 \citep{dubn98}, Cygnus X--1 \citep{gallo05,rus07}, 1E 1740--2942 \citep{mir92}, GRS 1758$-$258 \citep{marti02}, GRS 1915$+$105 \citep{kai04,rodmir98,chat01}, H1743--322 \citep{co05}, XTE J1550--564 \citep{cor02,karr3,migl17}, XTE J1748--288 \citep{br7}, GRO J1655--40 \citep{hjr95,hann00}, GX 339-4 \citep{gall04}, 4U 1755--33 \citep{kaa06}, XTE J1752--223 \citep{yang10,millerjones11,yang11,ratti12}, XTE J1650--500 \citep{corb04}, XTE J1908$+$094 \citep{rush17}, 4U 1630--47 \citep{neil14,kalem18}, LMC X--1 \citep{rus06b,coo07,hyd17}, and GRS 1009--45 \citep{rus06b}. 
From these past works, it is clear that finding and confirming interaction sites can be incredibly difficult and often observationally expensive (e.g., requiring deep, wide-field radio continuum observations). This difficulty mainly results from the fact that interaction sites can manifest with a wide variety of morphologies and emission properties, likely dependent on the properties of the BHXB (e.g., space velocity; \citealt{millerjones07,heinz08,wie09}) and/or local ISM properties (e.g., density; \citealt{heinz02,kai04}). Once identified, detailed calorimetric calculations of interaction sites are highly sensitive to the properties of the ISM (i.e., density, kinetic temperature, shock velocity; \citealt{rus07,sell15}), and thus require accurate constraints on the physical conditions in the interacting gas, which {cannot} be derived with continuum observations alone. Therefore, developing and implementing new methods that allow us to identify and place improved observational constraints on these parameters at multiple interaction sites, is crucial for taking full advantage of the diagnostic potential of these regions.

We expect jet-ISM interactions to alter the chemistry and excitation conditions of the gas in these regions.
Through observing molecular line emission we can trace the density, temperature, and the presence of a shock in the gas at interaction sites (see Table~\ref{table:trace} for a summary of the tracers sampled in this study; \citealt{molecular,gin15}). For example, CO and HCN emission track where most of the gas mass is located, allowing for the detection of a jet blown bubble or cavity structure in the molecular gas. {Moreover,} the detection of asymmetric line wings in the emission (on the scale of the interaction zone) provides kinematic evidence for the presence of a shock in the surrounding medium. Observing different isotopologues of CO probes the opacity (and density) of the gas. Observing different transitions of the same molecule can probe different temperature regimes (e.g., the minimum gas temperatures need for collisions to excite HNCO ($J=5_{0,5}-4_{0,4}$ vs HNCO ($J=4_{0,4}-3_{0,3}$) is $\sim15/10$ K). Further, observing emission from molecules like SiO, which trace dust-grain destruction, can be used to identify regions of shock excited gas. Telescopes like the Atacama Large Millimeter/Sub-Millimeter Array (ALMA) can observe several of these molecular lines in a single execution, making this technique much more observationally efficient than previous studies and more scaleable to larger surveys of interaction sites.

\begin{figure*}
\begin{center}
  \includegraphics[width=0.48\textwidth,height=7.3cm]{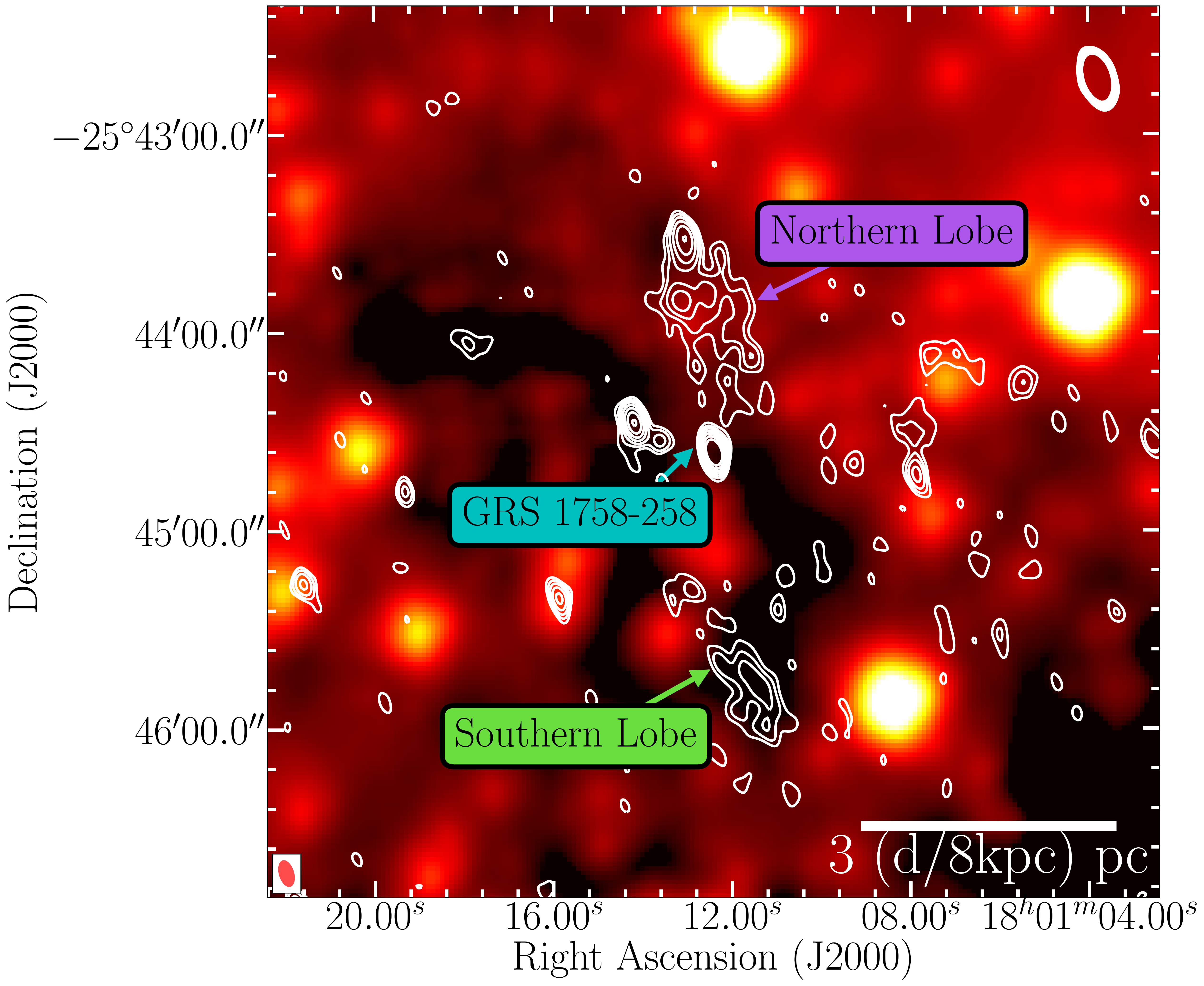}
  \includegraphics[width=0.46\textwidth,height=7.3cm]{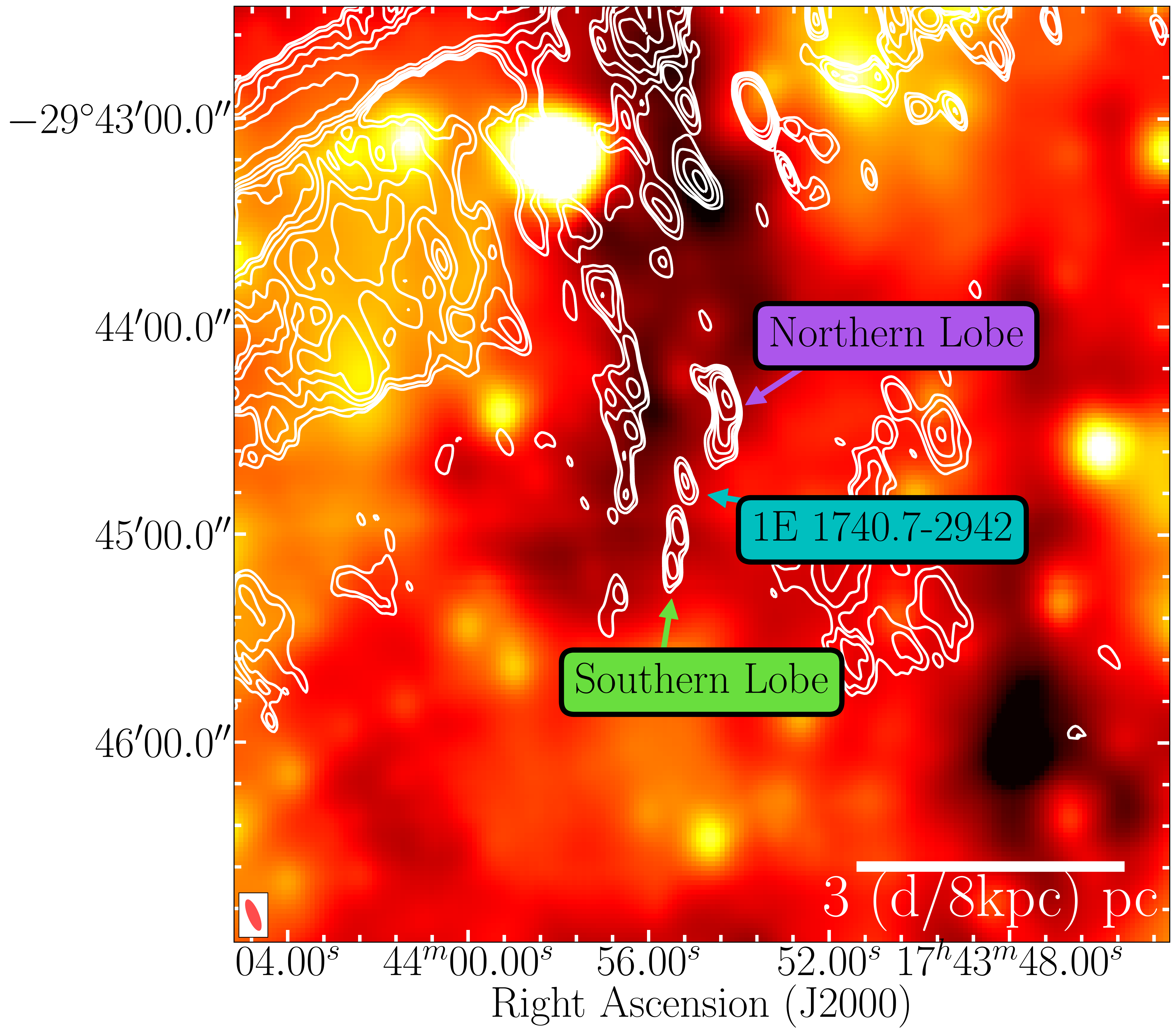}
 \caption{\label{fig:zonesrad2}Maps of the fields surrounding the BHXBs GRS 1758$-$258 (\textit{left}) and 1E 1740.7$-$2942 (\textit{right}), where each panel represents the 4.5 $\times$ 4.5 arcmin fields covered by the ALMA Compact Array.
 The background colour images show the WISE mid-infrared ${12\mu{\rm m}}$ band (W3 Band; see \S\ref{sec:mw}).
 The white contours represent 6 cm archival VLA C-configuration continuum radio maps from observations taken in 2016 March (see \S\ref{sec:vlarad}); contour levels are $2^{n}\,\times$ the rms noise of $3.5\mu{\rm Jy\,bm}^{-1}$ ($n=1.5, 2.0, 2.5, 3.0, 3.5, 4.0, 4.5$) for the \textit{left} panel and $2^{n}\times$ the rms noise of $8\mu{\rm Jy\,bm}^{-1}$ ($n=2.0, 2.5, 3.0, 3.5, 4.0, 5.0, 6.0$) for the \textit{right} panel. The red ellipses in the bottom left corners represent the VLA beams. A physical scale-bar, dependent on the assumed distance to both sources, is shown in both panels. Sources of interest are labelled, including the radio lobe interaction zones and the core jet emission. Bright, diffuse infrared emission is observed in the fields surrounding the candidate interaction zones (i.e., the radio lobes).}
\end{center}
 \end{figure*}

Recently, \cite{tet18i} performed a pilot study with ALMA, utilizing molecular line emission emitting across the sub-mm bands to characterize a candidate jet interaction zone near the BHXB GRS 1915$+$105.
In this study, we found new kinematic and morphological evidence  in the molecular gas supporting an association between the suspected impact site and the GRS 1915$+$105 jet, and we estimated the time-averaged power carried in the GRS 1915$+$105 jets from the ISM conditions at this site. This study also showed that molecular emission may allow us to distinguish between the effects of different forms of feedback in the gas; (i.e., BHXB jet versus the star formation process).
Therefore, by mapping molecular line emission near BHXBs, we can develop several lines of evidence to conclusively identify jet-ISM interaction regions, and accurately probe the ISM conditions at these sites.

Building on the success of this GRS 1915$+$105 study, we obtained new ALMA observations of a suite of molecular lines in the candidate interaction zones near the BHXBs GRS 1758$-$258 and 1E 1740.7$-$2942. With these data, we aim to characterize the properties of the molecular gas at these sites, test whether these candidate zones are consistent with being powered by a BHXB jet, and estimate the energy input into the ISM from these BHXB jets, thereby helping to better quantify the level of BHXB jet feedback in our Galaxy.

\subsection{GRS 1758--258 \& 1E 1740.7--2942}
GRS 1758$-$258 and 1E 1740.7$-$2942 are BHXBs that were discovered in X-rays by the GRANAT and Einstein satellites, respectively \citep{her84,mand90}.
The radio counterparts to these X-ray sources were identified with the Very Large Array (VLA; \citealt{rod92,mirbel2}), both of which revealed double-sided jet structures, leading to their classification as micro-quasars.
The donor stars in both systems have not been clearly identified, due to the high extinction in these regions. However, infrared imaging/spectroscopy observations have suggested several candidates; low-mass stars for GRS 1758$-$258 \citep{marti98} and high-mass stars for 1E 1740.7$-$2942 \citep{marti00,eik01}.
Recently, X-ray reflection model fitting \citep{ste20} has placed constraints on the inclination angle ($>50$ deg) and black hole mass ($5 M_\odot$) for 1E 1740.7$-$2942.
GRS 1758$-$258 and 1E 1740.7$-$2942 are particularly unique in the BHXB population, as they tend to remain persistently in bright outburst states, accreting near their peak luminosity most of the time \citep{tetb16}.

GRS 1758$-$258 and 1E 1740.7$-$2942 are  ideal  sources to target with our molecular tracer technique.  Reminiscent of AGN jet impact sites, they display radio lobes, with a small separation from the central source ($<3$ pc, see Figure~\ref{fig:zonesrad2}; \citealt{mir92,marti02,hard05}). These radio lobes have undergone significant evolution in brightness and morphology over the last 10--15 years \citep{ped15,marti15}, suggesting that the lobes are being powered by a variable source, such as the relativistic jet in the nearby BHXBs.  For instance, near GRS 1758$-$258, a hot-spot in the northern radio lobe moves at a rate of $\sim2\,{\rm arcsec\,yr}^{-1}$ (implying a jet speed of $\sim0.25\, c$; \citealt{marti15}). The lobe evolution in GRS 1758$-$258 has been attributed to hydrodynamic instabilities in the jet, while in 1E 1740.7$-$2942, it is believed that a precessing jet has contributed to the lobe evolution \citep{ped15,marti15}. Through stacking many archival VLA radio data sets, \cite{marti15} have presented evidence of a potential jet-blown cocoon structure surrounding GRS 1758$-$258 and its radio lobes. Further,  \citet{marti17} has also detected a low-surface-brightness Z-shaped feature in the GRS 1758$-$258 lobes (reminiscent of winged radio galaxies), which may be produced by a hydrodynamic back-flow of the jet material in the suspected cocoon structure.
Additionally, both targets have existing sub-mm detections of molecular lines near the candidate impact sites (e.g., CO, HCO+, HCN), which can allow for strong ISM constraints, and 1E 1740.7$-$2942 is believed to be located within or near the edge of a molecular cloud  \citep{lin00,phillaz95,mir91}.
Lastly, these two sources are conveniently close enough on the sky ($<5$ degrees) that they can be observed in a single ALMA observation.

To study the candidate interaction sites near GRS 1758$-$258 and 1E 1740.7$-$2942, we obtained ALMA Compact Array (ACA) and 12m array observations mapping the molecular line emission in the fields surrounding these BHXBs. The ACA was ideal for a first pass at imaging of these sources, as it closely matches the resolution of previous radio observations (${\sim15}\arcsec$; \citealt{ped15,marti15,marti17}) and can cover the largest angular scales in the radio lobes ($50\arcsec$ in extent), while the 12m array provided higher resolution followup observations of key molecular emission regions identified in the ACA data.
In \S\ref{sec:data}, we describe the data collection, reduction process, and imaging procedure. In \S\ref{sec:results}, we present maps of the continuum and molecular line emission in the fields (density tracers: CO, HCN, HCO+, HNCO; and shock tracers: SiO, CS, CH$_3$OH), as well as outline the morphological, spectral, and kinematic properties of this emission.
In \S\ref{sec:discuss}, we discuss the ISM conditions in these regions, what these conditions reveal about the presence of a jet-ISM interaction at these sites, and how these conditions help place constraints on jet properties from both BHXBs. We also present a comparison between these sites and our earlier molecular tracer work on GRS 1915$+$105, in an effort to identify a unique molecular signature of BHXB jet feedback. A summary of our work is presented in \S\ref{sec:sum}.

\section{Observations and data analysis}
\label{sec:data}

\subsection{ALMA sub-mm observations}
\label{sec:almasub}
\subsubsection{ALMA Compact Array (ACA)}
We observed the fields surrounding the BHXBs GRS 1758$-$258 and 1E 1740.7$-$2942 (Project Code: 2017.1.00928.S, PI: A.~Tetarenko) using two different spectral setups with the ACA on 2017 October 31 (22:06:16.2 -- 23:07:55.6 UTC, MJD 58057.9210 -- 58057.9638; hereafter Execution 1) and 2017 December 27 (14:46:43.6 -- 15:37:07.4 UTC, MJD 58114.6157 -- 58114.6507; hereafter Execution 2, see Table~\ref{table:almasetup}). The Band 3 receiver (84--116 GHz) was used for each of these two tunings.
During our observations, the array contained 11/10 antennas, and spent $\sim$22/19 min total on the target sources in Execution 1/2, respectively. We observed two 4.5 arcmin $\times$ 4.5 arcmin mosaic fields centred on the coordinates: (Equatorial J2000) RA $=$ 18:01:12.40, Dec $=$ $-25$:45:01.53 for GRS 1758$-$258 and RA $=$ 17:43:54.83, Dec $=$ $-2$9:44:42.60 for 1E 1740.7$-$2942; (Galactic) $l=4.51$, $b=-1.36$ deg and $l=359.12$, $b=-0.11$ deg, respectively.  Each mosaic field consisted of 27/39 pointings in Execution 1/2. The correlator was set up to yield $4\times2$ GHz wide base-bands, within which we defined 5/3 individual spectral windows centered on our target molecular lines in Execution 1/2, and one continuum spectral window for each execution (see Table~\ref{table:almasetup} for the central frequencies, bandwidth, and resolution of these spectral windows).

All of the data were reduced and imaged within the Common Astronomy Software Application package (\textsc{casa}, version 5.3; \citealt{mc07}).
Flagging and calibration of the data were performed with the ALMA pipeline.
We used J1924--2914 / J1517--2422 as both bandpass and flux calibrators, as well as J1826--2924 / J1744--3116 as phase calibrators for Execution 1/2.
All line imaging was done with the \texttt{tclean} task in \textsc{casa}, using natural weighting to maximize sensitivity, and pixel sizes of $1.8\arcsec$/$1.5\arcsec$ for Execution 1/2 (beam FWHM for Execution 1/2 were $14.4\arcsec$/$11.4\arcsec$). Additionally, we utilized the auto-masking algorithm, \textsc{auto-multithresh}\footnote{See \url{https://casaguides.nrao.edu/index.php/Automasking\_Guide} for details on the \textsc{auto-multithresh} algorithm.} \citep{multithresh}, within the \texttt{tclean} task to automatically mask regions during the cleaning process. When running the \textsc{auto-multithresh} algorithm, we set the following parameter values: ${\rm sidelobethreshold}=1.25$, ${\rm noisethreshold}=3.5$; ${\rm minbeamfrac}=0.1$; ${\rm lownoisethreshold}=2.0$; and ${\rm negativethreshold}=0.0$.
To image the continuum emission, we split out the continuum spectral window in each Execution's data set (using the \textsc{casa} \texttt{split} task) and combined them into a single data set with the \texttt{concat} task. We flagged any channels with clear line emission (determined by examining the data in \texttt{plotms}) and then performed 
multi-frequency synthesis imaging on the flagged continuum data using the \texttt{tclean} task, with natural weighting to maximize sensitivity.

\renewcommand\tabcolsep{1.5pt}
 \begin{table*}
\caption{ALMA Correlator Setup}\quad
\centering
\begin{tabular}{ ccccccc }
 \hline\hline
{\bf Execution}&{\bf Target Line}&{\bf Rest LSRK}&{\bf ACA}&{\bf 12m }&{\bf ACA }&{\bf 12m }\\
&  {\bf }&{\bf Freq. (GHz)}&{\bf Bandwidth}&{\bf Bandwidth}&{\bf Resolution}&{\bf Resolution}\\[0.15cm]
&  {\bf }&{\bf }&{\bf (${\rm \bf km\, s}^{-1}$)}&{\bf (${\rm \bf km\, s}^{-1}$)}&{\bf (${\rm \bf km\, s}^{-1}$)}&{\bf (${\rm \bf km\, s}^{-1}$)}\\[0.15cm]
  \hline

2&HCN ($\nu=0$, $J=1-0$)&88.631601&396&793&0.955&0.826 \\[0.1cm]
2&HCO$^+$ ($\nu=0$, $J=1-0$)&89.18853&396 &788&0.949&0.821\\[0.1cm]
2&SiO ($\nu=0$, $J=2-1$)&86.84696&809 &809&0.974 &0.843\\[0.1cm]
2&HNCO ($\nu=0,\,J=4_{0,4}-3_{0,3}$)&87.92524&799 &799&0.962 &0.832\\[0.1cm]
2&CS ($\nu=0$, $J=2-1$)&97.98095&1434 &1434&0.863 &0.747\\[0.1cm]

1&HNCO ($\nu=0,\,J=5_{0,5}-4_{0,4}$)&109.90575&320 &320&0.770 &0.666\\[0.1cm]
1&C$^{18}$O ($\nu=0$, $J=1-0$)&109.78218&320 &320&0.761 &0.667\\[0.1cm]
1&$^{13}$CO ($\nu=0$, $J=1-0$)&110.20135&320 &320&0.768 &0.664\\[0.1cm]
1&CH$_3$OH ($J=2_{1,1}-1_{1,0}$) / Continuum$^\dagger$&96.75550100 / 98.0&\dots&6118&\dots &47.798\\[0.1cm]
1, 2&Continuum$^\dagger$&100.0&5996&5996&46.843 &46.843\\[0.1cm]
&&108.0&5552&5552& 43.373&43.373\\[0.1cm]
 \hline
\end{tabular}\\
\begin{flushleft}
$^\dagger$ For the continuum, we state the central frequency for each spw in the Rest LSRK Frequency column.
\end{flushleft}
\label{table:almasetup}
\end{table*}
\renewcommand\tabcolsep{6pt}

\subsubsection{ALMA 12m Array}
Following our ACA observations, we obtained higher resolution ALMA 12m array observations of the regions in which we detected significant molecular emission in the ACA data (Project Code: 2019.1.01266.S, PI: A.~Tetarenko).
We observed these fields using two different spectral setups with the ALMA 12m array on 2019 November 24 (18:36:25.2 -- 19:33:58.8 UTC, MJD 58811.7753 -- 58811.8153; hereafter Execution 1) and 2019 December 10 (18:45:17.4 -- 19:33:42.0 UTC, MJD 58827.7815 -- 58827.8151; hereafter Execution 2). As with the ACA observations, the Band 3 receiver (84--116 GHz) was used for each of these two 12m array tunings.
During our 12m observations, the array contained 43 antennas and was in the C43-2/C43-1 array configurations for Execution 1/2. We spent $\sim$16/36 min total on GRS 1758$-$258/1E 1740.7$-$2942 in Execution 1, and $\sim$15/28 min total on GRS 1758$-$258/1E 1740.7$-$2942 in Execution 2. We observed a 2 arcmin $\times$ 2 arcmin mosaic field centred on the coordinates: (J2000) RA $=$ 18:01:11.95, Dec $=$ $-25$:45:33.5 for GRS 1758$-$258 (Galactic: $l=4.49$, $b=-1.37$ deg) and a 1.3 arcmin $\times$ 1.6 arcmin mosaic field centred on the coordinates: (J2000) RA $=$ 17:43:56.1770, Dec $=$ $-2$9:44:48.092 for 1E 1740.7$-$2942 (Galactic: $l=359.12$, $b=-0.11$ deg).  The GRS 1758$-$258 mosaic field consisted of 17/27 pointings in Execution 1/2, while the 1E 1740.7-2942 mosaic field consisted of 7/14 pointings in Execution 1/2. The correlator was set up with a nearly identical setup to the ACA observations; $4\times2$ GHz wide base-bands, within which we defined 5/3 individual spectral windows centered on our target molecular lines in Execution 1/2, and 1/3 continuum spectral windows for Execution 1/2 (see Table~\ref{table:almasetup} for the central frequencies, bandwidth, and resolution of these spectral windows).

All of the 12m array data were reduced and imaged within \textsc{casa}.
Flagging and calibration of the data were performed with the ALMA pipeline.
We used J1924--2914 as a bandpass and flux calibrator, as well as J1744--3116 as a phase calibrator for both executions.
All line imaging was done with the \texttt{tclean} task in \textsc{casa}, through combining the ACA and 12m data, using natural weighting to maximize sensitivity, a pixel size of 0.5\arcsec (beam FWHM for Execution 1/2 were $2.3\arcsec$ / $3.4\arcsec$), and the multi-scale algorithm (scales of $[0,5,10,20,40]\times$ the pixel size). Additionally, similar to the ACA only imaging, we utilized the auto-masking algorithm, \textsc{auto-multithresh}, within the \texttt{tclean} task to automatically mask regions during the cleaning process. When running the \textsc{auto-multithresh} algorithm on the combined the ACA and 12m data, we set the following parameter values: ${\rm sidelobethreshold}=2.0$; ${\rm noisethreshold}=3.5$; ${\rm minbeamfrac}=0.3$, ${\rm lownoisethreshold}=1.5$; and ${\rm negativethreshold}=0.0$.
To image the continuum emission, we followed the same procedure as with the ACA data alone, but this time combined the ACA and 12m data in the \texttt{tclean} task.

\subsection{VLA radio observations}
\label{sec:vlarad}

\renewcommand\tabcolsep{6pt}
 \begin{table}
\caption{VLA Archival Observations Log}\quad
\centering
\begin{tabular}{ lcc}
 \hline\hline
 {\bf Date}&{\bf UTC Time Range}& {\bf MJD Time Range}\\[0.15cm]
  \hline
  2016 March 4&14:55:20 -- 15:55:00&57451.6217 -- 57451.6631\\[0.1cm]
  2016 March 10&15:10:10 -- 16:09:55&57457.6320 -- 57457.6735\\[0.1cm]
  2016 March 11&15:39:36 -- 16:39:20& 57458.6525 -- 57458.6939\\[0.1cm]
  2016 March 21&13:36:55 -- 15:06:30& 57468.5673 -- 57468.6295\\[0.1cm]
  2016 March 22&13:16:45 -- 14:46:25& 57469.5529 -- 57469.6155\\[0.1cm]
 \hline
\end{tabular}\\
\label{table:vlaobs}
\end{table}
\renewcommand\tabcolsep{6pt}

We downloaded and reduced public archival VLA observations of GRS 1758$-$258 and 1E 1740.7$-$2942 (Project Code: 16A-005, PI: J.~Marti). These observations were taken over a series of five days (see Table~\ref{table:vlaobs}) in 2016 March, and consisted of scans on source in the C ($4$--$8\,{\rm GHz}$) band. The array was in its C configuration during the observations. All observations were made with an 8-bit sampler, comprised of 2 base-bands, each with 8 spectral windows of $64\times2$ MHz channels, giving a total bandwidth of 1.024 GHz per base-band. Flagging, calibration, and imaging of the data were carried out within \textsc{casa} using standard procedures outlined in the \textsc{casa} Guides\footnote{\url{https://casaguides.nrao.edu/index.php/Karl\_G.\_Jansky_VLA_Tutorials}.} for VLA data reduction (i.e., a priori flagging, setting the flux density scale, initial phase calibration, solving for antenna-based delays, bandpass calibration, gain calibration, scaling the amplitude gains, and final target flagging). For all VLA  observations, J1331$+$3030 was used as a flux and bandpass calibrator, and J1751--2524 was used as a phase calibrator. We imaged the target sources by combining the data from all executions and using the multi-frequency synthesis algorithm within the \texttt{tclean} task in \textsc{casa}. During imaging, we used two Taylor terms to account for the wide bandwidth, Briggs weighting with a robust parameter of 0.5 to balance sensitivity and angular resolution, the multi-scale clean algorithm (scales of [0, 10, 40] $\times$ the pixel size for GRS 1758$-$258 and [0, 10, 20, 40, 60] $\times$ the pixel size for 1E 1740.7$-$2942, where the pixel size was $0.7\arcsec$ for both) to effectively deconvolve extended emission, and a \textit{uv}-taper of $25\,{\rm k}\lambda$ for GRS 1758$-$258 to better enhance the extended emission (as was done in \citealt{marti17}).
VLA radio maps of GRS 1758$-$258 and 1E 1740.7$-$2942 are shown in Figure~\ref{fig:zonesrad2}.

\subsection{Multi-wavelength observations}
\label{sec:mw}
We searched for observations of our ALMA covered fields surrounding GRS 1758$-$258 and 1E 1740.7$-$2942 taken at other wavelengths with large continuum surveys, finding infrared coverage with NASA WISE  \citep{wri10} and Spitzer GLIMPSE/MIPSGAL \citep{ben03,church09,carey09}. These infrared maps (24, 12, 8, and 5.8 $\mu m$) of the fields surrounding GRS 1758$-$258 and 1E 1740.7$-$2942 are shown in Figures~\ref{fig:zonesrad2}, \ref{fig:mwgrs}, and \ref{fig:mw1e}.

\section{Results}
\label{sec:results}

\subsection{Continuum Emission}
\label{sec:res_cont}
The radio continuum emission in both the GRS 1758$-$258 and 1E 1740.7$-$2942 regions shows a point-source component (coincident with the BHXB X-ray positions) and two extended structures to the north and south of the core (see Figure~\ref{fig:zonesrad2}). These features are interpreted as an unresolved compact jet (also known as a core) and two jet driven lobes in each source.
Additionally, other (likely) unrelated radio emission is seen in both fields; the GRS 1758$-$258 field displays other point sources (the brightest of which is located to the north-east of the core) and the 1E 1740.7$-$2942 field displays extended structures to the north-east and south-west of the core.
The radio continuum emission from the lobe structures in both sources shows a steep spectral index ($\alpha<0$, where $f_\nu\propto\nu^\alpha$),
which is distinct from the flat ($\alpha\sim0$) spectral index observed in the core jet emission \citep{mir92,marti02,hard05}. The steep/flat spectral indices are consistent with partially self-absorbed, non-thermal synchrotron emission from a relativistic plasma, commonly observed from BHXB jets \citep{fen06}.

In GRS 1758$-$258, the radio lobes extend up to $\sim 1.5$ arcmin (corresponding to $\sim3.5$ pc at a distance of 8.0 kpc\footnote{\label{fnote1}See \S\ref{sec:dist} for details on our distance estimates.}) from the central source, and show varying brightness and morphology between the two. In particular, the northern lobe has a higher surface brightness (peak of $\sim57\mu {\rm Jy\,bm}^{-1}$ vs. $\sim28\mu {\rm Jy\,bm}^{-1}$ for north and south lobes, respectively) and is more extended ($\sim55\arcsec\times75\arcsec$ vs. $\sim40\arcsec\times58\arcsec$ for north and south lobes, respectively) when compared to the southern lobe, leading to a higher integrated brightness ($\sim514\mu {\rm Jy}$ vs. $\sim360\mu {\rm Jy}$ for north and south lobes, respectively).

In 1E 1740.7$-$2942, the radio lobes extend up to $35\arcsec$ (corresponding to $\sim1.4$ pc at a distance of 8.0 kpc\textsuperscript{\ref{fnote1}}) from the central source. While both lobes display a similar wavy structure (previously suggested to be indicative of precessional motion; \citealt{ped15}), the northern lobe has a higher surface brightness (peak of $\sim300\mu {\rm Jy\,bm}^{-1}$ vs. $\sim90\mu {\rm Jy\,bm}^{-1}$ for north and south lobes, respectively), is more extended ($\sim15\arcsec\times30\arcsec$ vs. $\sim8\arcsec\times25\arcsec$ for the northern and southern lobes, respectively), and has a higher integrated brightness ($\sim715\mu {\rm Jy}$ vs. $\sim206\mu {\rm Jy}$ for north and south lobes, respectively) when compared to the southern lobe. The varying lobe properties could be an indication of an asymmetric density distribution in the local ISM for both regions (as previously suggested in \citealt{marti15,marti17}).

We do not detect any significant sub-mm continuum emission in our combined ALMA ACA + 12m data, placing $3\sigma$ upper limits of $\sim120\,\mu {\rm Jy\,bm}^{-1}$ and $210\,\mu {\rm Jy\,bm}^{-1}$ for the GRS 1758$-$258 and 1E 1740.7$-$2942 fields, respectively. These non-detections are expected, given that an extrapolation of the radio lobe synchrotron spectrum to sub-mm frequencies predicts flux densities of a few to tens of $\mu {\rm Jy\,bm}^{-1}$ (well below our ALMA continuum detection limits).

The mid-infrared continuum emission ($12-24\mu m$) shows bright, diffuse structures across the entire fields near GRS 1758$-$258 and 1E 1740.7$-$2942 (see Figures~\ref{fig:zonesrad2}, \ref{fig:mwgrs}, and \ref{fig:mw1e}). This type of emission is indicative of the presence of heated gas and dust. 
The near-infrared continuum emission ($\leq8\mu m$) in both fields is dominated by compact point sources, likely arising from stellar emission unrelated to our targets.

\subsection{Molecular Emission}
\label{sec:res_molec}
The radio continuum emission in both the GRS 1758$-$258 and 1E 1740.7$-$2942 fields display intriguing radio lobe structures located on either side of the core jet emission. The location and structure of these lobes suggest that they may be impact sites between the BHXB jet and the local ISM. Depending on the scale of the interaction, we expect to observe signatures of a jet impact in the molecular gas coincident with these lobes, or surrounding the lobes if a cavity is present. Therefore, to search for interaction signatures, we first mapped the molecular emission in a $4.5\arcmin \times4.5\arcmin$ field surrounding each BHXB with the ACA, followed by high resolution followup of the ACA detected emission with the 12m array.

\begin{figure*}
\begin{center}
  \includegraphics[width=0.34\textwidth,height=5.8cm]{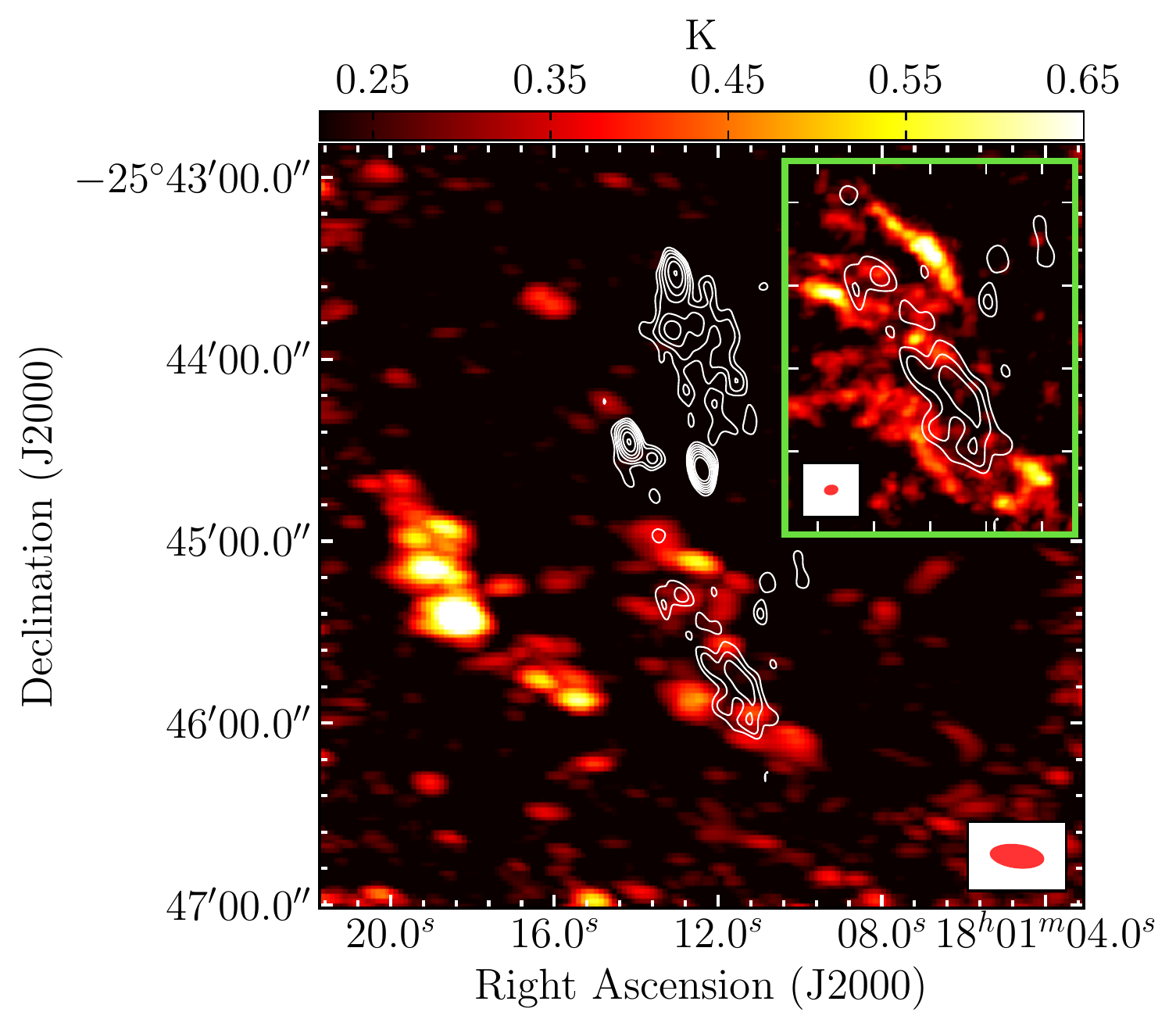}
  \includegraphics[width=0.31\textwidth,height=5.8cm]{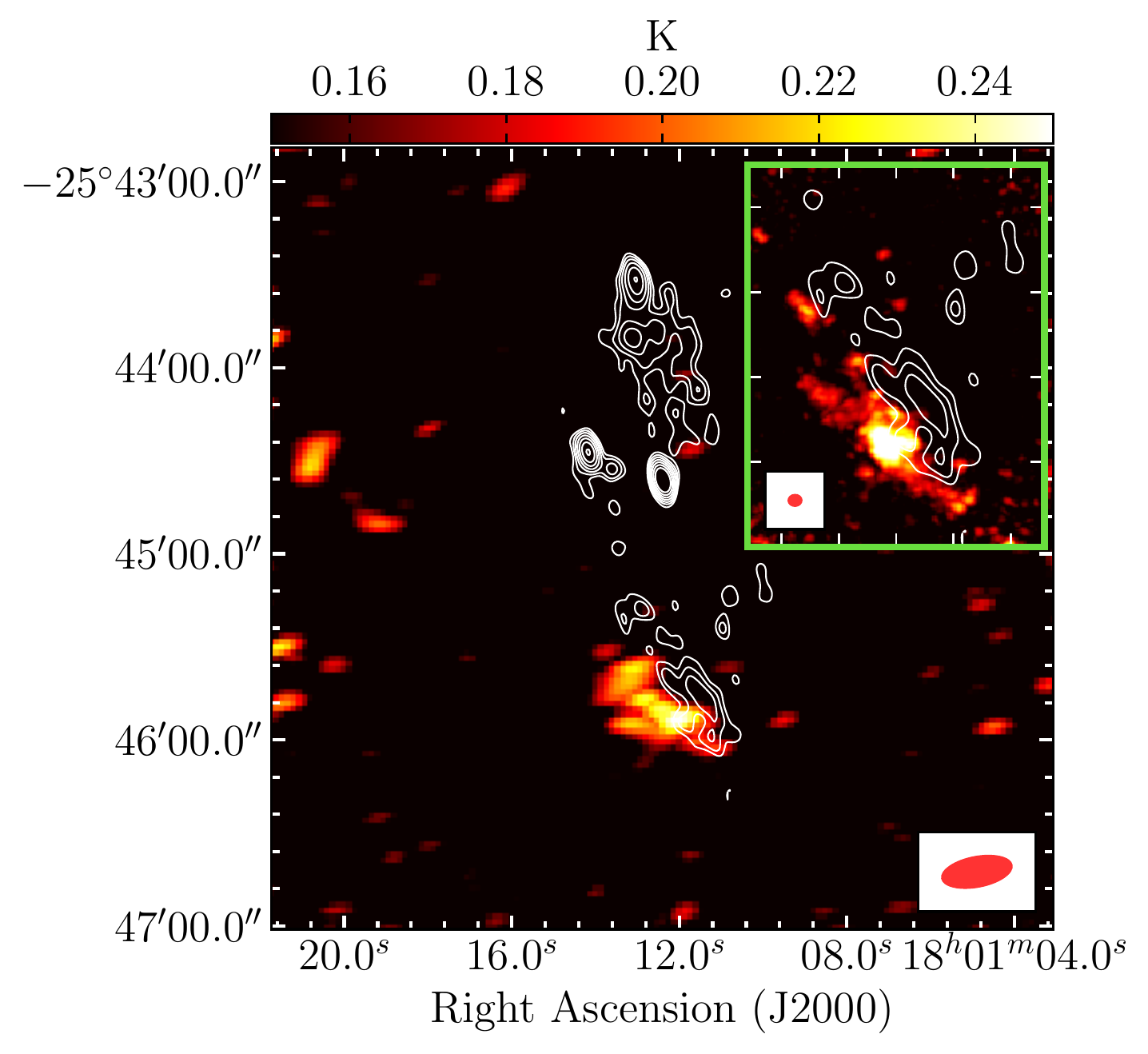}
  \includegraphics[width=0.31\textwidth,height=5.8cm]{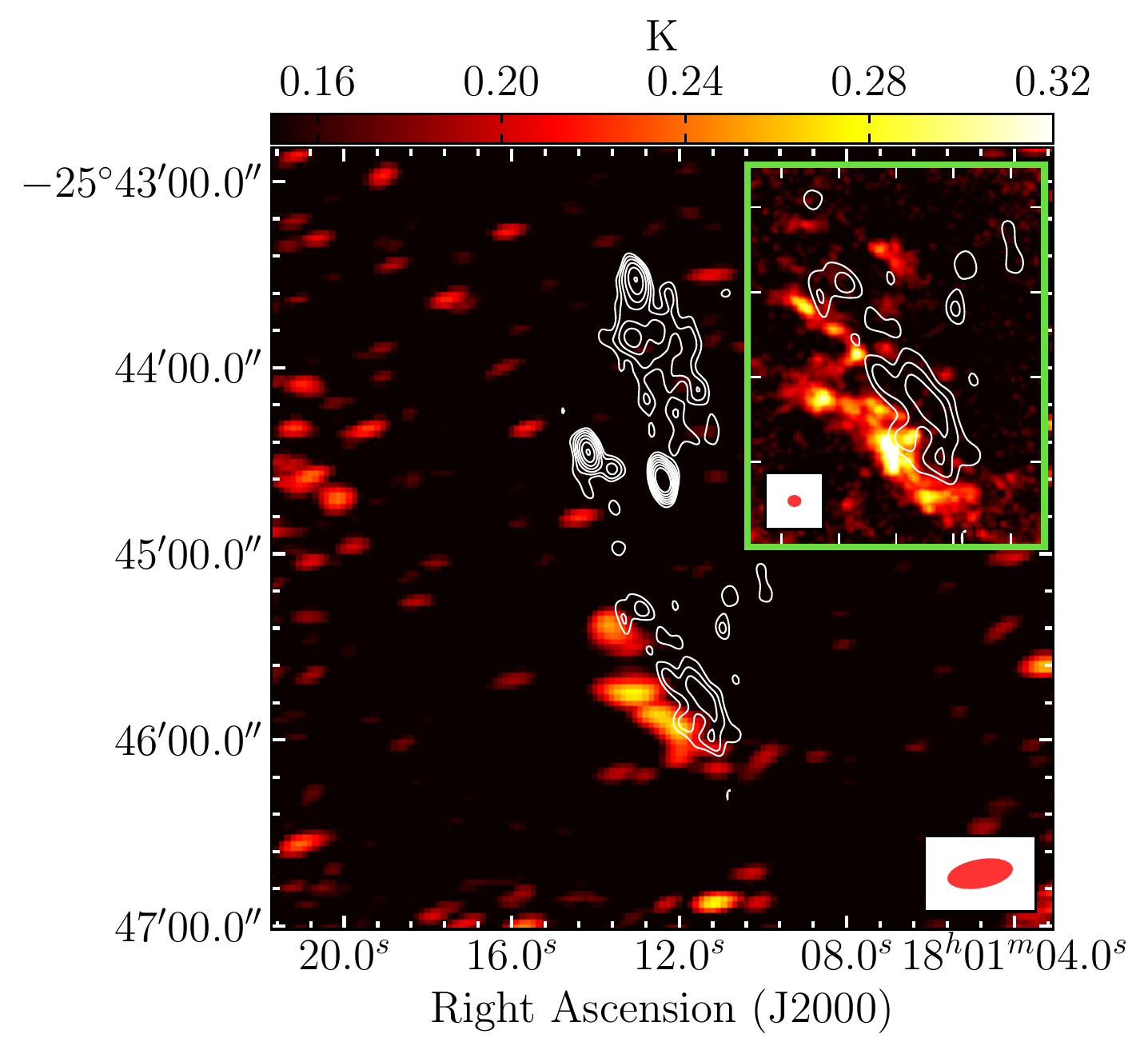}
 \caption{\label{fig:tmax_grs}Maximum intensity maps of the field surrounding the BHXB GRS 1758$-$258. The main panels display the ACA data alone, while the inset panels display the combination of ACA + 12m array data for the southern lobe region. Here we show emission from the $^{13}$CO (density tracer; \textit{left}), HCN (density tracer; \textit{middle}), and CS (shock and density tracer; \textit{right}) molecules, in the velocity range $50-100\,{\rm km\,s}^{-1}$ (in units of Kelvin).
The colour scale represents the intensity of the molecular emission (the colour bar range for the inset panels have the same lower limits as the main panel, but upper limits of 1.5, 0.45, and 0.8 K for panels \textit{left} through \textit{right}, respectively), while the white contours represent continuum radio emission (masked to only show the radio lobes, contour levels are $2^{n}\times$ the rms noise of $3.5\mu{\rm Jy\,bm}^{-1}$, where $n=1.5, 2.0, 2.5, 3.0, 3.5, 4.0, 4.5$; see \S\ref{sec:vlarad}). The red ellipses indicate the ALMA beams.
In these ALMA data, while we do not detect any significant molecular emission in the northern lobe, we identify a bright molecular structure in both density and shock tracing molecules at the southern lobe, possibly tracing the edge of a jet-blown cavity.}
\end{center}
 \end{figure*}

In the GRS 1758$-$258 field, we detect emission from the $^{13}$CO (density tracer), HCN (density tracer), CS (shock tracer), and SiO (shock tracer; only seen in the combined ACA + 12m data) molecules in the velocity range of 50--100 ${\rm km\, s}^{-1}$. Figures~\ref{fig:tmax_grs} and \ref{fig:specgrs_SiO} display the maximum intensity maps of these molecules. 
In these maps, we observe a bright molecular structure, in both density and shock tracing molecules, coincident with the eastern edge of the southern lobe. This molecular emission appears to be aligned along the edge of the potential jet-blown cavity suggested by \citet{marti15}, and possibly along one end of the Z-shaped back-flow identified by \citet{marti17}. We do not detect any significant molecular emission coincident with or surrounding the northern lobe. Nonetheless, we note that \cite{marti17} report a molecular cloud (in the \cite{dame01} Galactic plane CO survey) located near the northern lobe, with a central velocity of 220 ${\rm km\, s}^{-1}$ (unfortunately outside the velocity range covered by our ALMA observations). However, we also note that this cloud could simply be a chance alignment, located in the foreground or background relative to the BHXB.
Our new ALMA line detections are consistent, in terms of central velocities (70--80 ${\rm km\, s}^{-1}$) and peak intensities ($<1$ K), with past lower resolution IRAM/SEST molecular detections achieved from averaging spectra over many positions along the jet axis \citep{lin00}.

\begin{figure*}
\begin{center}
  \includegraphics[width=0.34\textwidth,height=5.8cm]{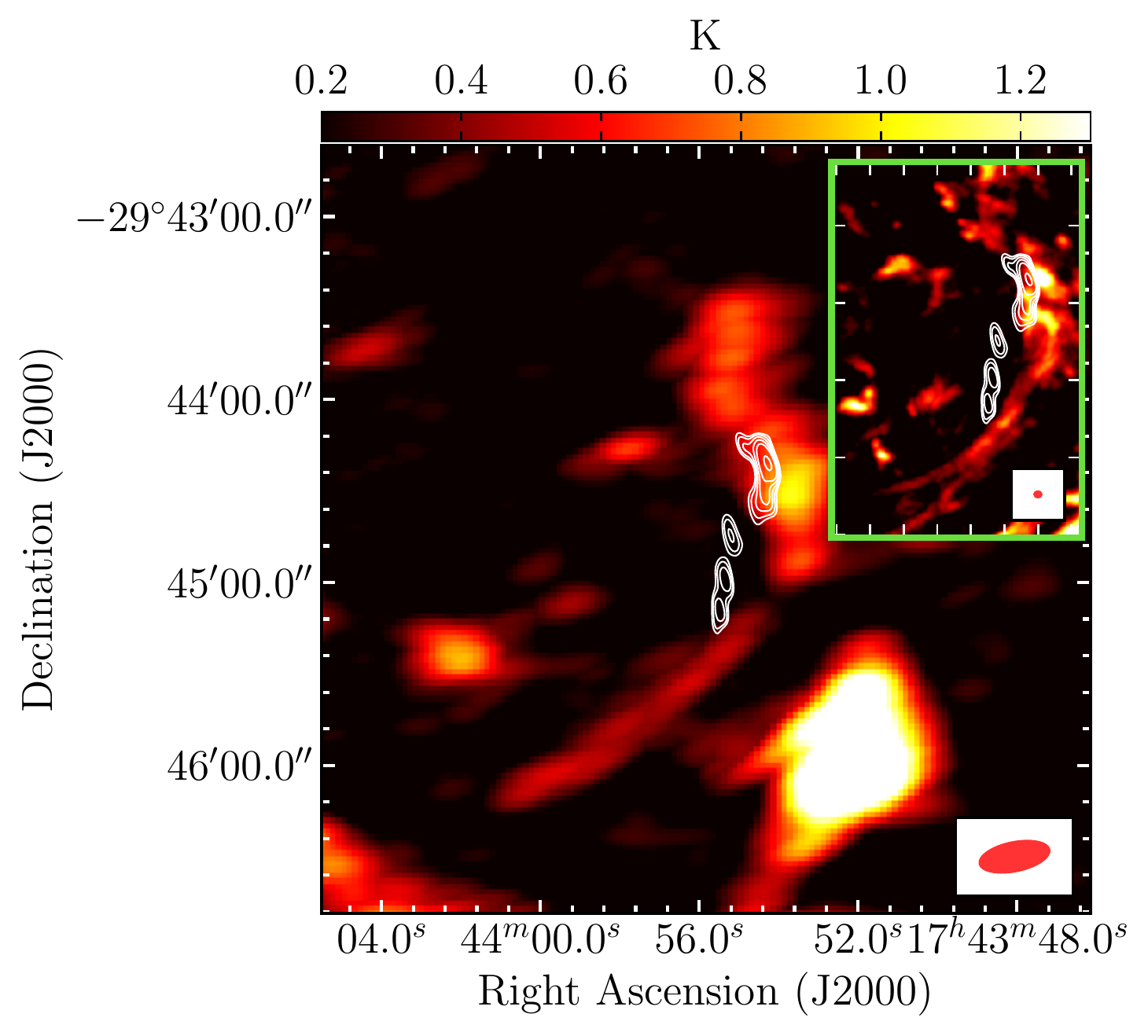}
  \includegraphics[width=0.31\textwidth,height=5.8cm]{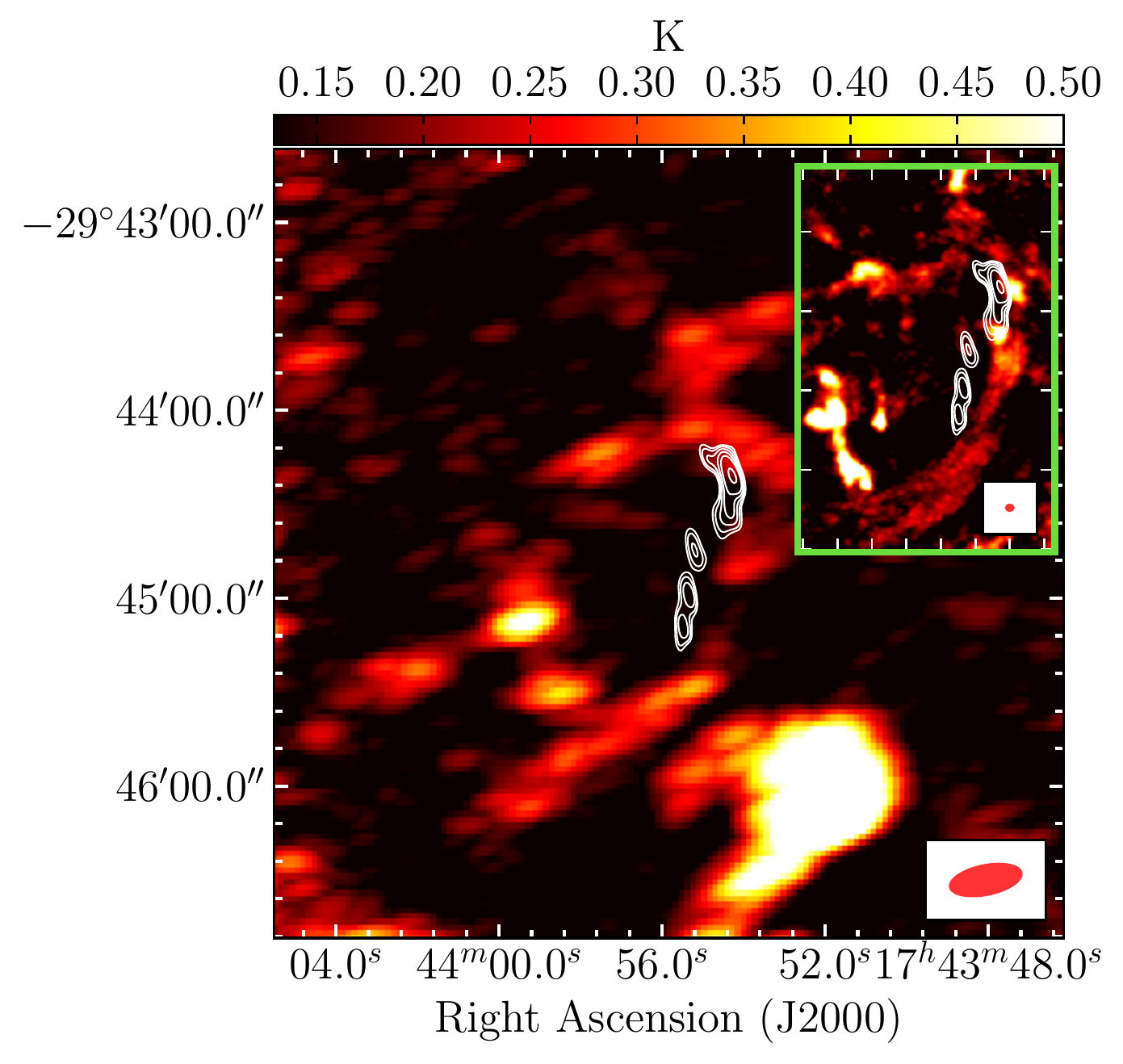}
  \includegraphics[width=0.31\textwidth,height=5.8cm]{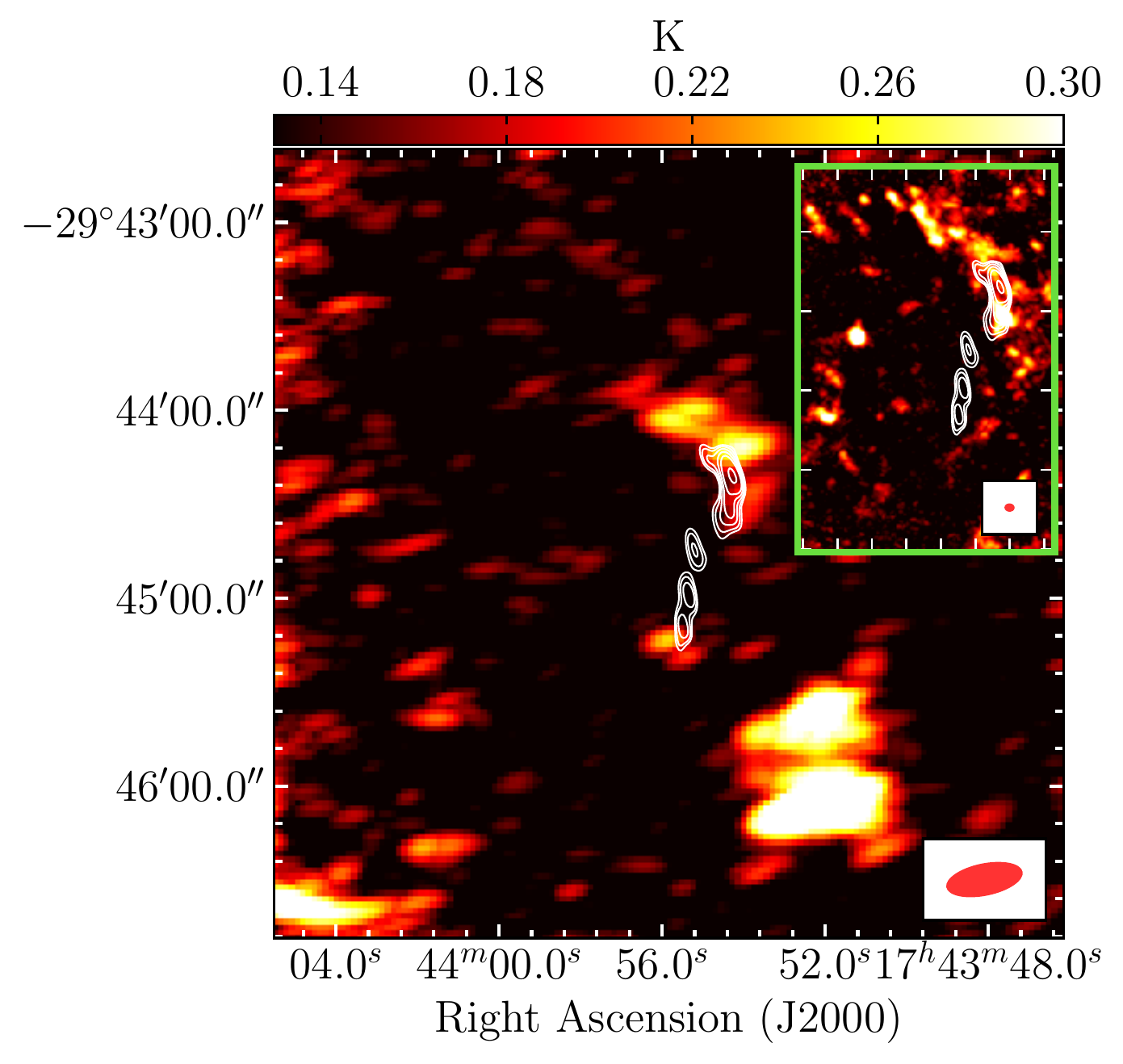}
 \caption{\label{fig:tmax_1E}Maximum intensity maps of the field surrounding the BHXB 1E 1740.7$-$2942. The main panels display the ACA data alone, while the inset panels display the combination of ACA + 12m array data. Here we show the HCN (density tracer; \textit{left}), HCO$+$ (density tracer; \textit{middle}), and SiO (shock tracer; \textit{right}) molecules, in the velocity range $0-60\,{\rm km\,s}^{-1}$ (in units of Kelvin). 
The colour scale represents the intensity of the molecular emission (the colour bar range for the inset panels have the same lower limits as the main panel, but upper limits of 2.0, 0.85, and 0.5 K for panels \textit{left} through \textit{right}, respectively), while the white contours represent continuum radio emission (masked to only show the radio lobes, contour levels are $2^{n}\times$ the rms noise of $8\mu{\rm Jy\,bm}^{-1}$, where $n=2.0, 2.5, 3.0, 3.5, 4.0, 5.0, 6.0$; see \S\ref{sec:vlarad}). The red ellipses indicate the ALMA beams.
In these ALMA data, we identify an elongated molecular ring structure surrounding the central BHXB, possibly tracing a jet-blown cavity, and shock tracing emission coincident with the radio lobes aligned with the jets.}
\end{center}
 \end{figure*}

In the 1E 1740.7$-$2942 field, we detect emission from the HCN (density tracer), HCO$^+$ (density tracer), SiO (shock tracer), CS (shock tracer), CO (density tracer), HNCO (density tracer), and CH$_3$OH (shock tracer; only seen in the 12m data, as the ACA did not sample the corresponding frequency range) molecules in the velocity range of 0--60 ${\rm km\, s}^{-1}$. Figure~\ref{fig:tmax_1E} displays maximum intensity maps of the HCN, HCO$^+$, and SiO molecules, with the remainder of the molecules shown in Figures~\ref{fig:spec1E_CS} -- \ref{fig:spec_CH3OH} in Appendix~\ref{sec:appenmol}. In these maps, we observe a ring-shaped molecular structure encircling the central BHXB in density tracing molecules (HCN, HCO$^+$, and HNCO, but oddly this structure is not seen in CO emission), which could be tracing the edges of a jet-blown cavity in the molecular gas. 
We also note that the BHXB is not centrally located in this ring structure, but rather offset to the west of centre. This offset could be due to the peculiar velocity of the BHXB over its lifetime (see \S\ref{sec:calorimetry} and \S\ref{sec:others} for details). We also observe shock tracing molecules (SiO, CS, CH$_3$OH) in this velocity range coincident with the brightest radio hot-spots in the northern and southern lobes, and possibly along the ring-shaped structure as well.
Further, in addition to the 0--60 ${\rm km\, s}^{-1}$ emission, we identify an isolated molecular cloud (in the HCO$+$, HCN, HNCO, and CS molecules) at a central velocity of $-140 {\rm \, km\, s}^{-1}$, located to the north-east of the radio lobe structures (see Figure~\ref{fig:spec_alt} in Appendix~\ref{sec:appenmol}). Our new ALMA line detections are consistent with past lower resolution IRAM molecular detections in this region, which also show HCO$+$ and CS components in the 0--60 ${\rm km\, s}^{-1}$ velocity range and near $-140 {\rm \, km\, s}^{-1}$, all with peak intensities $<1$ K \citep{mir91}.


\subsection{Properties of the Molecular Gas}
\label{sec:res_molec_prop}
\begin{figure*}
\begin{center}
  \includegraphics[width=0.7\textwidth]{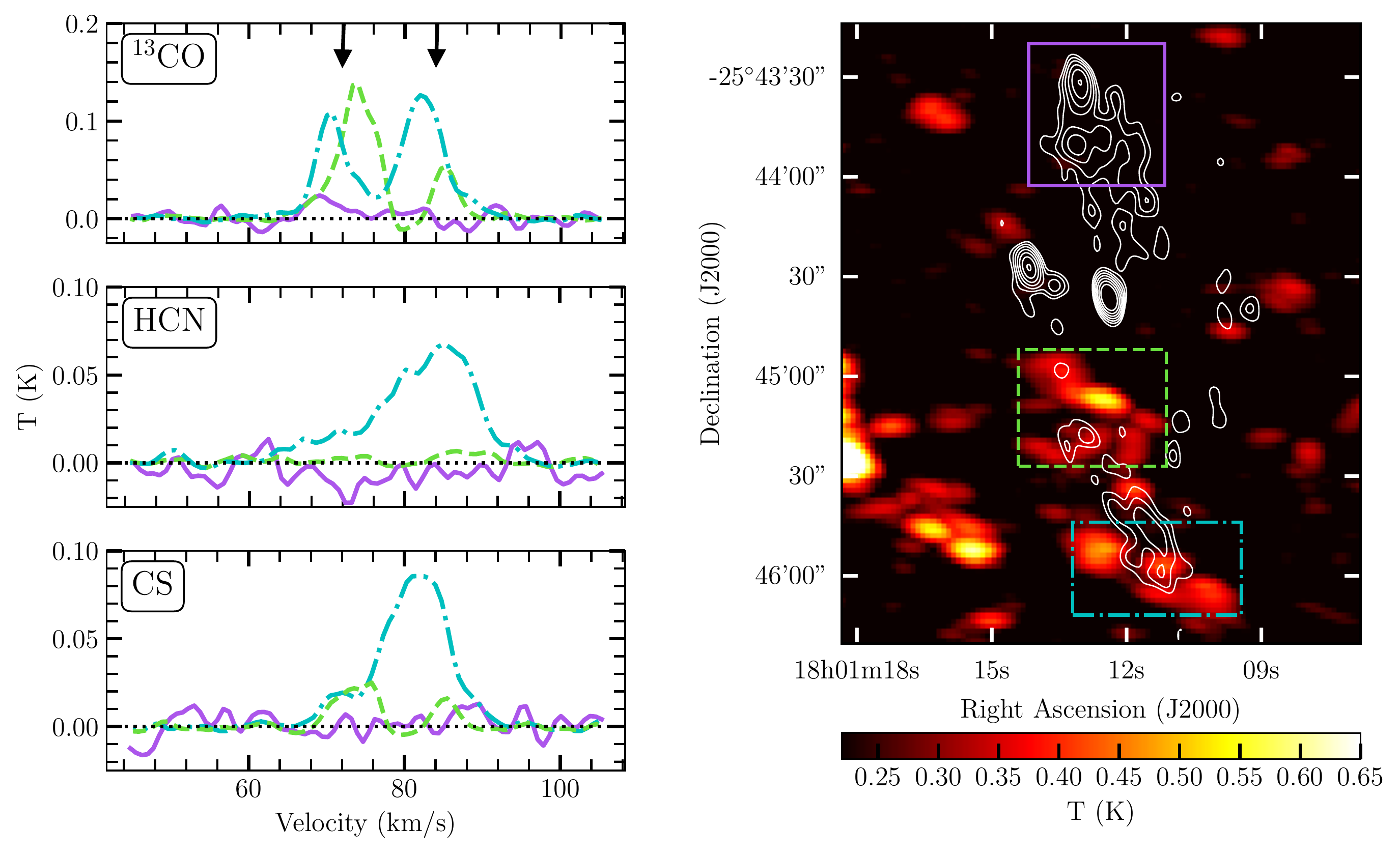}
 \caption{\label{fig:spec_grs} Spectra of the molecular emission detected in the field surrounding the BHXB GRS 1758$-$258. The \textit{left} three panels display spectra of the $^{13}$CO, HCN, and CS molecules. The \textit{right} panel displays the spectral extraction regions (boxes), on top of a maximum intensity map of the $^{13}$CO emission from ACA data only (background colourmap and colourbar indicate intensity in Kelvin) and radio continuum emission (white contours masked to only show the radio lobes; levels are $2^{n}\times$ the rms noise of $3.5\mu{\rm Jy\,bm}^{-1}$, where $n=1.5, 2.0, 2.5, 3.0, 3.5, 4.0, 4.5$; see \S\ref{sec:vlarad}). The spectra in the northern lobe region (purple/solid line box) are extracted from ACA data only (as our 12m data do not cover the northern lobe), while the spectra in the southern lobe regions (green/dashed and cyan/dash-dotted line boxes) are extracted from the combined ACA + 12m data.  We identify two multi-peaked components in the southern lobe (green/dashed and cyan/dash-dotted line boxes, marked by the black arrows), with no molecular emission detected in the northern lobe (purple/solid line box). }
\end{center}
 \end{figure*}

\begin{figure*}
\begin{center}
  \includegraphics[width=0.7\textwidth]{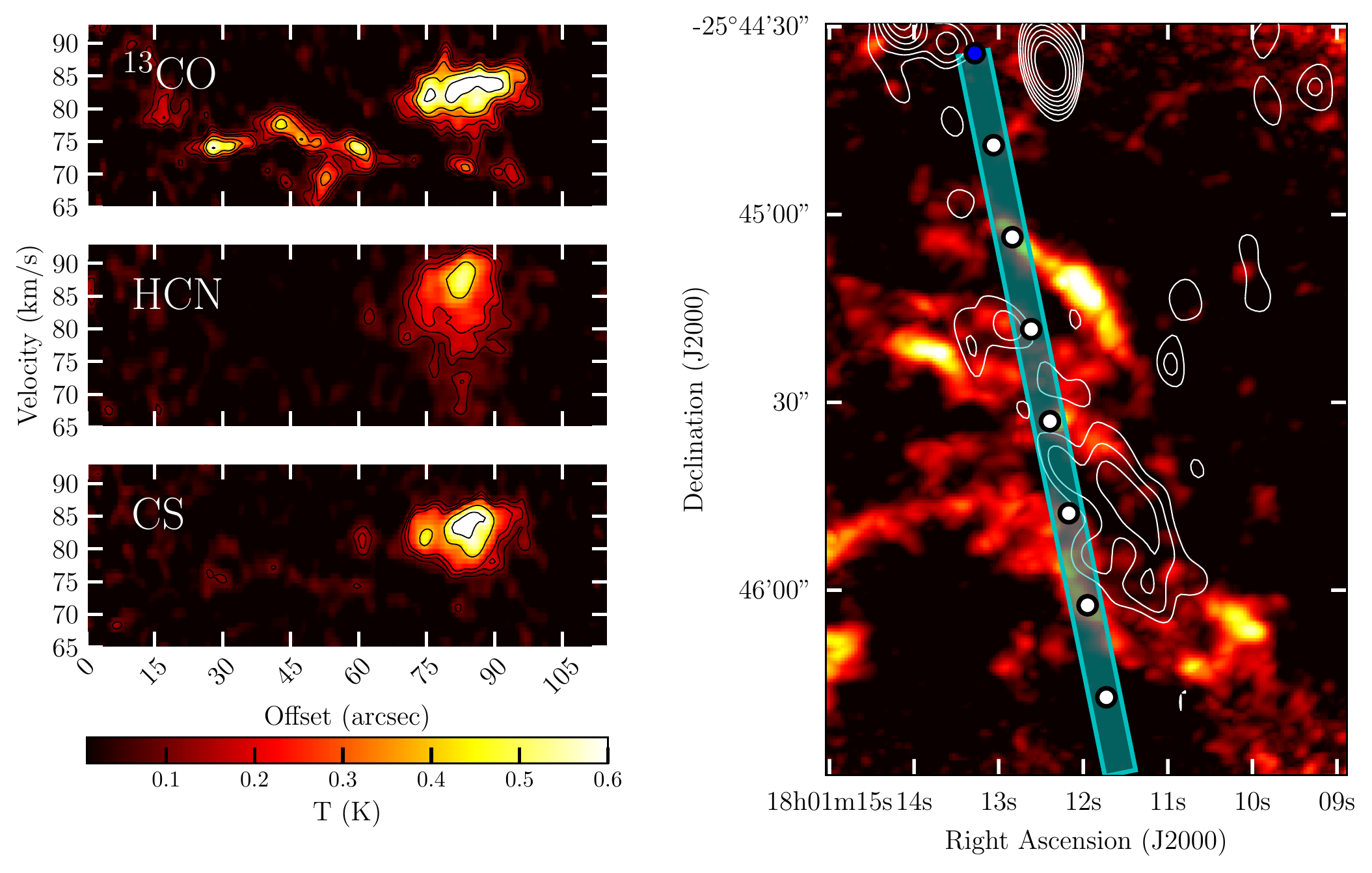}
 \caption{\label{fig:pv_grs} Kinematic analysis of the molecular emission through a slice along the eastern edge of the southern radio lobe in the field surrounding GRS 1758$-$258. The \textit{left} three panels display position-velocity diagrams of the $^{13}$CO (contour levels of 0.1, 0.15, 0.2, 0.3, 0.4, 0.6 K), HCN (contour levels of 0.1, 0.15, 0.2, 0.4 K), and CS (contour levels of 0.1, 0.15, 0.2, 0.4, 0.6 K) molecules, using the ACA + 12m data combined. The \textit{right} panel indicates the position-velocity slice (cyan rectangle; the blue dot indicates the zero point of the position-velocity slice, and the white dots indicate the offset along the slice in increments of $15\arcsec$) on top of the $^{13}$CO maximum intensity map (ACA + 12m data) and the radio continuum emission (white contours masked to only show the radio lobes; levels are $2^{n}\times$ the rms noise of $3.5\mu{\rm Jy\,bm}^{-1}$, where $n=1.5, 2.0, 2.5, 3.0, 3.5, 4.0, 4.5$; see \S\ref{sec:vlarad}). At an offset of 45--$75\arcsec$ along the slice, the $^{13}$CO emission appears to be spectrally displaced, being pushed to lower velocities when compared to the neighbouring emission. Additionally, the HCN and CS emission in this same region span a wide range in velocity space.}
\end{center}
 \end{figure*}

\begin{figure}
\begin{center}
\quad\quad\quad\includegraphics[width=0.38\textwidth]{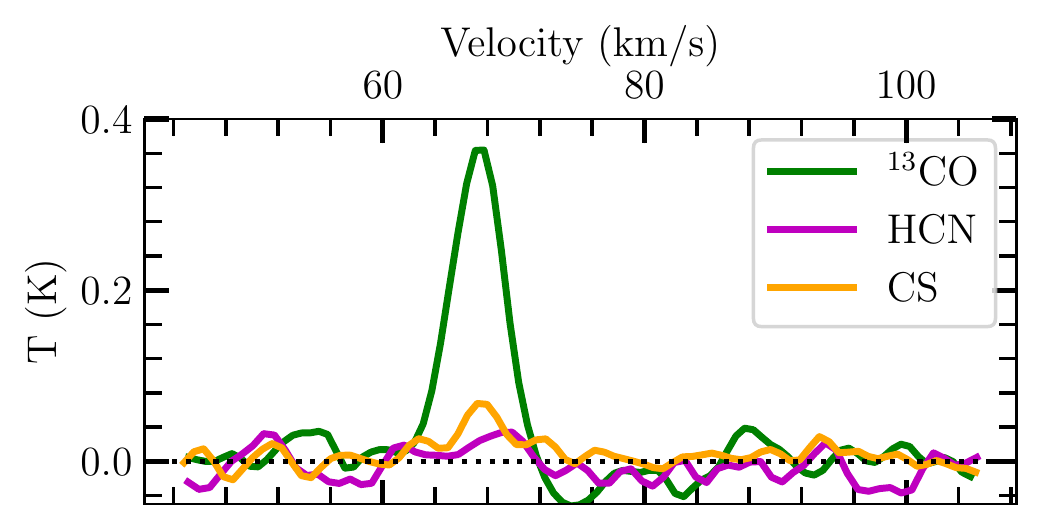}
  \includegraphics[width=0.45\textwidth]{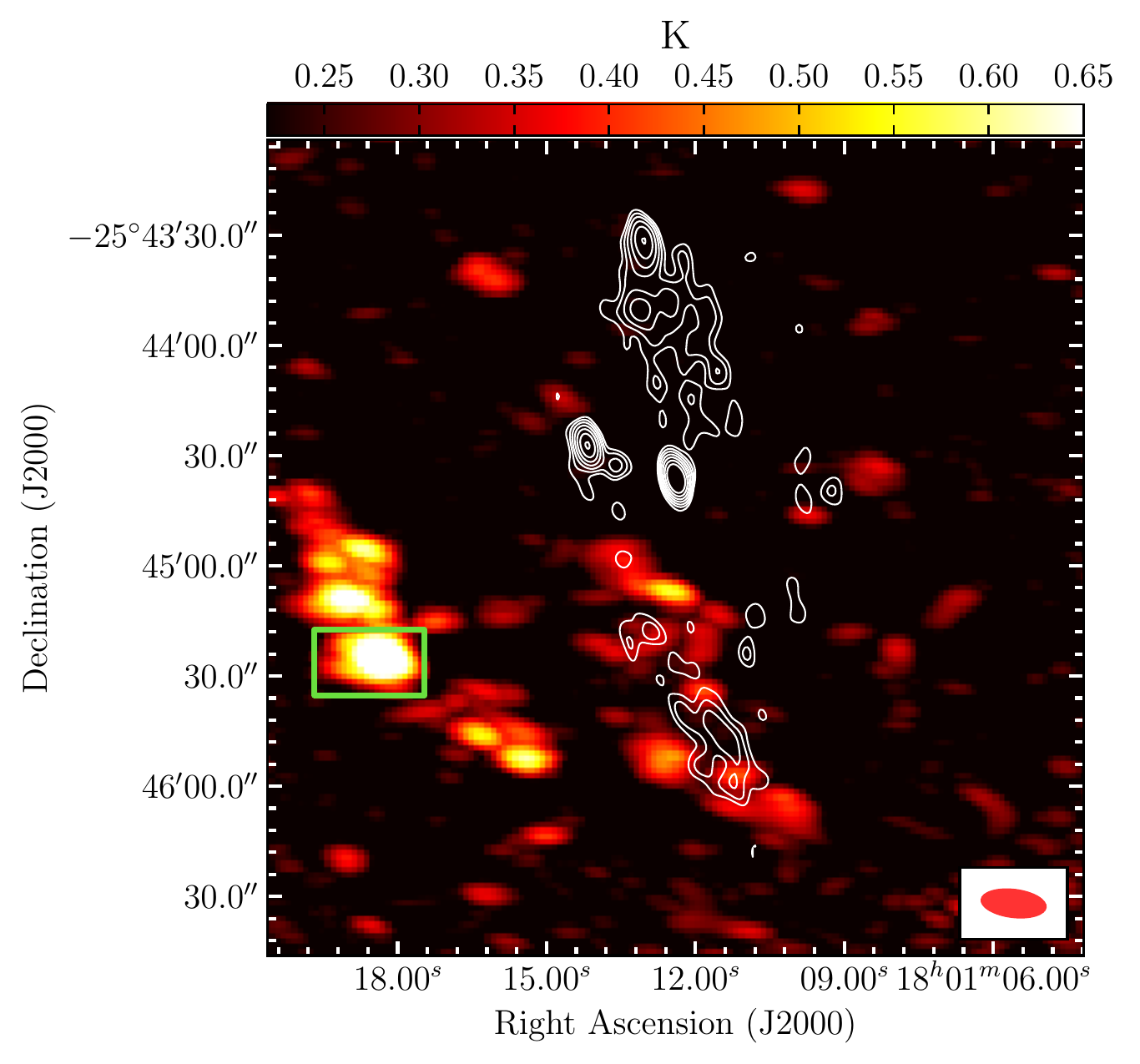}\\
 \caption{\label{fig:grs_alt} Comparison spectra of molecular emission in the GRS 1758$-$258 field (in the velocity range $40$ to $110\,{\rm km\,s}^{-1}$), away from the suspected interaction site and unlikely associated with GRS 1758$-$258. The \textit{bottom} panel displays a maximum intensity map of the $^{13}$ CO molecule (ACA data only), where the colour scale represents the intensity of the molecular emission, while the white contours represent continuum radio emission (masked to only show the radio lobes; contour levels are $2^{n}\times$ the rms noise of $3.5\mu{\rm Jy\,bm}^{-1}$, where $n=1.5, 2.0, 2.5, 3.0, 3.5, 4.0, 4.5$; see \S\ref{sec:vlarad}). The red ellipse indicates the ACA beam. The \textit{top} panel displays the spectra of the emission in the spectral extraction region indicated by the green square, from ACA data only (as our 12m data does not cover this region).
 $^{13}$CO is the only molecule detected away from the suspected interaction site, and this emission shows a single component line structure.}
\end{center}
 \end{figure}

\begin{figure*}
\begin{center}
  \includegraphics[width=0.7\textwidth]{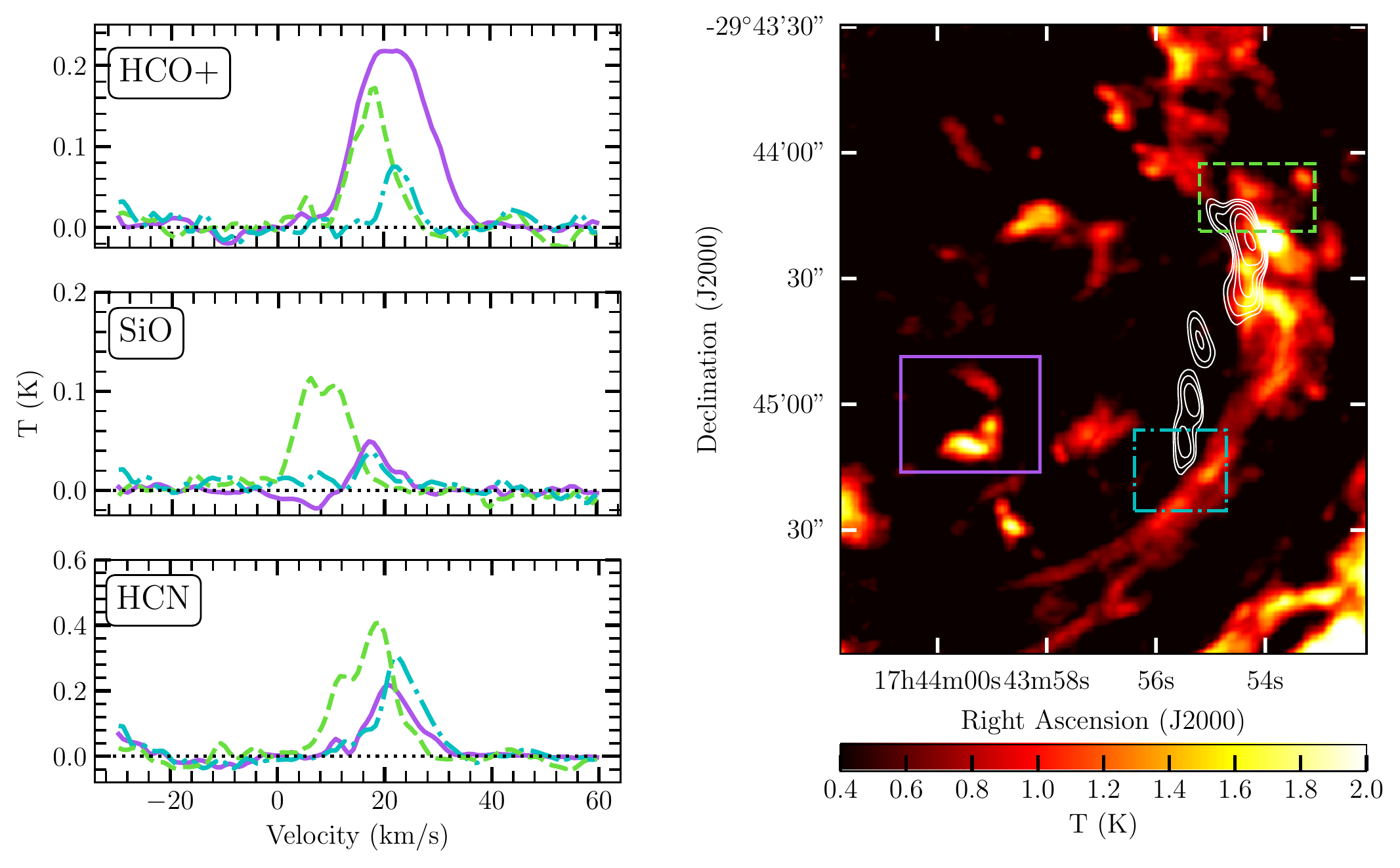}
 \caption{\label{fig:spec_1E}Spectra of the molecular emission detected in the field surrounding the BHXB 1E 1740.7$-$2942. The \textit{left} three panels display spectra of the HCO$+$, SiO, and HCN molecules. The \textit{right} panel displays the spectral extraction regions (boxes), on top of a maximum intensity map of the HCN emission from ACA + 12m data (background colourmap and colourbar indicate the intensity in Kelvin) and radio continuum emission (white contours masked to only show the radio lobes; levels are $2^{n}\times$ the rms noise of $8\mu{\rm Jy\,bm}^{-1}$, where $n=2.0, 2.5, 3.0, 3.5, 4.0, 5.0, 6.0$; see \S\ref{sec:vlarad}). The spectra in all regions are extracted from the ACA + 12m data. In all of the molecules, the lines peak at a lower velocity on the northern edge of the molecular ring (green/dashed line box) and at a higher velocity on the southern edge of the ring (purple/solid and cyan/dash-dotted line boxes), possibly indicating a velocity gradient across the ring and radio lobe structures. }
\end{center}
 \end{figure*}

\begin{figure*}
\begin{center}
  \includegraphics[width=0.7\textwidth]{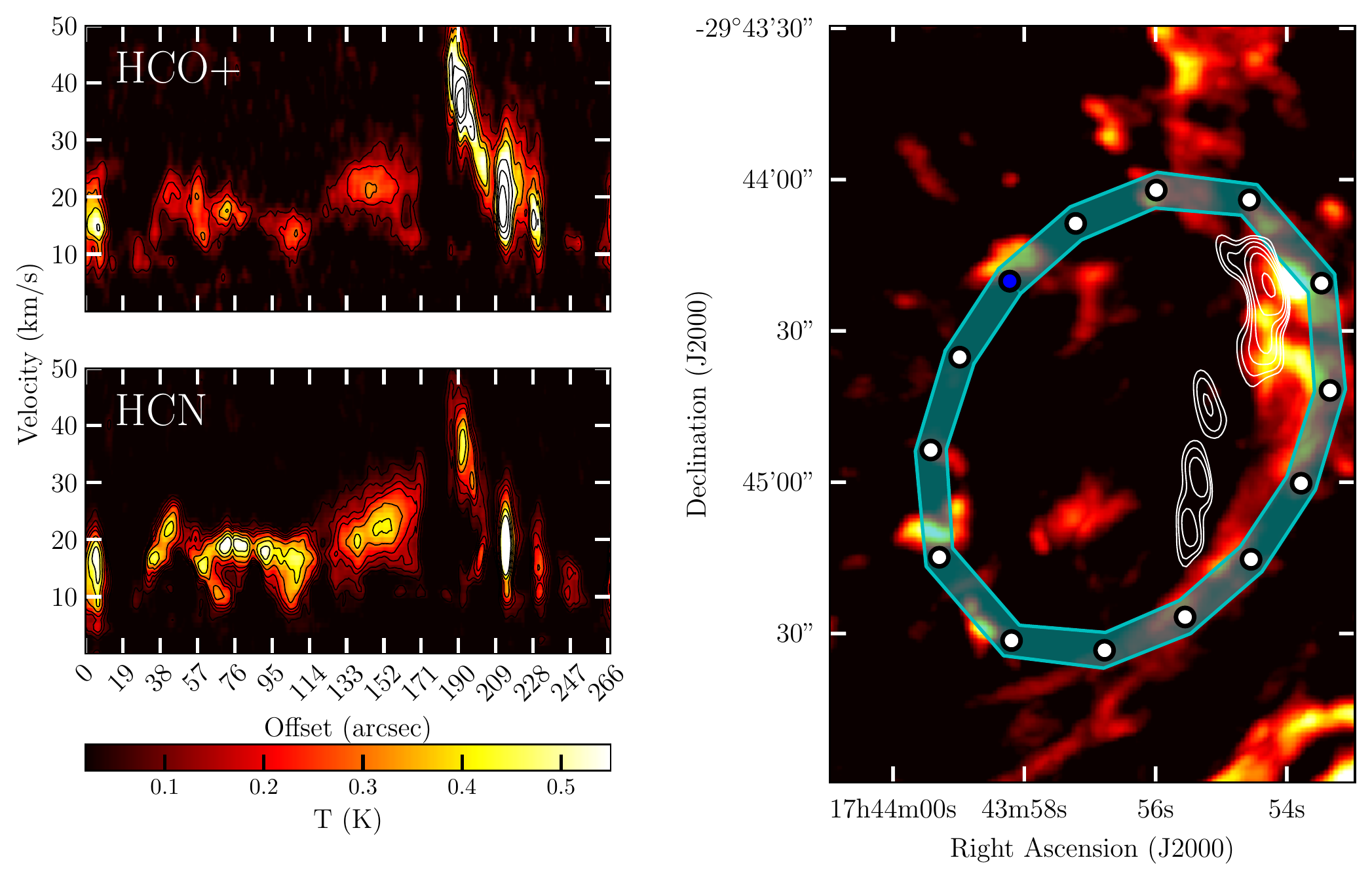}
 \caption{\label{fig:pv_1E} Kinematic analysis of the molecular emission through a slice around the ring structure surrounding 1E 1740.7$-$2942. The \textit{left} two panels display position-velocity diagrams of the HCO$+$ (contour levels of 0.1, 0.2, 0.3, 0.4, 0.6, 0.8, 1 K) and HCN (contour levels of 0.2, 0.4, 0.6, 0.8, 0.12, 0.16, 2 K; intensity colour map is scaled down by a factor of 2 here) molecules, using the ACA + 12m data combined. The \textit{right} panel indicates the position-velocity slice (cyan ring; the blue dot indicates the zero/end point of the position-velocity slice, and the white dots indicate the offset along the slice moving clockwise in increments of $19\arcsec$) on top of the HCN maximum intensity map (ACA + 12m data) and the radio continuum emission (white contours masked to only show the radio lobes; levels are $2^{n}\times$ the rms noise of $8\mu{\rm Jy\,bm}^{-1}$, where $n=2.0, 2.5, 3.0, 3.5, 4.0, 5.0, 6.0$; see \S\ref{sec:vlarad}). These position-velocity diagrams display a velocity gradient across the ring, where the gas on the northern edge of the ring is being pushed to lower velocities (U shape at $19\arcsec$ offset) and the gas on the southern edge is being pushed to higher velocities (upside down U shape at $190\arcsec$ offset). This kinematic structure may be consistent with the ring tracing an expanding bubble structure.}
\end{center}
 \end{figure*}

\begin{figure}
\begin{center}
\quad\quad\quad\includegraphics[width=0.38\textwidth]{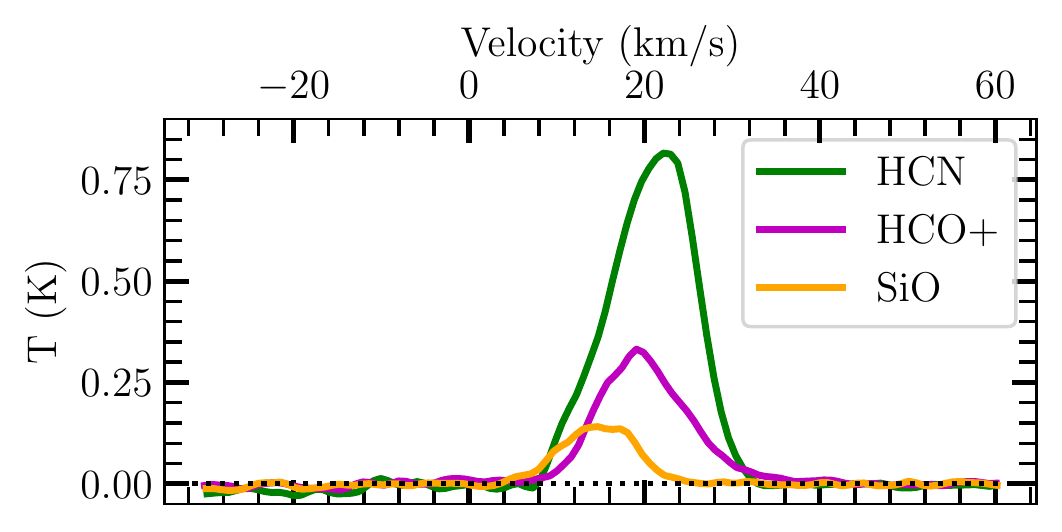}
  \includegraphics[width=0.45\textwidth]{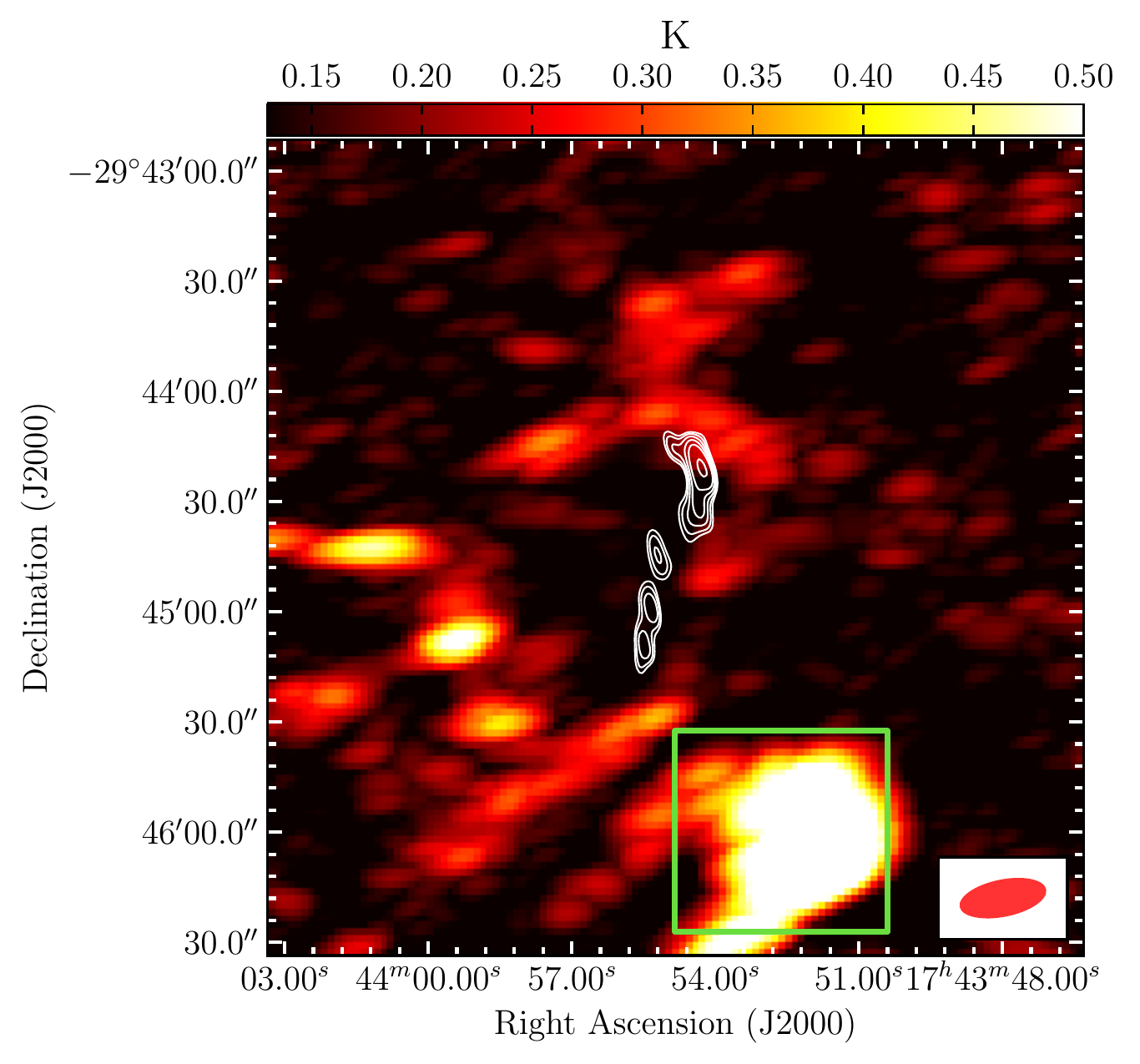}\\
 \caption{\label{fig:1e_alt}Comparison spectra of molecular emission in the 1E 1740.7$-$2942 field (in the velocity range $-30$ to $60\,{\rm km\,s}^{-1}$), away from the suspected interaction site. The \textit{bottom} panel displays a maximum intensity map of the HCO$+$ molecule (ACA data only), where the colour scale represents the intensity of the molecular emission, while the white contours represent continuum radio emission (masked to only show the radio lobes; contour levels are $2^{n}\times$ the rms noise of $8\mu{\rm Jy\,bm}^{-1}$, where $n=2.0, 2.5, 3.0, 3.5, 4.0, 5.0, 6.0$; see \S\ref{sec:vlarad}). The red ellipse indicates the ACA beam. The \textit{top} panel displays the spectra of the emission in the spectral extraction region indicated by the green square, from ACA data only (as our 12m data does not cover this region).
All the molecules detected away from the suspected interaction site show a single component line structure, with the HCN emission showing much brighter peak temperatures than HCO$+$ or SiO.}
\end{center}
 \end{figure}

In \S\ref{sec:res_molec}, we identified several intriguing molecular structures that may be connected to jet impact sites near the BHXBs GRS 1758$-$258 and 1E 1740.7$-$2942. 
The GRS 1758$-$258 region displays two molecular components along our line of sight; 50--100 ${\rm km\, s}^{-1}$ structures coincident with the southern lobe (identified in our new ALMA data), and an isolated 220 ${\rm km\, s}^{-1}$ cloud coincident with the northern lobe (identified in \citealt{marti17}). Similarly, the 1E 1740.7$-$2942 region displays 0--50 ${\rm km\, s}^{-1}$ structures surrounding and coincident with the radio lobes, and an isolated $-140  {\rm \, km\, s}^{-1}$ cloud to the north-east of the radio lobes (both identified in our new ALMA data). Here we present a more detailed analysis of the spectral line characteristics and kinematics of the GRS 1758$-$258 (Figures~\ref{fig:spec_grs}, \ref{fig:pv_grs}, and \ref{fig:grs_alt}) and 1E 1740.7-2942 (Figures~\ref{fig:spec_1E}, \ref{fig:pv_1E}, and \ref{fig:1e_alt}) molecular structures, to investigate which emission structures are most likely to be related to the BHXB and its radio lobes.

Figure~\ref{fig:spec_grs} displays molecular line spectra across three different regions of interest near GRS 1758$-$258 (northern lobe, southern lobe ``top'', southern lobe ``bottom''). As with the maximum intensity maps of this region (Figure~\ref{fig:tmax_grs}), we do not detect any clear line emission in the northern lobe, but the southern lobe shows lines with multi-peaked profiles, and intensities of tens to hundreds of mK. In particular, we can identify two multi-peaked $^{13}$CO components across the southern lobe (central velocities of $\sim$ 72 and $\sim$84 ${\rm km\, s}^{-1}$; marked by black arrows in Figure~\ref{fig:spec_grs}). The HCN emission displays a similar pattern to $^{13}$CO, where we see lower velocity and higher velocity components in the bottom portion of the southern lobe, although no HCN emission is detected in the top portion of the southern lobe. In the CS emission, we detect both the lower velocity and higher velocity components throughout the whole southern lobe. Multi-peaked line profiles indicate the presence of multiple gas components at different velocities (i.e., ordered flow in different directions), and wide line profiles indicate a continuum of velocities consistent with a turbulent molecular medium, both of which could be induced by a collision between the jet and molecular gas. Thus the distinct lower and higher velocity components (especially seen in $^{13}$CO) may be consistent with gas flowing in different directions, where gas is being pushed both towards and away from us along our line of sight.
In contrast, the spectra of the emission located away from the suspected interaction site (Figure~\ref{fig:grs_alt}) displays a clear single component line profile, and thus no evidence of a collision in the gas.
Moreover, Figure~\ref{fig:spec_grs} also shows that {the HCN/$^{13}$CO intensity ratio}\footnote{\label{fnote2}Note that since we do not have single dish data, it is possible that spatial filtering is preventing us from recovering CO emission on larger scales, and thus contributing to the large HCN/CO  and HCO$+$/CO ratios seen here. Self-absorption from foreground gas may also suppress the brightness of the CO emission. See \S~\ref{sec:dist} for details.} {in the southern lobe approaches ${\sim0.6}$}, which is quite atypical for the Galactic ISM \citep[e.g., the ratio is 0.025 in the Orion B molecular cloud,][]{pety17}, and could suggest an external process is enhancing the HCN abundance in this region.

To further investigate the dynamics of the gas across the southern lobe region in GRS 1758$-$258, we created position-velocity diagrams of the $^{13}$CO, HCN, and CS molecules through a slice along the eastern outer edge of the lobe (see Figure~\ref{fig:pv_grs}). In these position-velocity diagrams, we observe that the $^{13}$CO emission at an offset of 45--$75\arcsec$ along the slice appears to be spectrally displaced, being pushed to lower velocities. Further, the HCN and CS emission appears to show gas in this same region with a broad spread in velocity space, that could indicate multiple different velocity components blended together. These kinematic features may be consistent with the hydrodynamic backflow theory presented by \citet{marti17}, where jet particles hit the terminating end of a cavity and are reflected back along the cavity edge. Such a reflection would likely push molecular gas in different directions and create molecular components at different velocities.

Figures ~\ref{fig:spec_1E}, \ref{fig:spec1E_CS}, \ref{fig:spec1E_hnco4}, \ref{fig:spec1E_hnco5}, and \ref{fig:spec_CH3OH} display molecular line spectra across different regions of interest near 1E 1740.7$-$2942 in the 0--60 ${\rm km\, s}^{-1}$ velocity range (south-east edge of the molecular ring structure, northern lobe and north-west edge of the molecular ring structure, and the southern lobe). Similar to our ALMA detections in the GRS 1758$-$258 field, the spectra in the 0--60 ${\rm km\, s}^{-1}$ velocity range here show many lines with wide and/or multi-peaked profiles, which is indicative of a collision occurring in the molecular gas. Through comparing the HCO$+$ and HCN emission in this velocity range across the different spectral extraction regions, the north-west edge of the molecular ring shows a lower velocity component when compared to the south-east edge of the ring. A similar pattern is observed in the SiO emission, where the northern lobe shows lower velocity gas when compared to the southern lobe. These results suggest a velocity gradient across the molecular ring structure and between shock-tracing emission in the radio lobes. In contrast, the spectra of the emission away from the suspected interaction site (Figure~\ref{fig:1e_alt}) and the isolated cloud in the -180 to -100 ${\rm km\, s}^{-1}$ velocity range (Figure~\ref{fig:spec_alt}) display clear single component line profiles, and thus no evidence of a collision in the gas. Additionally, Figures \ref{fig:spec1E_13CO} and \ref{fig:spec_18CO} show a distinct lack of CO emission coincident with the HCN/HCO$+$ ring emission (atypical for the Galactic ISM; \citealt{pety17}), suggesting that, similar to the GRS 1758$-$258 results, an external process is enhancing the HCN/HCO$+$ abundance in this region\textsuperscript{\ref{fnote2}}.

To further investigate the dynamics of the molecular gas near 1E 1740.7$-$2942, we created position-velocity diagrams of the HCO$+$ and HCN molecules through an elliptical slice directed clockwise around the ring structure (see Figure~\ref{fig:pv_1E}). In these position-velocity diagrams, we confirm a velocity gradient between the northern and southern edges of the molecular structures, where the gas on the northern side of the ring/lobes is moving at slower velocities along our line of sight when compared to the gas on the southern side, and the most extreme velocities are seen on opposite sides of the ring (19 and $190\arcsec$ offsets).
These kinematic features may be consistent with the ring tracing an expanding bubble structure, where the SiO/CH$_3$OH emission could in turn be tracking outward moving (relative to the central BHXB) shocks in the radio lobe regions, both potentially driven by the BHXB jets.

\section{Discussion}
\label{sec:discuss}

\subsection{Distance constraints}
\label{sec:dist}
Currently, no independent distance constraints exist for either GRS 1758$-$258 or 1E 1740.7$-$2942.
For a jet-ISM interaction to be occurring, both the BHXBs and the molecular gas must be located at the same distance.
Therefore, our detection of molecular gas structures that are consistent with being powered by the GRS 1758$-$258 and 1E 1740.7$-$2942 jets, allows us to constrain the distance to both objects.  The molecular emission provides kinematic information, where we observe emission with velocities near $v_\mathrm{LSR} = (78\pm10)~\mathrm{km~s^{-1}}$ for GRS~1758$-$258 and $v_\mathrm{LSR} = (20\pm6) \,{\rm km \, s}^{-1}$ for 1E 1740.7-2942. This information can be converted into a kinematic distance, but kinematic distances are unreliable at these Galactic longitudes near the Galactic centre ($l=4.51$ and $l=359.12$ for GRS 1758$-$258 and 1E 1740.7$-$2942, respectively; \citealt{bal15}). For example, using the methodology of \cite{wag18}, we find distances ranging from 3 to 18 kpc in the case of 1E 1740.7$-$2942, and better constrained distances for GRS 1758$-$258 between 8.0 and 8.7 kpc.  

\begin{figure}
    \centering
    \includegraphics[width=\columnwidth]{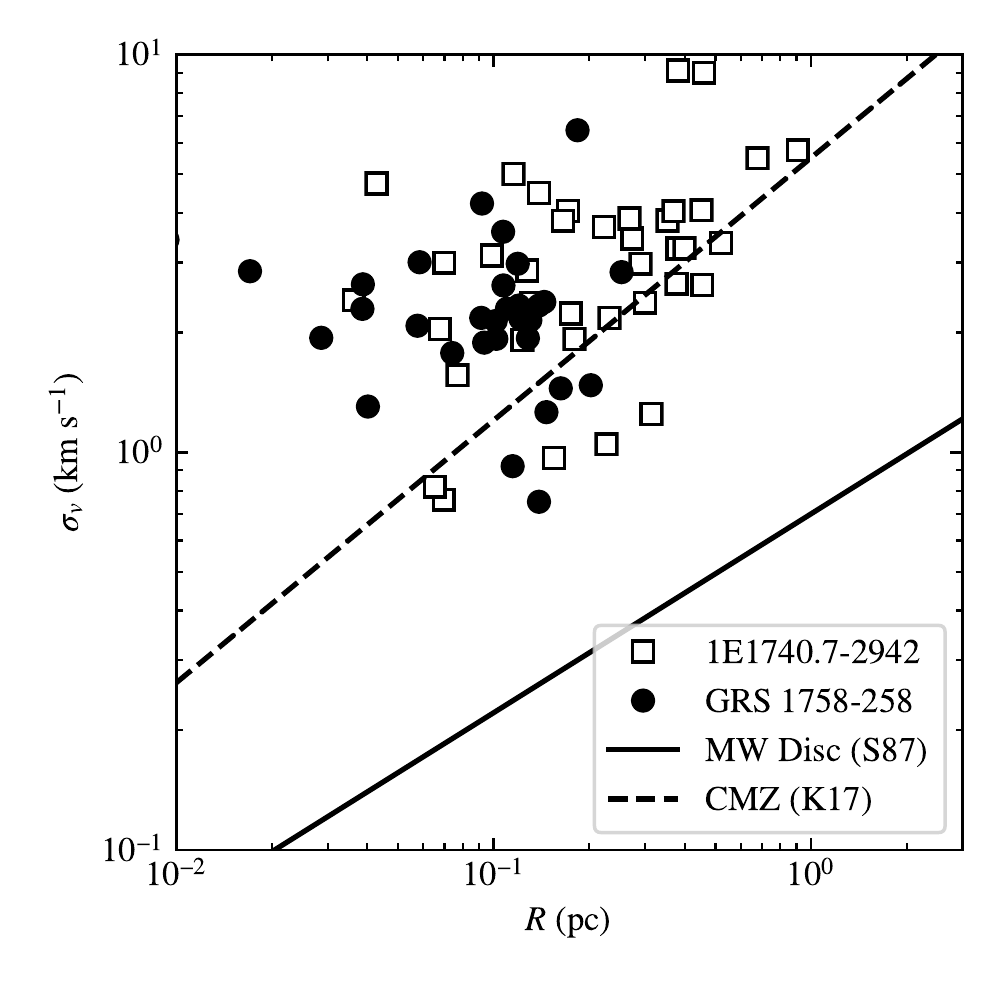}
    \caption{The size-line width relationship for HCO$^+$ emission structures in our ALMA data for both BHXBs.  Here we show the velocity dispersion as a function of the beam-deconvolved effective radius of molecular gas emission regions, determined from a dendrogram analysis (where the sizes assume a distance of $D=8~\mathrm{kpc}$). The solid line shows the relationship of \citetalias{sol87} for the Milky Way disc, and the dashed line shows the more turbulent clouds presented in \citetalias{kauff17}, that are found in the CMZ. The characteristic HCO$^+$ line widths on a given scale observed in our data are consistent with the molecular emission (and presumably the BHXBs as well) being found in the CMZ.}
    \label{fig:rdv}
\end{figure}

Additionally, kinematic distance methods assume objects are moving on disc-like orbits, and are not in the central molecular zone (CMZ) of the Galaxy.  In our data, the characteristic line widths of the molecular emission are quite large on relatively small angular scales ($\sigma_v\sim 5~\mathrm{km~s^{-1}}$ on $30''$ scales; see Figures \ref{fig:grs_alt} and \ref{fig:1e_alt}), and
we can use these line widths to characterize the distance to the molecular gas in both fields.  In particular, the molecular ISM is turbulent, where one of the manifestations of these turbulent flows is a relationship between the size ($R$) of molecular gas structures compared to their line widths, here expressed in terms of the velocity dispersion $\sigma_v$.  Thus, $\sigma_v = \sigma_0 (R/\mathrm{1~pc})^\beta$, where typical molecular clouds in the disc of the Milky Way show $\sigma_0\sim 0.7~\mathrm{km~s}^{-1}$ and $\beta=0.5$ \citep[][hereafter \citetalias{sol87}]{sol87}, while the CMZ has notably larger line widths of molecular gas features on a given spatial scale, when compared to gas found in the disc of the Galaxy. 

In Figure \ref{fig:rdv}, we show the size-line width relationship for the molecular gas structures seen in the ALMA HCO$^{+}$ emission for both fields. The structures are measured from the analysis size structures found in the dendrogram of the emission, following the methods outlined in \citet{roso08} and \citet[][hereafter \citetalias{kauff17}]{kauff17}. {Briefly, the dendrogram method measures the contour levels (isosurfaces) in the position-position-velocity data cube of emission.  The dendrogram summarizes how those contour levels merge together as a function of changing contour level, identifying the contour levels just above the values where objects blend together. Then, the algorithm measures the properties of emission for that contour including the radius (spatial extent) and the line width (velocity extent) of the emission contour. The radius and line width of the emission structures will be related by the velocity power spectrum of the turbulent flow \citep{maclow04}.} The radius of structures is corrected for the telescope beam and scaled to match the measurements used in \citetalias{sol87}.  We measure a physical size for the regions by assuming a fiducial distance of $D=8$~kpc, where the sizes would scale linearly with distance, and the velocity dispersions would not change with assumed distance.  We compare our observed data to the size-line width relationships deduced by \citetalias{sol87} for the disc of the Galaxy and by \citetalias{kauff17} for clouds in the CMZ. Assuming $D=8~\mathrm{kpc}$, the gas around our candidate jet-ISM regions clearly shows turbulent velocity scalings consistent with being found in the high-pressure CMZ.  If the molecular gas structures were considerably closer or farther, then they would no longer be in the CMZ and their size-line width scalings would be anomalous. Thus, assuming these molecular emission structures represent a true jet-ISM interaction, this constrains the originating source to be in the CMZ, and thus the distance of the BHXBs to be $D=8.0\pm 1.0\,\mathrm{kpc}$.

If these sources are indeed in the CMZ, we must show care when interpreting the line ratios between different species. In particular, transitions to the ground state may be heavily self-absorbed, and molecular emission is found throughout the medium.  The absence of single dish data in our observations may indicate that a significant amount of the emission has been spatially filtered out by the interferometer. Therefore, to counteract these issues, in the following section when discussing evidence for jet-ISM interactions, we focus on the kinematic structures in the emission and on the comparison of emission between the dense gas tracers.

\subsection{Evidence for jet-ISM interactions}
In both the GRS 1758$-$258 and 1E 1740.7$-$2942 fields, we have identified new molecular structures that could be associated with the jets in these BHXBs. We summarize our key evidence for each target in this section.

  \begin{figure}
\begin{center}
  \includegraphics[width=0.5\textwidth]{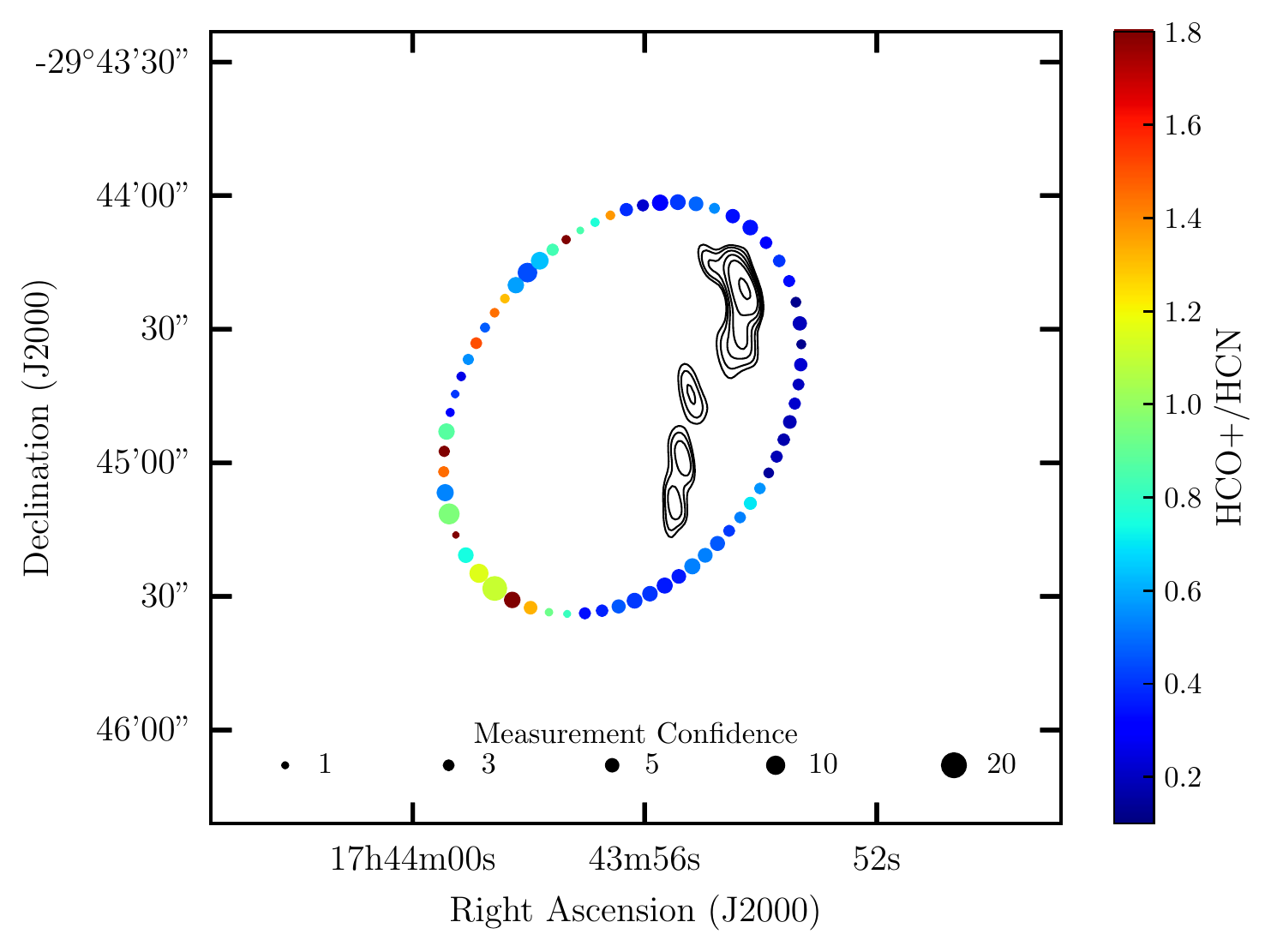}\\
 \caption{\label{fig:1e_hcophcn}HCO$+$/HCN intensity ratio along the molecular ring structure surrounding the BHXB 1E 1740.7$-$2942. Here we compute the ratio by extracting spectra of both molecules along the center of the elliptical shaped slice from Figure~\ref{fig:pv_1E}. The colour scale/bar indicates the value of the HCO$+$/HCN ratio, the size of the markers represents the measurement confidence (${\rm ratio}/\sigma_{\rm ratio}$, as shown by the legend at the bottom of the panel), and the black contours represent continuum radio emission (masked to only show the radio lobes; contour levels are $2^{n}\times$ the rms noise of $8\mu{\rm Jy\,bm}^{-1}$, where $n=2.0, 2.5, 3.0, 3.5, 4.0, 5.0, 6.0$; see \S\ref{sec:vlarad}). The HCO$+$/HCN ratio varies along the elliptical path tracing the molecular ring structure, with the higher values located on the side of the ring farthest from the BHXB. For comparison, the HCO$+$/HCN ratio in the off emission region (see Figure~\ref{fig:1e_alt}), located to the south of the suspected interaction site, is $\sim0.4$.}
\end{center}
 \end{figure}

{\bf GRS 1758$-$258:} Our ALMA observations revealed a molecular structure and shock tracing emission in the velocity range 50--100 ${\rm km\,s}^{-1}$, detected with the $^{13}$CO, HCN, CS, and SiO molecules. This molecular gas is spatially coincident with the eastern edge of the southern radio lobe, and appears to be aligned along the edge of the previously suggested cavity and back-flowing material identified in deep radio continuum data \citep{marti15,marti17}. We observe intriguing chemistry and kinematics from this molecular emission towards the southern lobe. In particular, the spectrum of the $^{13}\mathrm{CO}$ emission shows two velocity components, which may indicate displacement of the gas by the jet. However, these two components could alternatively indicate self-absorption of the profile. To try to resolve this ambiguity, we can examine the CS, SiO, and HCN line properties. The CS and SiO lines are particularly helpful, as these lines represent transitions between excited states, and thus are less subject to self-absorption in the ground state.
Both the CS and SiO lines show single components near $v_\mathrm{LSR}\sim 83\mathrm{~km~s^{-1}}$, while the HCN emission has an emission maximum at $v_\mathrm{LSR}\sim 89\mathrm{~km~s^{-1}}$, with a much broader line profile. As the HCN line is actually made up of several hyper-fine components, we attempted to fit a hyper-fine model to these lines of sight. This fitting process yielded a central velocity for the HCN emission closer to that of the CS/SiO emission of $v_\mathrm{LSR} = 85\mathrm{~km~s^{-1}}$; however, matching the amplitudes of the HCN emission with our model required anomalous excitation of the hyper-fine structure. {Specifically, the hyperfine components of HCN have fixed ratios of optical depths.  The only way to fit the observed hyperfine ratios is for the different components to have different excitation temperatures.  Such conditions can occur under in non-LTE models \citep[e.g.,][]{vandertak07}, but assessing whether the excitation conditions are physical would require full three-dimensional radiative transfer modelling beyond the scope of this work. Practically, we determined the velocity by fitting the HCN hyperfine emission pattern to the line allowing the ratios between the components to vary as free parameters.} Therefore, we conclude that this molecular structure, located exactly on the limb of the non-thermal radio emission from the jet, is consistent with a knot of dense molecular gas that has been compressed\footnote{The effective critical densities for CS and HCN are \citep[\(8000\mathrm{~cm^{-3}}\),][]{shir15}.} and displaced by 7~km~s$^{-1}$ with respect to the ambient gas in the region.  
Lastly, we note that the non-detection of HCO$^+$ emission in this region, is consistent with the expected brightness ratios of $I$(HCN)/$I$(HCO$^+$)$>10$, common in many parts of the CMZ \citep{pound18}.

In the scenario where the southern lobe represents the receding jet\footnote{It is currently unknown which side of the jet in either GRS 1758$-$258 or 1E 1740.7$-$2942 is approaching or receding. While there is a clear difference in the radio brightness of the lobes in both systems (the northern lobe is brighter in both cases), the lobes are likely non-relativistic (as seen from our shock speed estimate in \S\ref{sec:calorimetry} and the motion resolved by \citealt{marti15}). This implies that the brightness difference cannot be due to Doppler boosting effects and thus does not distinguish which jet is  approaching/receding.}, gas in this lobe flowing in the same direction of the jet would appear to be at higher velocities along our line of sight, and gas flowing in the direction opposite to the jet would appear to be at lower velocities along our line of sight. Therefore, we postulate that the following scenario produces the observed molecular gas features: Over time the jet has blown out a cavity in the surrounding gas. As the jet continues to propagate into the cavity, it will hit the southern end of the cavity (driving molecular gas away from us, thereby creating higher velocity molecular components) and be reflected back toward the BHXB (driving molecular gas towards us along the edge of evacuated cavity, thereby creating lower velocity molecular components). In this scenario, the lack of molecular emission in the northern interaction region could simply indicate a lack of molecular gas in that region.

{\bf 1E 1740.7$-$2942:} Our ALMA observations revealed a molecular ring structure and shock tracing emission in the velocity range 0--60 ${\rm km\,s}^{-1}$, detected with the HCN, HCO$^+$, SiO, CS, HNCO, and CH$_3$OH molecules. The spectra and kinematic analysis of this molecular emission suggest that there is a velocity gradient across the ring structure, and between the SiO hot-spots, where gas on the northern side of the structures is moving more slowly along our line of sight than gas on the southern side of the structures. Similar to GRS 1758$-$258, if the northern lobe arises from the approaching jet and the southern lobe arises from the receding jet, we expect gas moving in the same direction of the jet to be moving faster along our line of sight in the southern lobe when compared to the northern lobe. The direction of the velocity gradient we observe across the molecular ring, and the two extreme gas velocities observed on opposite sides of the ring, are consistent with such a scenario.

Additionally, the detection of the ring shaped cavity in the HCO$+$/HCN molecules is another clue pointing towards a connection between the BHXB source and the molecular gas. In particular, enhancement of the HCO$+$ molecule in the ISM has been predicted to occur near high energy sources. \cite{krolik83} studied how different ionization rates affect HCO$+$ abundance, finding that while ionization rates close to the X-ray source would be high enough to inhibit HCO$+$ formation, lower rates farther away from the X-ray source could actually result in an increased abundance of this molecule.  This is particularly striking when compared to the weak HCO$^+$ emission in GRS 1758$-$258 and other sources in the CMZ. In fact, \cite{phillaz95} modelled ionization rates and cloud densities near 1E 1740.7$-$2942, finding that at a $\sim 1$ pc distance from the X-ray source (consistent with the location of our detected ring structure), the ionization rate would reach the value needed to enhance HCO$+$ abundances.  We observe a pattern consistent with this prediction in our data. In particular, Figure~\ref{fig:1e_hcophcn} displays the HCO$^+$/HCN intensity ratio along the ring structure, where we clearly observe a variation in the HCO$^+$/HCN ratio across the ring, with a larger value in the portion of the ring located farthest away from the central X-ray source.
Additionally, this process may also explain why we observe strong HCO$^+$/HCN emission in the absence of CO in this region.

Given the above evidence, we postulate that the molecular ring represents an expanding bubble structure, blown out by the BHXB jets and X-ray radiation from the central BHXB, and the SiO hot-spots represent terminating shocks at the edges of this bubble structure. As the molecular cloud we detected at $-140 {\rm \, km\, s}^{-1}$ (FWHM line width of 5 ${\rm km\, s}^{-1}$) shows no kinematic or spectral features that would indicate a collision in the gas, we postulate this emission represents an isolated cloud near the Galactic centre.

\subsection{Deriving jet properties}
\label{sec:calorimetry}
In GRS 1758$-$258 and 1E 1740.7$-$2942, we have identified molecular structures that appear to be jet blown cavities surrounding the BHXB systems. Using a calorimetric approach, we can estimate the time averaged power that the BHXB jets would need to carry in order to create and maintain such structures in the local ISM. Following the model of \citealt{kai97} (with the formalism outlined in \S 5.4 and Appendix C in \citealt{tet18i}), the total jet power (averaged over the jet's lifetime) as a function of ISM properties at the impact sites can be represented as

\begin{equation}
    Q_{\rm jet}=\left(\frac{5}{3}\right)^3\frac{\rho_0}{C_1^5}L_J^2v^3.
    \label{eq:pow}
\end{equation}
Here $L_J$ represents the length of the jet path, $\rho_0$ represents the density of the molecular medium with which the jet is colliding, $v$ represents the velocity of the shocked gas at the interaction site, and $C_1$ is a constant dependent on the adiabatic indices of the material in the jet, cavity, and external medium, as well as the jet opening angle.

To estimate the length of the jet path, we use the angular extents of the radio lobes ($l_{\rm arcsec}$, measured in the direction parallel to the jet axis), the distance to the source ($D$ in kpc), and the inclination angle of the jet axis ($i$, to account for the projection effects), to yield,
\begin{equation}
    L_J=(1.5\times10^{16})\, D \,\frac{l_{\rm arcsec}}{\sin i} \, \,{\rm cm}.\nonumber
\end{equation}
While the jet opening angle can be expressed as,
\begin{equation}
\theta={\rm arctan}(w_{\rm arcsec}/l_{\rm arcsec}),\nonumber
\end{equation}
where $w_{\rm arcsec}$ represents the angular extents of the radio lobes at the terminating edges, measured in the direction perpendicular to the jet axis.

To estimate the density of the molecular medium, we first compute the H$_2$ column density (using the formulation in \citealt{mag15}, assuming optically thin emission and fractional abundance ratios of $^{13}$CO/H$_2=6\times10^{-5}$, HCO$+$/H$_2=3\times10^{-9}$; \citealt{wilsonrood,gerin2019}). From this we can estimate the mass of molecular gas in each region. 
This procedure yields estimates of $7.4\,D^2 M_\odot$ in the southern molecular structure of GRS 1758$-$258 and $210\,D^2 M_\odot$ in the ring structure surrounding 1E 1740.7$-$2942. 
To check that our H$_2$ column density measurements are reasonable, we compare the range of column densities that we calculated in each region to the column density obtained independently from X-ray spectral fitting of the target sources, although we note the latter may include extinction near the accretion disc, but not near the radio lobes. In GRS 1758$-$258, X-ray spectral fits find $N_{\rm H}=1.6\times10^{22} \,{\rm cm}^{-2}$ \citep{soriaf11}, which lies within the broad range we observe in our molecular column density maps, $N_{\rm H}=3.2\times10^{18}-4.8\times10^{22}\,{\rm cm}^{-2}$. In 1E 1740.7-2942, X-ray spectral fits find $N_{\rm H}=1.05\times10^{23}\,{\rm cm}^{-2}$ \citep{gallofen02}, which lies within the broad range we we observe in our molecular column density maps\footnote{We note that a previous study \citep{vilhu97} calculated the column density from lower spectral and angular resolution observations of the CO molecule in this region to be $N_{\rm H}=(3-11)\times10^{22}\,{\rm cm}^{-2}$, which is also in line with our measurements.}, $N_{\rm H}=1.0\times10^{19}-4.4\times10^{23}\,{\rm cm}^{-2}$.

We then model the GRS 1758$-$258 molecular structure as an ellipsoid (major and minor axes of $95\arcsec$ and $40\arcsec$; $V=2.6\times10^{53} D^3 f {\rm cm}^{3}$), and the 1E 1740.7$-$2942 molecular structure as a partially hollow ellipsoid (major and minor axes of $118\arcsec$/$90\arcsec$ and $75\arcsec$/$50\arcsec$ for outer/inner portions, respectively; $V=7.6\times10^{53} D^3 f {\rm cm}^{3}$), where $f$ represents the volume filling factor\footnote{$f\ll 1$ would represent a completely hollow ellipsoid.}. Combining the mass and volume estimates yields a density for GRS 1758$-$258 of, 
$    \rho_{0,1758}=5.7\times10^{-20} D^{-1} f^{-1} \,\,{\rm g\,cm}^{-3} $ and a density for 1E 1740.7$-$2942 of $
\rho_{0,1740}=5.5\times10^{-19} D^{-1} f^{-1}\,\, {\rm g\,cm}^{-3}.$
As with the column density, to check that these density measurements are reasonable, we can compare our estimates to the critical densities needed for our highest density tracing molecules detected to form; $n_{\rm crit}$ of the HCO$+$ (1-0), HCN (1-0), and CS (2-1) molecules are equivalent to $10^4-10^{5} {\rm cm}^{-3}$ \citep{shir15}. Using $D=8.0$ kpc and $f=0.1$ (see \S\ref{sec:dist} and below for details on these choices), we estimate that the number densities for the gas in the GRS 1758$-$258/1E 1740.7$-$2942 regions are equivalent to $\rho_0/m_{\rm H}=2.1\times10^{4}/2.1\times10^{5}\, {\rm cm}^{-3}$, both of which are in line the expected range of densities needed for emission from these molecules to be excited.

To calculate the time-averaged jet power for each system, we use Equation~\ref{eq:pow} with inputs of $D=7.0-9.0$ kpc for GRS 1758$-$258 and 1E 1740.7$-$2942 (see \S\ref{sec:dist}), $v=15 \,{\rm km\,s}^{-1}$ (equivalent to the approximate FWHM of the shock-tracing CS and SiO lines), $l_{\rm arcsec}=90 \arcsec/25\arcsec$ and $w_{\rm arcsec}=30\arcsec$/$8\arcsec$ for GRS 1758$-$258/1E 1740.7$-$2942 (the dimensions of the southern radio lobes), filling factor of $f=0.1$ (a reasonable assumption given that the jet is displacing molecular gas), $i=30$ deg or $i=80$ deg (as inclination angle is unknown for both systems\footnote{We note that a previous radio study \citep{ped15} and a recent X-ray reflection study \citep{ste20} of 1E 1740.7$-$2942, both favour a higher inclination angle ($>50$ degrees) in this particular system.}, we consider two extremes here), and adiabatic indices of the jet, cavity, and external medium of $\Gamma_{j} =\Gamma_{c}=\Gamma_{x}=5/3$. If the jets are pointing closer to our line of sight ($i=30$ deg), these calculations yield a time averaged jet power of $(4.4-5.7)\times10^{36}{\rm erg\,s}^{-1}$ over 0.24--0.31 Myr for GRS 1758$-$258 and $(2.7-3.5)\times10^{37}{\rm erg\,s}^{-1}$ over 0.20--0.26 Myr for 1E 1740.7$-$2942. If the jets are aligned closer to the plane of the sky ($i=80$ deg), these calculations yield a time averaged jet power of $(1.1-1.5)\times10^{36}{\rm \, erg\,s}^{-1}$ over 0.12--0.16 Myr for GRS 1758$-$258 and $(7.1-9.1)\times10^{36}{\rm \, erg\,s}^{-1}$ over 0.10--0.13 Myr for 1E 1740.7$-$2942. These jet power estimates lie close to the distribution of estimated jet powers in the BHXB population ($\sim10^{36} - 10^{38} {\rm erg\,s}^{-1}$; \citealt{curr14}), reinforcing our hypothesis that the BHXB jets are driving the molecular structures we have detected in the fields surrounding GRS 1758$-$258 and 1E 1740.7$-$2942.

\subsection{Other mechanisms driving the excited molecular emission}
\label{sec:others}
In this Section, we explore alternative scenarios (namely star formation, supernova explosions, stellar winds, and accretion disc winds) that could create the molecular emission structures we have identified in the fields surrounding GRS 1758$-$258 and 1E 1740.7$-$2942.

In our previous study of the jet interaction sites near GRS 1915$+$105, we found that the energy released by the high mass star formation process, in addition to a BHXB jet, was powering the radio lobe structure to the south of the central source \citep{tet18i}. However, in the cases of both GRS 1758$-$258 and 1E 1740.7$-$2942, there is a distinct lack of extended near-infrared emission ($8\mu m$  and $5.8\mu m$; see Spitzer GLIMPSE maps in Figures~\ref{fig:mwgrs} and \ref{fig:mw1e}) in the fields surrounding the BHXBs, likely indicating a lack of high mass star formation activity occurring in these regions.

In 1E 1740.7$-$2942, we identified a ring-like molecular structure surrounding the central BHXB, whose morphology could also be consistent with that of a supernova remnant (i.e., similar to XRB Circinus X-1 surrounded by its natal supernova remnant; \citealt{heinz13}). This opens up the possibility that the ring structure was not created by the BHXB jets, but rather represents the supernova remnant associated with the 1E 1740.7$-$2942 black hole or a chance projection of an unrelated supernova remnant. The typical kinetic energies produced by supernova explosions are $\sim10^{51}$ erg \citep{chev77,korpi99,rubin16}, which is close to the energy range that we estimate was needed to create the molecular ring structure ($\sim10^{49}-10^{50}$ erg; see \S\ref{sec:calorimetry}). However, while this scenario is plausible from an energetics perspective, there are several inconsistencies with the supernova theory; we estimate a shock velocity of 15 ${\rm km \, s}^{-1}$ from the SiO emission in the region (see \S\ref{sec:calorimetry}), which is much slower than velocities expected from a typical supernova driven forward shock (thousands of ${\rm km \, s}^{-1}$; \citealt{rubin16}), and there is no detection of extended radio/X-ray emission (originating from non-thermal synchrotron/thermal bremstrahlung emission) consistent with the location of the molecular ring, as is typical for supernova forward shocks. {Furthermore, if we interpret the offset in position of 1E 1740.7$-$2942 from the centre of the molecular ring as due to the BHXBs peculiar motion\footnote{This peculiar motion offset scenario could possibly be tested with a proper motion measurement from Very Long Baseline Interferometry (VLBI) observations.} over its lifetime (driven by the natal kick the system received at birth), given the estimated ages from \S\ref{sec:calorimetry}, this implies a very small peculiar velocity\footnote{This small peculiar velocity estimate is consistent with 1E 1740.7$-$2942 being located close to the Galactic plane.} of $\sim5\, {\rm km \, s}^{-1}$. Such a small peculiar velocity would be indicative of the black hole in 1E 1740.7$-$2942 forming by direct collapse rather than a supernova explosion at all \citep{atri2019}.} Given the above arguments, we hypothesize that the molecular ring structure near 1E 1740.7$-$2942 is unlikely to represent a supernova remnant.

Lastly, other forms of outflows originating in the BHXB systems could drive the molecular structures we observed near GRS 1758$-$258 and 1E 1740.7$-$2942; e.g., either an accretion disc wind or a stellar wind from a high mass donor star. There are currently no detections\footnote{Although, we note that accretion disc winds are only observed in more edge-on BHXB systems.  Therefore, if either system has a more face-on orientation (i.e., lower inclination angle), a disc wind could be present, but not be detected.} of accretion disc winds in either source, with \cite{ponti12} only providing an upper limit for the presence of a disc wind in GRS 1758$-$258. Furthermore, as both sources are known to spend the majority of their lifetimes in a hard accretion state, in which disc winds are not expected to be present \citep{ponti12}, we find it unlikely that accretion disc winds are significantly contributing to driving the molecular structures. The current best candidates for the donor star in 1E 1740.7$-$2942 are all high mass stars, implying that the strong stellar wind from the donor may drive or contribute to driving the molecular ring structure near this system (e.g., similar to Cygnus X-1, where a combination of the stellar wind and BHXB jet is believed to be driving a bubble structure in the local ISM; \citealt{sell15}, see also the simulations by \citealt{RR16} showing the significant effect that stellar feedback can have on molecular clouds). However, in GRS 1758$-$258, the potential detection of radio continuum structures moving at $\sim 2\,{\rm arcsec\,yr}^{-1}$ (equivalent to a speed of 0.25 c, at a distance of 8.0 kpc; \citealt{marti15}), disfavours a scenario where the donor stellar wind plays a significant role, as the speeds of these stellar winds are much slower ($\sim$ thousands of km/s or $<0.01$ c). Ultimately, limited knowledge about the donor stars in both systems prevents us from completely ruling out that those stars contribute to the formation of the molecular structures detected near these systems.

\subsection{Identifying a molecular signature for jet feedback}
In our past molecular tracer study of the field surrounding the BHXB GRS 1915+105 \citep{tet18i}, we discovered that another feedback process, namely high mass star formation, can manifest in an similar way as BHXB feedback in molecular gas. As such, it is of interest to try and identify a molecular signature that is unique to BHXB jet feedback. To this end, we compare the molecular tracer properties of the GRS 1915+105 study and the two target sources presented in this work. While three sources  is admittedly a small sample, the molecular line properties between the interaction sites near these sources seem to be vastly different (e.g., GRS 1915+105 shows higher intensities, narrower line widths, and different emission region morphology over different physical scales). Alternatively, perhaps the specific molecules detected in a suspected interaction site, rather than the line properties, could be more unique to BHXB feedback effects. For example, as discussed above, HCO $+$ emission (and possibly other molecules as well, such as HCN, CS; \citealt{yan97}) may be enhanced in the vicinity of high energy X-ray sources. These specific molecules also all have high critical densities, and thus are expected to be present in more extreme environments (as expected near BHXB sources), when compared to other molecules such as CO (which is abundant throughout the Galaxy). Therefore, we postulate that understanding the chemistry, in addition to line properties and morphology, of molecular gas in potential BHXB jet interaction regions may be key in uniquely identifying these sites in our Galaxy. As such, sampling a suite of molecular transitions, including those tracing high density, ionization, and shocks (e.g., HCN and HCO$+$, SiO, CS), may be the best way to continue to advance these efforts.

\section{Summary}
\label{sec:sum}
In this paper, we present the results of our ALMA observations of the fields surrounding the BHXBs GRS 1758$-$258 and 1E 1740.7$-$2942. Both of these sources are powerful micro-quasars, which consistently accrete near their peak luminosity, and have double-sided radio lobe structures within a few pc of the central BHXB. As the brightness and morphology of these radio lobes have undergone significant evolution over the last decade, they are excellent candidates for the impact sites of the BHXB jet on the local ISM.

With our ALMA data, we mapped the molecular line emission in the fields surrounding GRS 1758$-$258 and 1 E1740.7$-$2942 and used the molecular gas properties to search for signs of jet interactions in these regions.
We detected emission from the HCN [$J=1-0$], HCO$^+$ [$J=1-0$], SiO [$J=2-1$], CS [$J=2-1$], $^{13}$CO [$J=1-0$], C$^{18}$O [$J=1-0$], HNCO [$J=4_{0,4}-3_{0,3}$], HNCO [$J=5_{0,5}-4_{0,4}$], and CH$_3$OH [$J=2_{1,1}-1_{1,0}$] molecules. This emission revealed new molecular structures in both regions that appear to trace jet-blown cavities in the surrounding gas. Using these molecular cavities as calorimeters, we estimate the time averaged jet powers from these systems, finding $(1.1-5.7)\times10^{36}{\rm \, erg\,s}^{-1}$ over $0.12-0.31$ Myr for GRS 1758$-$258 and $(0.7-3.5)\times10^{37}{\rm \, erg\,s}^{-1}$ over $0.10-0.26$ Myr for 1E 1740.7$-$294, dependent on the distance to the systems and the inclination angle of the jet axes.
Additionally, the spectral line characteristics of the molecular emission we detected indicate these structures are most likely located in the CMZ of our Galaxy, which constrains the distances to these structures (and both BHXBs) to be  $8.0\pm1.0$ kpc.

Overall, our work here demonstrates the efficacy of astro-chemistry at pinpointing jet-ISM interaction zones near BHXBs, and quantifying the effects of BHXB jet feedback in our Galaxy.

\section*{Acknowledgements}
The authors thank the anonymous referee for the time and effort put into reviewing this manuscript, especially in these complicated times during the COVID-19 pandemic.
We wish to thank Gerald Schieven for his help in configuring the ALMA observations, Adam Ginsburg for discussions on molecular emission from the CMZ, Eric Koch for discussions about molecular transitions, and Steve Mairs for discussions on stellar feedback. JCAMJ is the recipient of an Australian Research Council Future Fellowship (FT140101082), funded by the Australian government. EWR acknowledges the support of the Natural Sciences and Engineering Research Council of Canada (NSERC), funding reference number RGPIN-2017-03987.  GRS acknowledges support from an NSERC Discovery Grant (RGPIN-06569-2016). This paper makes use of the following ALMA data: ADS/JAO.ALMA\#2017.1.00928.S, ADS/JAO.ALMA\#2019.1.01266.S. ALMA is a partnership of ESO (representing its member states), NSF (USA) and NINS (Japan), together with NRC (Canada), MOST and ASIAA (Taiwan), and KASI (Republic of Korea), in cooperation with the Republic of Chile. The Joint ALMA Observatory is operated by ESO, AUI/NRAO and NAOJ. The National Radio Astronomy Observatory is a facility of the National Science Foundation operated under cooperative agreement by Associated Universities, Inc. In this work, we make extensive use of the radio astro tools \textsc{python} packages, including \textsc{spectralcube}, \textsc{pvextractor}, and \textsc{radio beam}, as well as the \textsc{aplpy} plotting package.

\section*{Data Availability}
The observational data presented in this work is available in the ALMA (\url{http://almascience.nrao.edu/aq/}) and VLA (\url{https://archive.nrao.edu/archive/}) online data archives.



\bibliography{ABrefList}



\appendix

\section{Multi-wavelength Coverage Maps}
\label{sec:appenmw}

\begin{figure*}
\begin{center}
  \includegraphics[width=0.7\textwidth,]{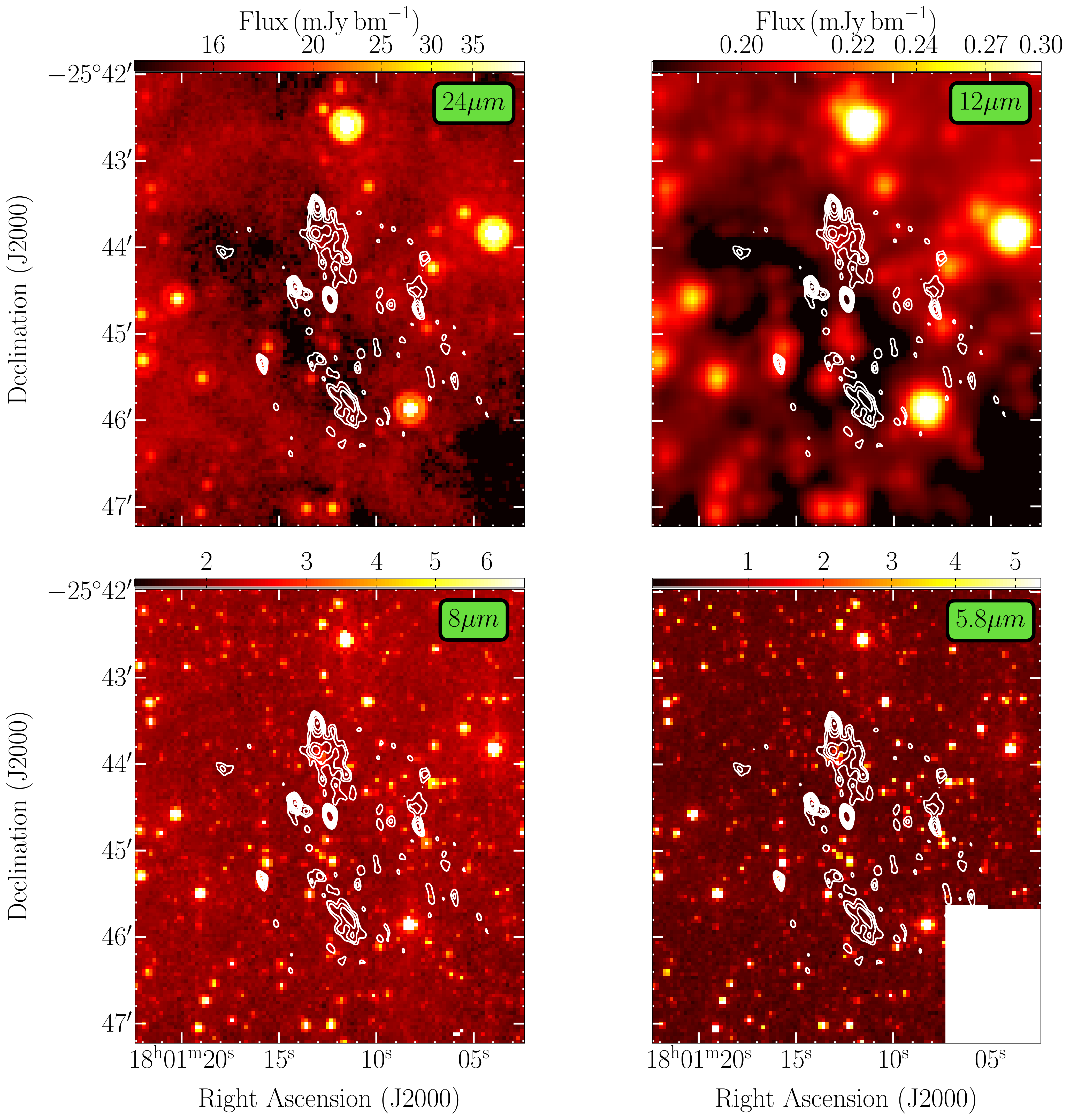}
 \caption{\label{fig:mwgrs}
 Multi-wavelength maps of the field surrounding the BHXB GRS 1758$-$258. The background colour images (in units of ${\rm mJy\, bm}^{-1}$ shown by the colour bars at the top of each panel) are taken with Spitzer MIPSGAL (${24\mu{\rm m}}$), NASA WISE (${12\mu{\rm m}}$), and Spitzer GLIMPSE (${8\mu{\rm m}}$ and ${5.8\mu{\rm m}}$); see \S\ref{sec:mw} for details.
 The white contours represent 6 cm archival VLA C-configuration continuum radio maps from observations taken in 2016 March (see \S\ref{sec:vlarad}); contour levels are $2^{n}\times$ the rms noise of $3.5\mu{\rm Jy\,bm}^{-1}$ ($n=1.5, 2.0, 2.5, 3.0, 3.5, 4.0, 4.5$). The diffuse mid-infrared emission (${24\mu{\rm m}}$ and ${12\mu{\rm m}}$) likely originates from heated gas and dust, while the near-infrared emission (${8\mu{\rm m}}$ and ${5.8\mu{\rm m}}$)
 appears to be dominated by stellar emission.}
\end{center}
 \end{figure*}

\begin{figure*}
\begin{center}
  \includegraphics[width=0.7\textwidth,]{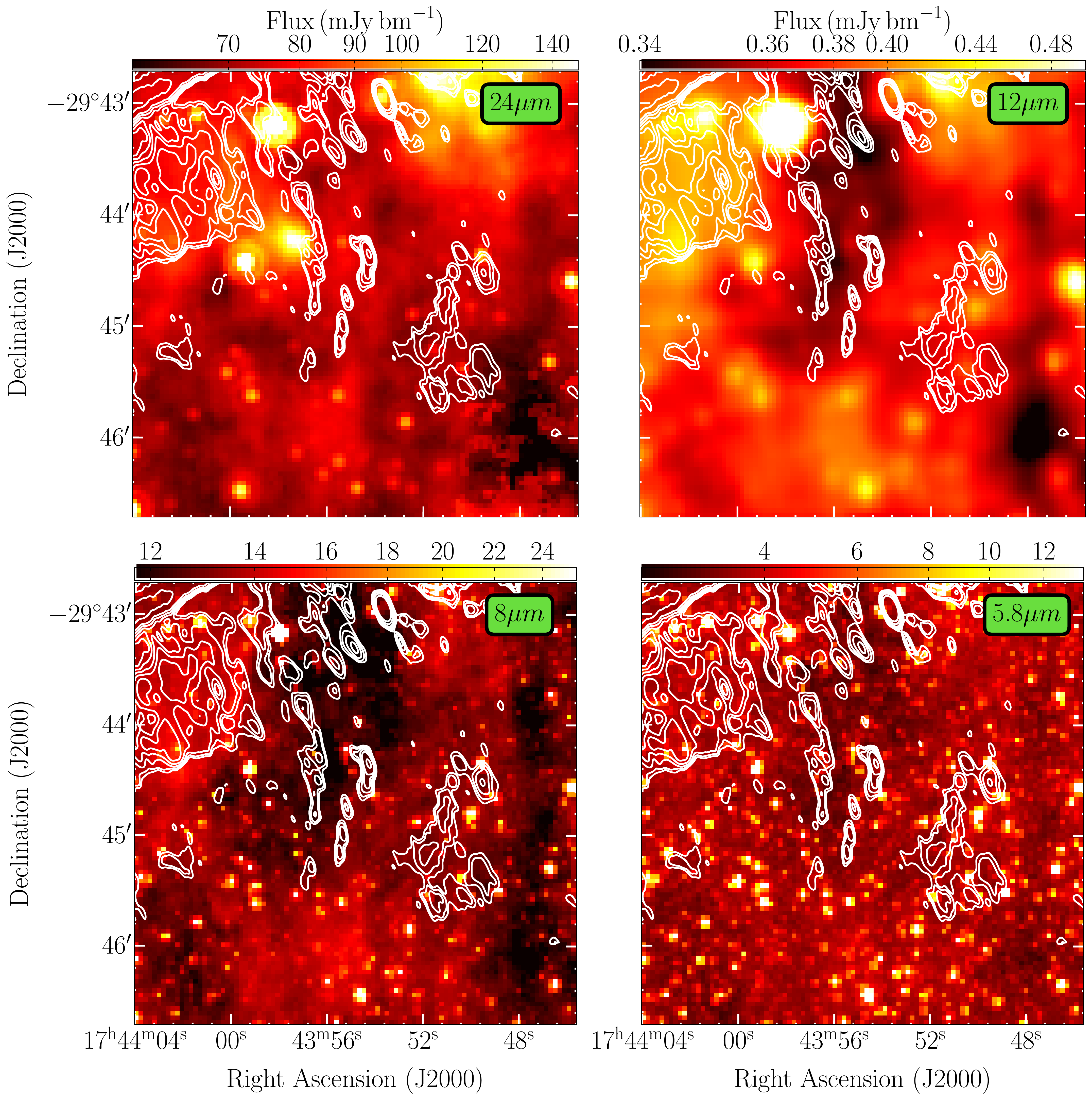}
 \caption{\label{fig:mw1e}Multi-wavelength maps of the field surrounding the BHXB 1E 1740.7$-$2942. The background colour images (in units of ${\rm mJy\, bm}^{-1}$ shown by the colour bars at the top of each panel) are taken with Spitzer MIPSGAL (${24\mu{\rm m}}$), NASA WISE (${12\mu{\rm m}}$), and Spitzer GLIMPSE (${8\mu{\rm m}}$ and ${5.8\mu{\rm m}}$); see \S\ref{sec:mw} for details.
 The white contours represent 6 cm archival VLA C-configuration continuum radio maps from observations taken in 2016 March (see \S\ref{sec:vlarad}); contour levels are $2^{n}\times$ the rms noise of $8\mu{\rm Jy\,bm}^{-1}$ ($n=2.0, 2.5, 3.0, 3.5, 4.0, 5.0, 6.0$). The diffuse mid-infrared emission (${24\mu{\rm m}}$ and ${12\mu{\rm m}}$) likely originates from heated gas and dust, while the near-infrared emission (${8\mu{\rm m}}$ and ${5.8\mu{\rm m}}$)
 appears to be mainly dominated by stellar emission.}
\end{center}
 \end{figure*}

In this Section, we show infrared maps (24, 12, 8, and 5.8 $\mu m$) of the fields surrounding GRS 1758$-$258 and 1E 1740.7$-$2942; Figures~\ref{fig:mwgrs} and \ref{fig:mw1e}.

 \section{Other molecular emission along the line of sight}
\label{sec:appenmol}

\begin{figure}
\begin{center}
\quad\quad\quad\includegraphics[width=0.42\textwidth]{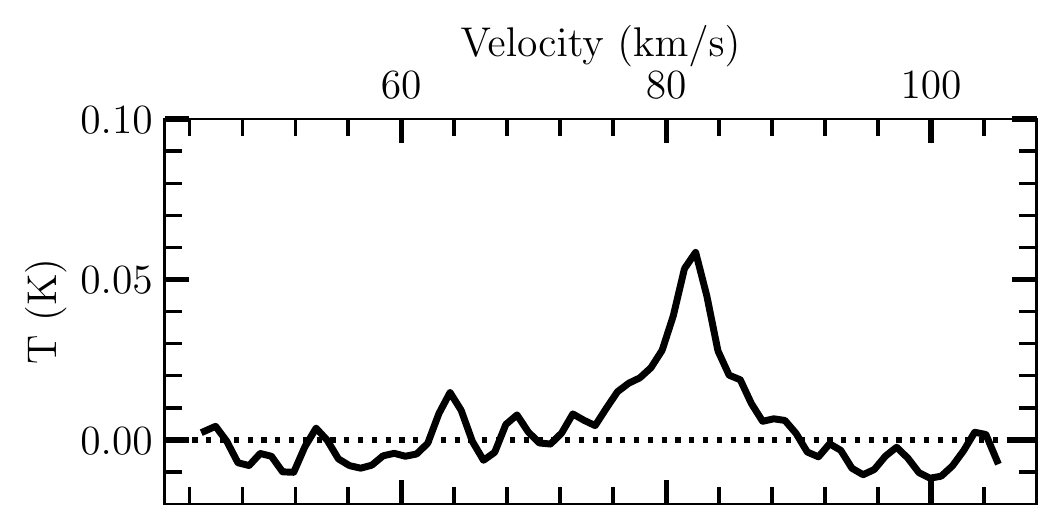}
  \includegraphics[width=0.48\textwidth]{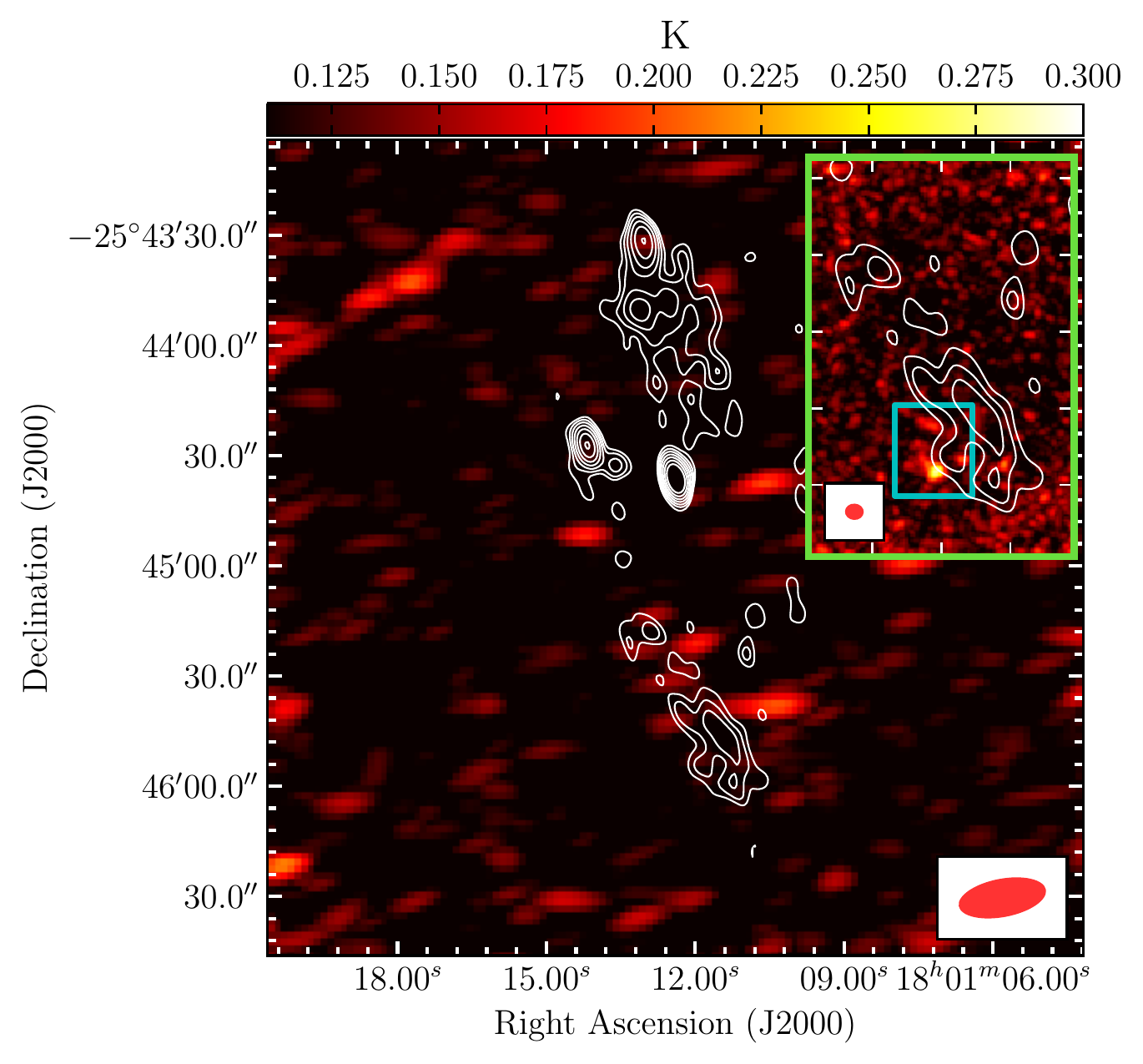}\\
 \caption{\label{fig:specgrs_SiO}SiO emission (shock tracer) detected near the BHXB GRS 1758$-$258 in the velocity range $45$ to $105\,{\rm km\,s}^{-1}$ (in units of Kelvin). The \textit{bottom} panels display maximum intensity maps, where the main panel displays the ACA data alone, while the inset panel displays the combination of ACA + 12m array data. The colour scale represents the intensity of the molecular emission (the colour bar range for the inset panel has the same limits as the main panel), while the white contours represent continuum radio emission (masked to only show the radio lobes; contour levels are $2^{n}\times$ the rms noise of $3.5\mu{\rm Jy\,bm}^{-1}$, where $n=1.5, 2.0, 2.5, 3.0, 3.5, 4.0, 4.5$; see \S\ref{sec:vlarad}). The red ellipses indicate the ALMA beams. The \textit{top} panel displays the spectrum of the SiO emission (ACA + 12m data) in the southern radio lobe (the cyan square in the \textit{bottom} inset panel represents the spectral extraction region). We detect SiO emission coincident with the peak of our other shock tracer, CS, only in the combined ACA + 12m array data.  The SiO spectra shows a wide line profile, consistent with a turbulent molecular medium.}
\end{center}
 \end{figure}

\begin{figure}
\begin{center}
\quad\quad\quad\includegraphics[width=0.38\textwidth]{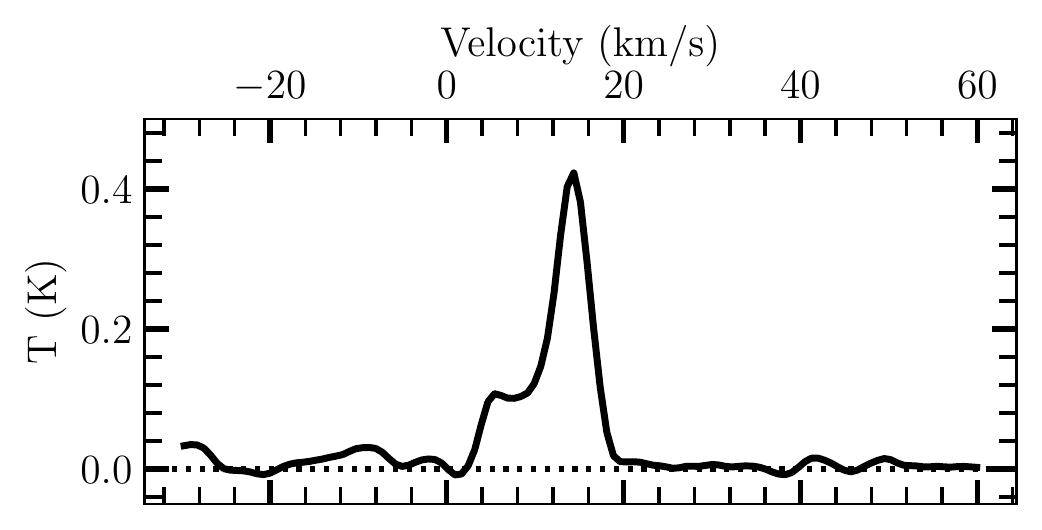}
  \includegraphics[width=0.48\textwidth]{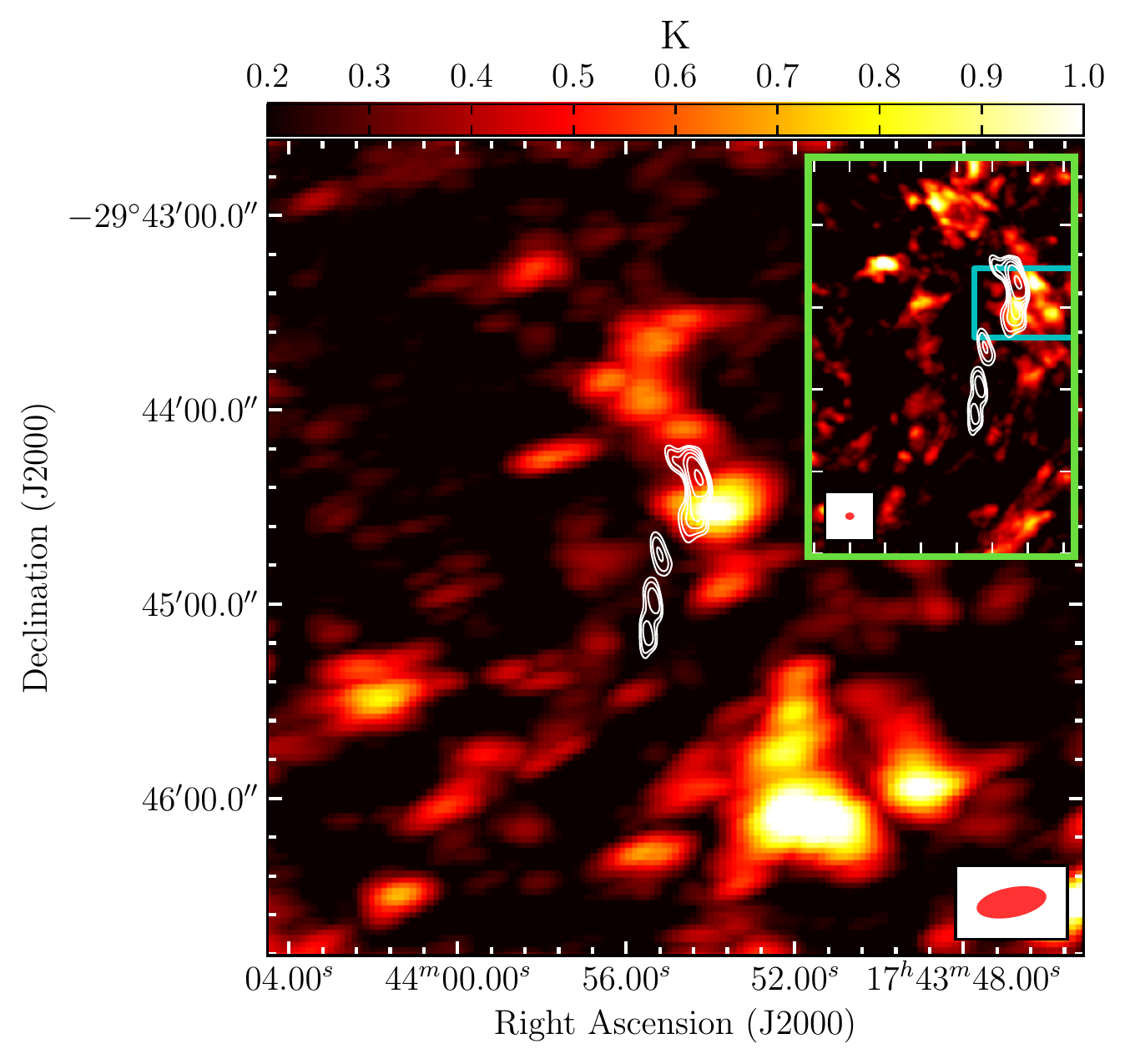}\\
 \caption{\label{fig:spec1E_CS}CS emission (shock tracer) detected near the BHXB 1E 1740.7$-$2942 in the velocity range $-30$ to $60\,{\rm km\,s}^{-1}$ (in units of Kelvin). The \textit{bottom} panels display maximum intensity maps, where the main panel displays the ACA data alone, while the inset panel displays the combination of ACA + 12m array data. The colour scale represents the intensity of the molecular emission (the colour bar range for the inset panel has the same lower limits as the main panel, but an upper limit of 2.0 K), while the white contours represent continuum radio emission (masked to only show the radio lobes; contour levels are $2^{n}\times$ the rms noise of $8\mu{\rm Jy\,bm}^{-1}$, where $n=2.0, 2.5, 3.0, 3.5, 4.0, 5.0, 6.0$; see \S\ref{sec:vlarad}). The red ellipses indicate the ALMA beams. The \textit{top} panel displays the spectrum of the CS emission (ACA + 12m data) coincident with the northern lobe (the cyan square in the \textit{bottom} inset panel represents the spectral extraction region). We detect CS emission coincident with the northern side of the ring structure identified in Figure~\ref{fig:spec_1E}, as well as additional CS emission to the south of the ring structure.  The CS spectrum shows a wide line profile, consistent with a turbulent molecular medium.}
\end{center}
 \end{figure}

\begin{figure}
\begin{center}
\quad\quad\quad\includegraphics[width=0.38\textwidth]{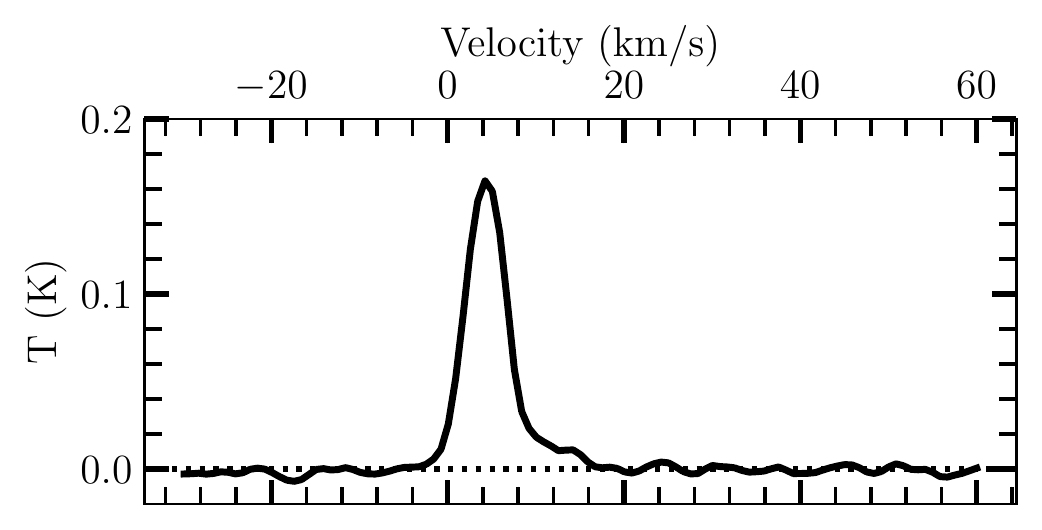}
  \includegraphics[width=0.48\textwidth]{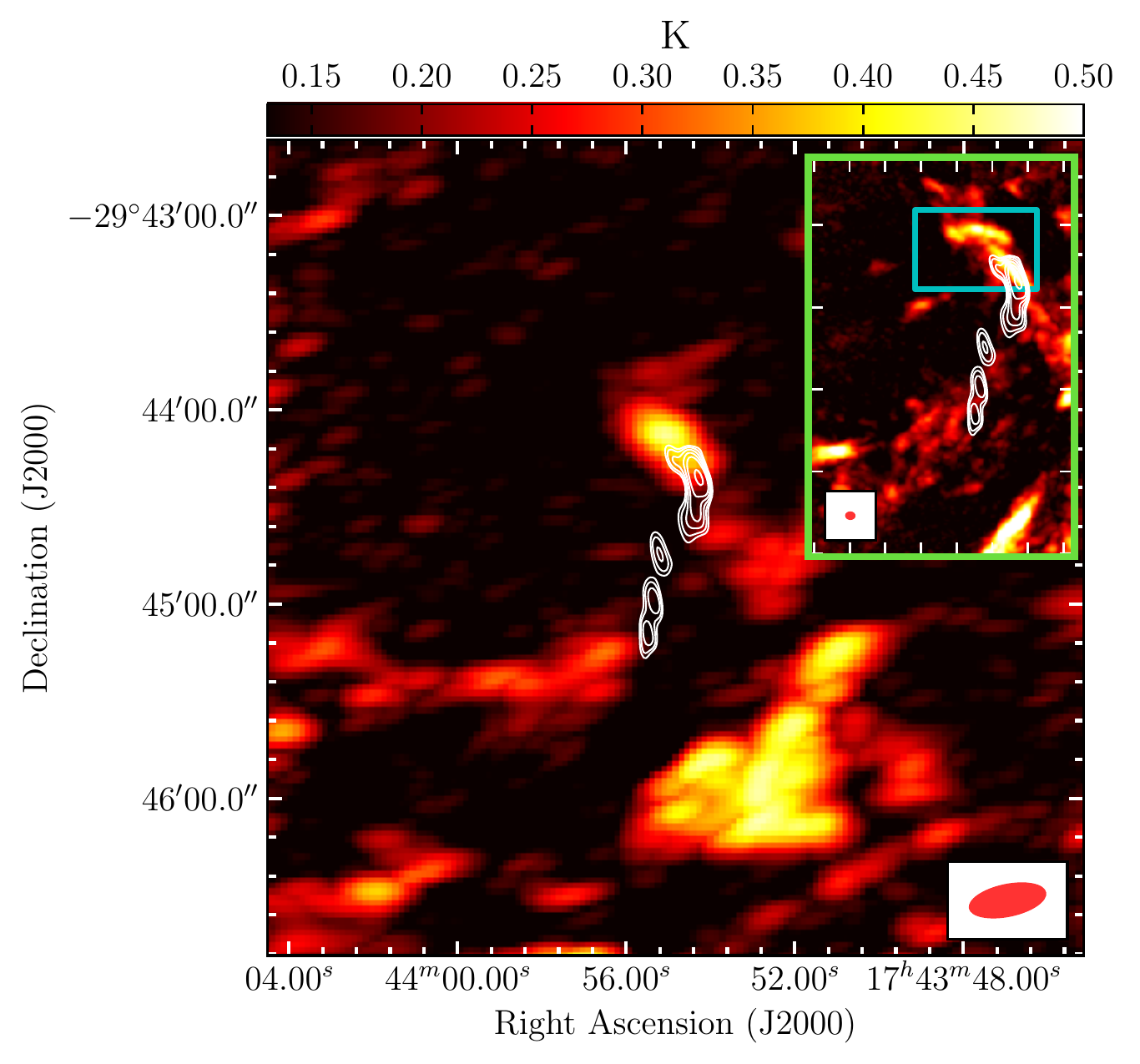}\\
 \caption{\label{fig:spec1E_hnco4}HNCO [$J=4-3$] emission (density and shock tracer) detected near the BHXB 1E 1740.7$-$2942 in the velocity range $-30$ to $60\,{\rm km\,s}^{-1}$ (in units of Kelvin). The \textit{bottom} panels displays maximum intensity maps, where the main panel displays the ACA data alone, while the inset panel displays the combination of ACA + 12m array data. The colour scale represents the intensity of the molecular emission (the colour bar range for the inset panel has the same lower limits as the main panel, but an upper limit of 1.0 K), while the white contours represent continuum radio emission (masked to only show the radio lobes; contour levels are $2^{n}\times$ the rms noise of $8\mu{\rm Jy\,bm}^{-1}$, where $n=2.0, 2.5, 3.0, 3.5, 4.0, 5.0, 6.0$; see \S\ref{sec:vlarad}). The red ellipses indicate the ALMA beams. The \textit{top} panel displays the spectrum of the HNCO [$J=4-3$] emission (ACA + 12m) coincident with the northern lobe (the cyan square in the \textit{bottom} inset panel represents the spectral extraction region). We detect HNCO [$J=4-3$] emission coincident with the northern and southern sides of the ring structure identified in Figure~\ref{fig:spec_1E}, as well as additional HNCO [$J=4-3$] emission to the south of the ring structure.  The HNCO [$J=4-3$] spectrum shows a wide line profile, consistent with a turbulent molecular medium.}
\end{center}
 \end{figure}

\begin{figure}
\begin{center}
\quad\quad\quad\includegraphics[width=0.38\textwidth]{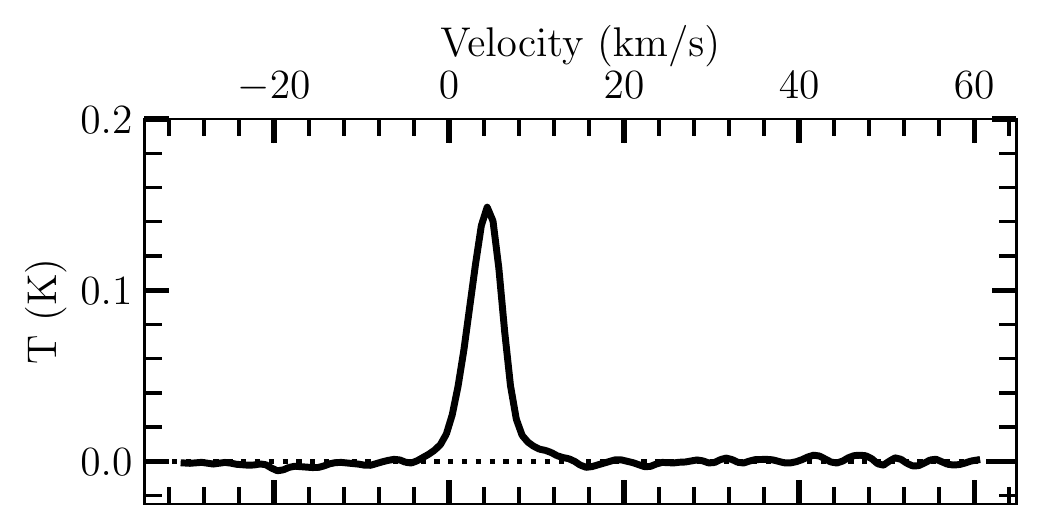}
  \includegraphics[width=0.48\textwidth]{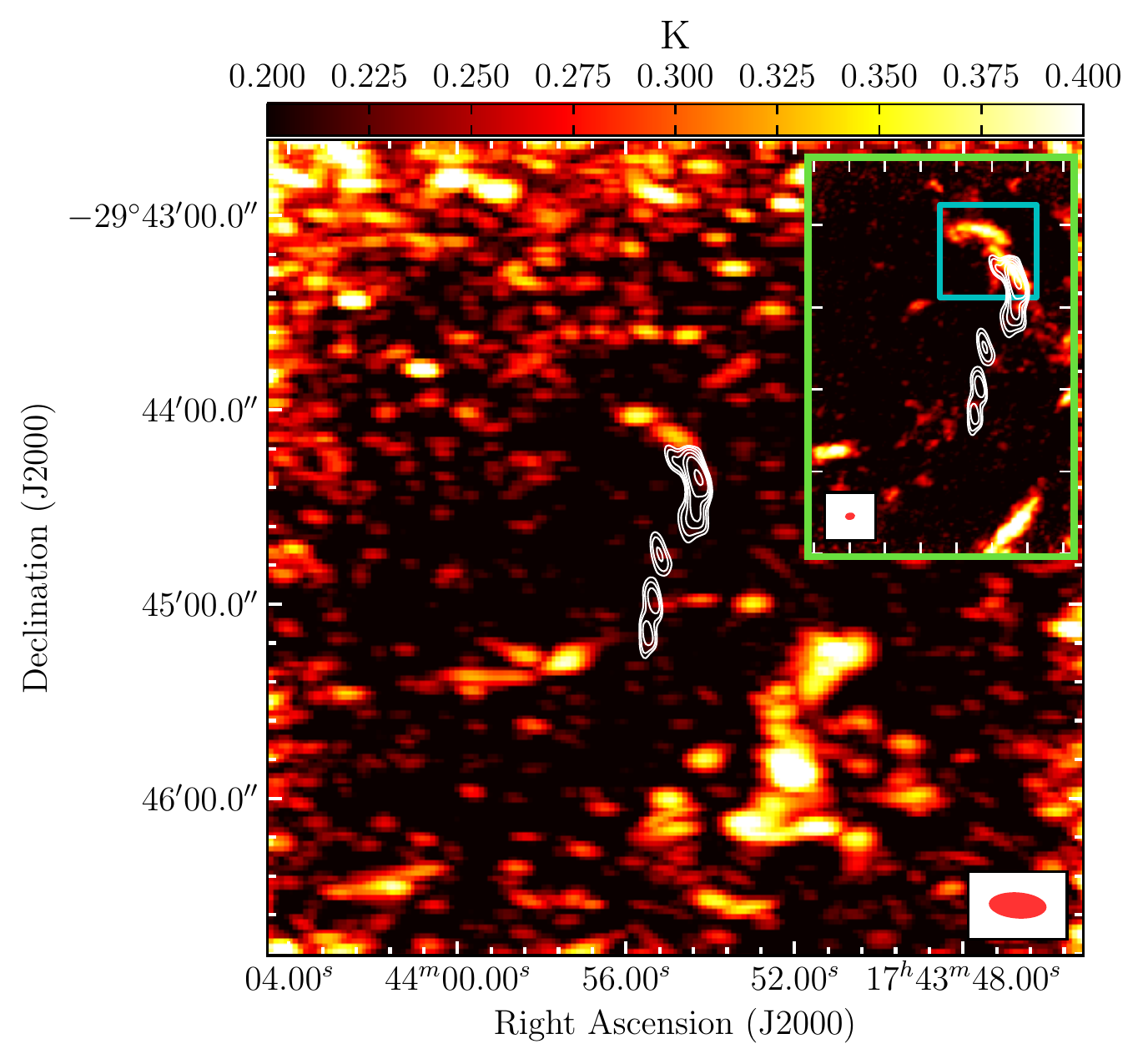}\\
 \caption{\label{fig:spec1E_hnco5}HNCO [$J=5-4$] emission (density and shock tracer) detected near the BHXB 1E 1740.7$-$2942 in the velocity range $-30$ to $60\,{\rm km\,s}^{-1}$ (in units of Kelvin). The \textit{bottom} panel displays maximum intensity maps, where the main panel displays the ACA data alone, while the inset panel displays the combination of ACA + 12m array data. The colour scale represents the intensity of the molecular emission (the colour bar range for the inset panel has the same lower limits as the main panel, but an upper limit of 0.8 K), while the white contours represent continuum radio emission (masked to only show the radio lobes; contour levels are $2^{n}\times$ the rms noise of $8\mu{\rm Jy\,bm}^{-1}$, where $n=2.0, 2.5, 3.0, 3.5, 4.0, 5.0, 6.0$; see \S\ref{sec:vlarad}). The red ellipses indicate the ALMA beams. The \textit{top} panel displays the spectrum of the HNCO [$J=5-4$] emission (ACA + 12m) coincident with the northern lobe (the cyan square in the \textit{bottom} inset panel represents the spectral extraction region).  We detect HNCO [$J=5-4$] emission coincident with the northern radio lobe, as well as additional HNCO [$J=5-4$] emission to the south of the radio lobe structures.  The HNCO [$J=5-4$] spectrum shows a wide line profile, consistent with a turbulent molecular medium.}
\end{center}
 \end{figure}

\begin{figure}
\begin{center}
\quad\quad\quad\includegraphics[width=0.38\textwidth]{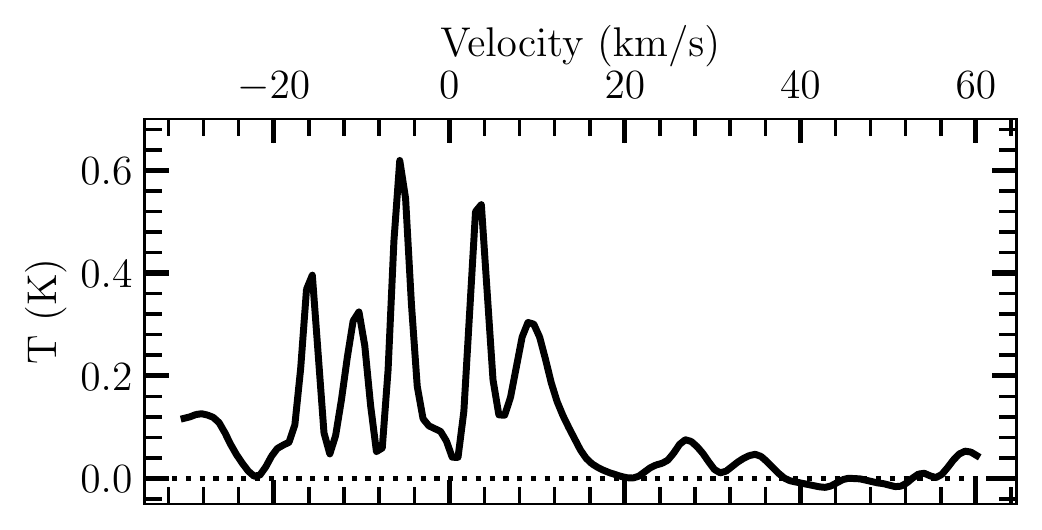}
  \includegraphics[width=0.48\textwidth]{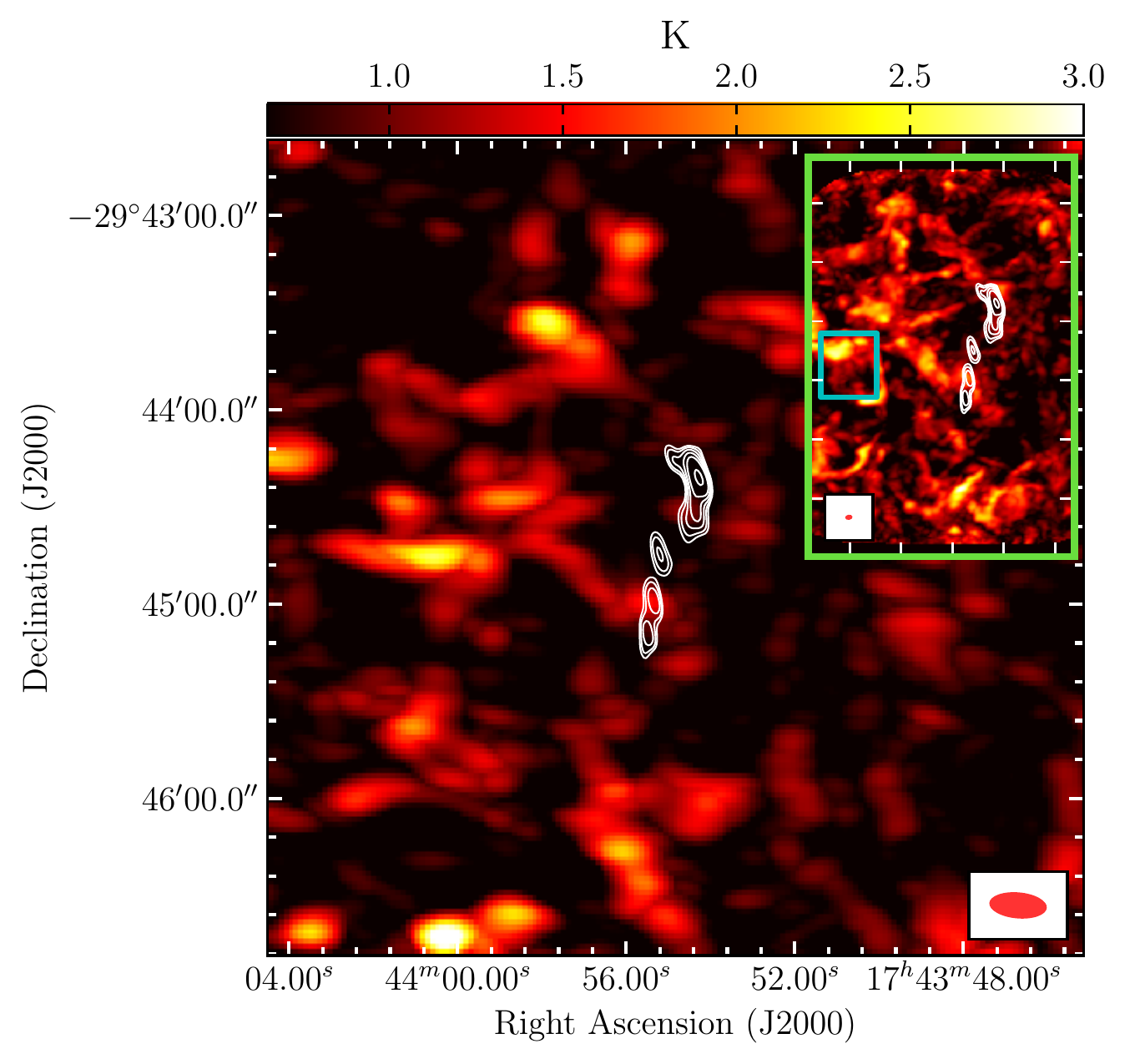}\\
 \caption{\label{fig:spec1E_13CO}$^{13}$CO emission (density tracer) detected near the BHXB 1E 1740.7$-$2942 in the velocity range $-30$ to $60\,{\rm km\,s}^{-1}$ (in units of Kelvin). The \textit{bottom} panel displays maximum intensity maps, where the main panel displays the ACA data alone, while the inset panel displays the combination of ACA + 12m array data. The colour scale represents the intensity of the molecular emission (the colour bar range for the inset panel has the same lower limits as the main panel, but an upper limit of 5.0 K), while the white contours represent continuum radio emission (masked to only show the radio lobes; contour levels are $2^{n}\times$ the rms noise of $8\mu{\rm Jy\,bm}^{-1}$, where $n=2.0, 2.5, 3.0, 3.5, 4.0, 5.0, 6.0$; see \S\ref{sec:vlarad}). The red ellipses indicate the ALMA beams. The \textit{top} panel displays the spectrum of the $^{13}$CO emission (ACA + 12m) coincident with the eastern edge of the molecular ring (the cyan square in the \textit{bottom} inset panel represents the spectral extraction region).  There is a distinct lack of $^{13}$CO emission coincident with the radio lobes or ring structure seen in HCO$+$ and HCN, with the majority of the $^{13}$CO emission on the outskirts of the field surrounding 1E 1740.7$-$2942.  The spectrum of the $^{13}$CO emission closest to the HCO$+$/HCN ring structure displays several closely spaced velocity components along the line of sight, and thus it is difficult to determine if this emission is related to the HCO$+$/HCN emission.}
\end{center}
 \end{figure}

\begin{figure}
\begin{center}
\quad\quad\quad\includegraphics[width=0.38\textwidth]{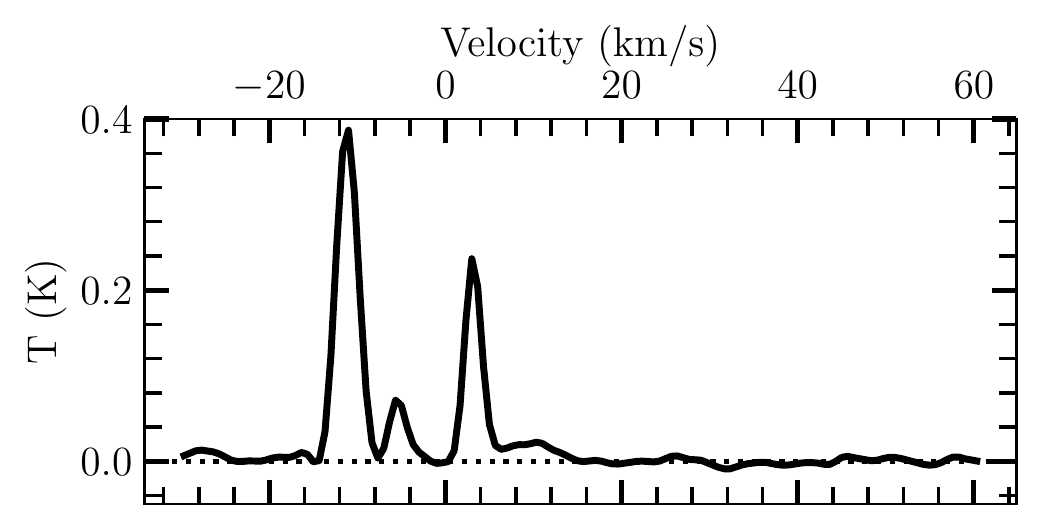}
  \includegraphics[width=0.48\textwidth]{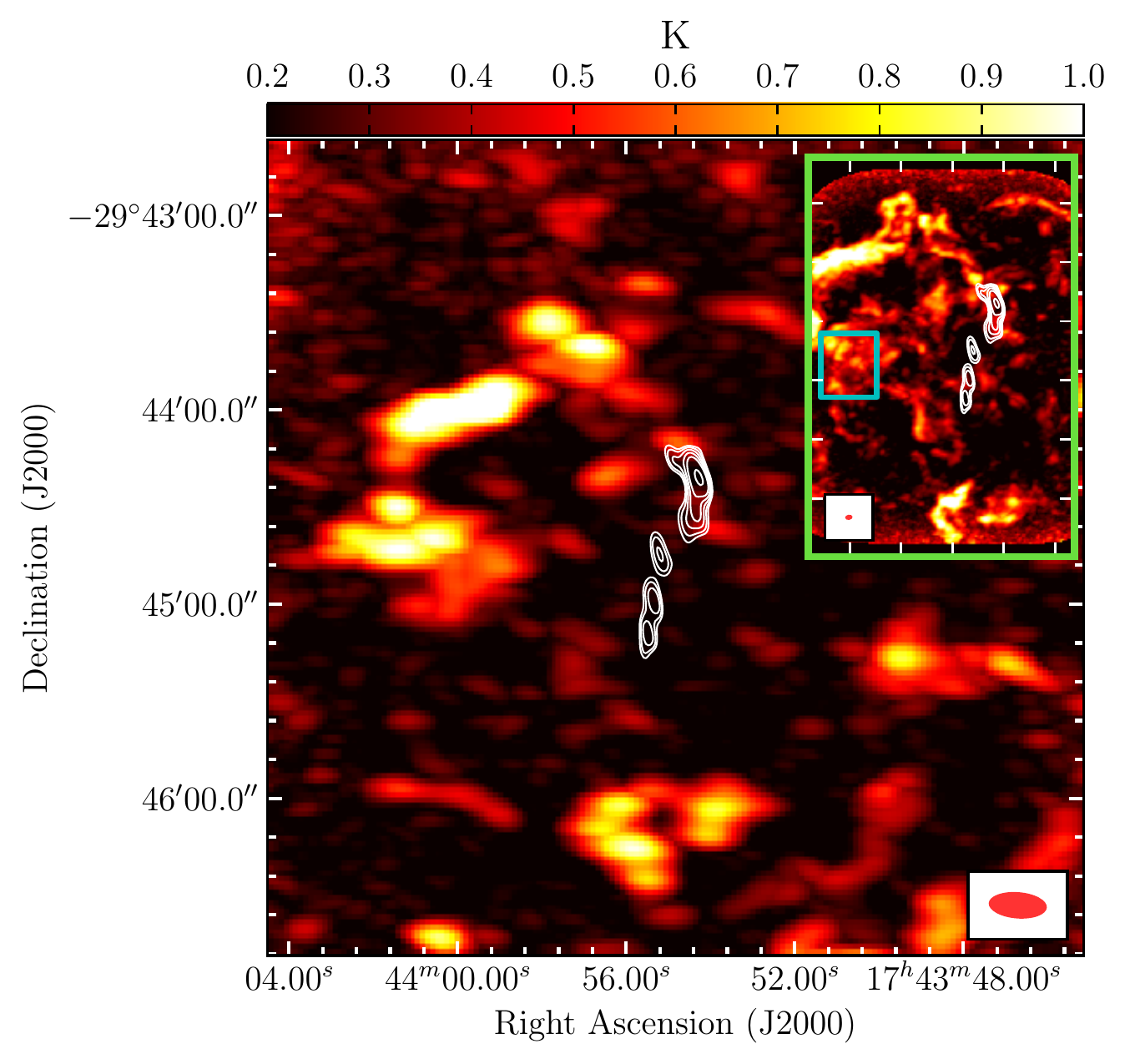}\\
 \caption{\label{fig:spec_18CO}C$^{18}$O emission (density tracer) detected near the BHXB 1E 1740.7$-$2942 in the velocity range $-30$ to $60\,{\rm km\,s}^{-1}$ (in units of Kelvin). The \textit{bottom} panel displays a maximum intensity map, where the main panel displays the ACA data alone, while the inset panel displays the combination of ACA + 12m array data. The colour scale represents the intensity of the molecular emission (the colour bar range for the inset panel has the same lower limits as the main panel, but an upper limit of 2.0 K), while the white contours represent continuum radio emission (masked to only show the radio lobes; contour levels are $2^{n}\times$ the rms noise of $8\mu{\rm Jy\,bm}^{-1}$, where $n=2.0, 2.5, 3.0, 3.5, 4.0, 5.0, 6.0$; see \S\ref{sec:vlarad}). The red ellipses indicate the ALMA beams. The \textit{top} panel displays the spectrum of the C$^{18}$O emission (ACA + 12m) coincident with the eastern edge of the molecular ring (the cyan square in the \textit{bottom} inset panel represents the spectral extraction region).  As with  $^{13}$CO, there is a distinct lack of C$^{18}$O emission coincident with the radio lobes or ring structure seen in HCO$+$ and HCN, with the majority of the C$^{18}$O emission on the outskirts of the field surrounding 1E 1740.7$-$2942.  The spectrum of the C$^{18}$O emission closest to the HCO$+$/HCN ring structure displays several velocity components along the line of sight, although these components appear to be at lower velocities when compared to the HCO$+$/HCN emission.}
\end{center}
 \end{figure}

\begin{figure}
\begin{center}
\quad\quad\quad\includegraphics[width=0.38\textwidth]{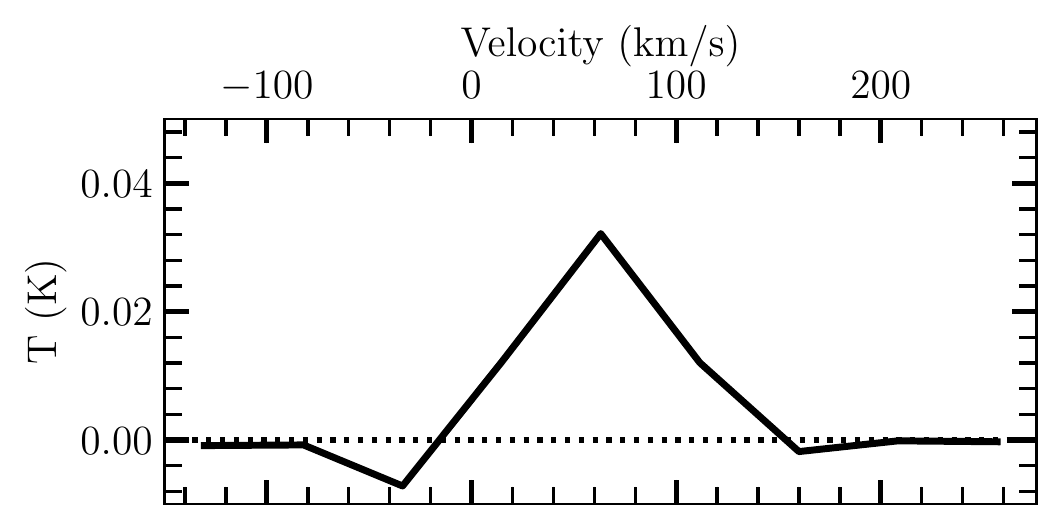}
  \includegraphics[width=0.52\textwidth]{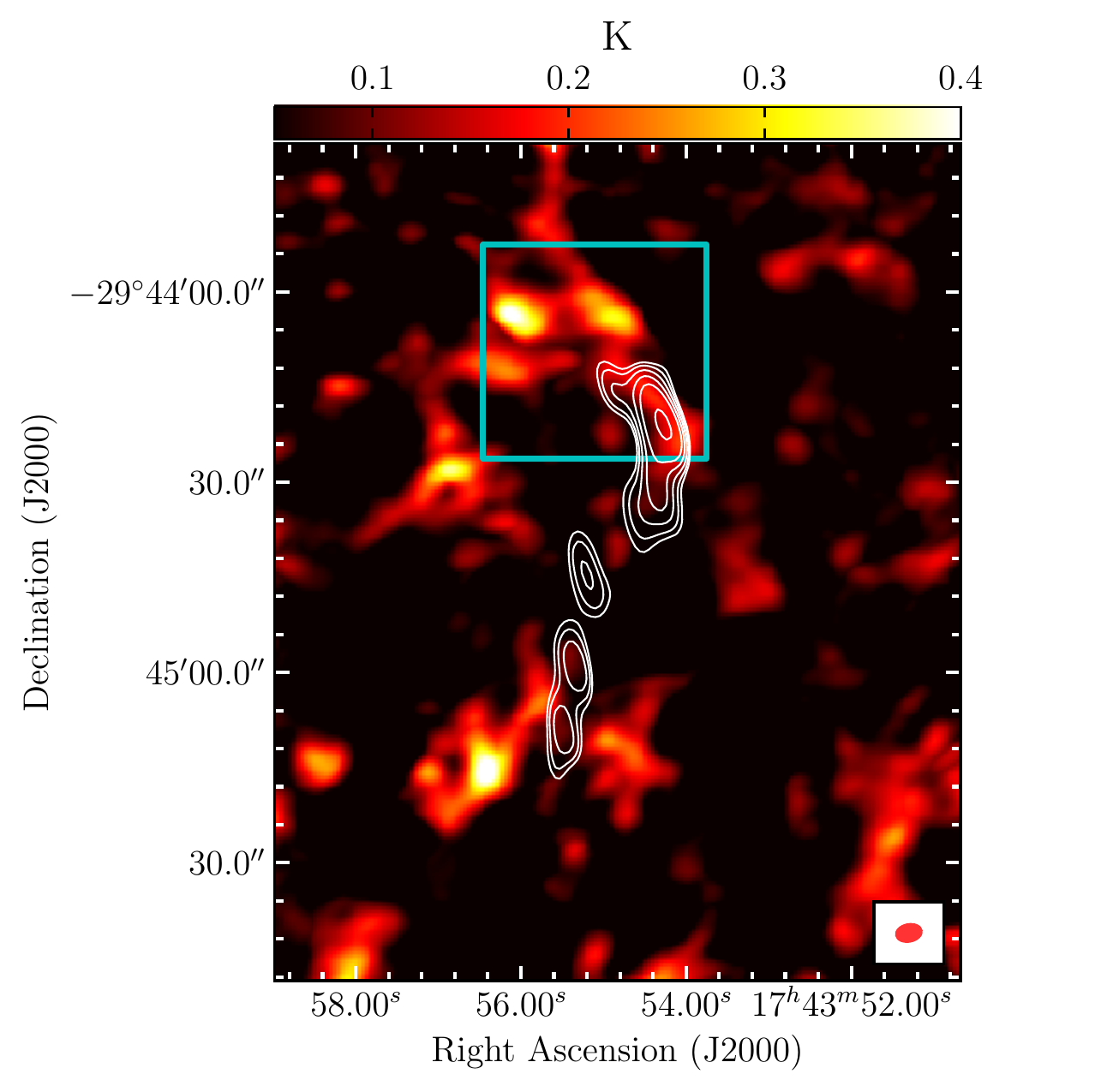}\\
 \caption{\label{fig:spec_CH3OH}CH$_3$OH emission (shock tracer) detected near the BHXB 1E 1740.7$-$2942 in the velocity range $-100$ to $200\,{\rm km\,s}^{-1}$ (in units of Kelvin). The \textit{bottom} panel displays a maximum intensity map created from 12m array data only, as we did not sample this frequency range with the ACA. The colour scale represents the intensity of the molecular emission, while the white contours represent continuum radio emission (masked to only show the radio lobes; contour levels are $2^{n}\times$ the rms noise of $8\mu{\rm Jy\,bm}^{-1}$, where $n=2.0, 2.5, 3.0, 3.5, 4.0, 5.0, 6.0$; see \S\ref{sec:vlarad}). The red ellipse indicates the ALMA 12m array beam. The \textit{top} panel displays the spectrum of the CH$_3$OH emission (12m only) coincident with the northern radio lobe (the cyan square in the \textit{bottom} panel represents the spectral extraction region).  We detect CH$_3$OH emission close to both radio lobes near 1E 1740.7$-$2942.  The spectrum of the CH$_3$OH emission is very low resolution ($\sim50\, {\rm km\,s}^{-1}$), as we detected this line in a spectral window set up to measure the continuum emission.}
\end{center}
 \end{figure}

\begin{figure}
\begin{center}
\quad\quad\quad\includegraphics[width=0.38\textwidth]{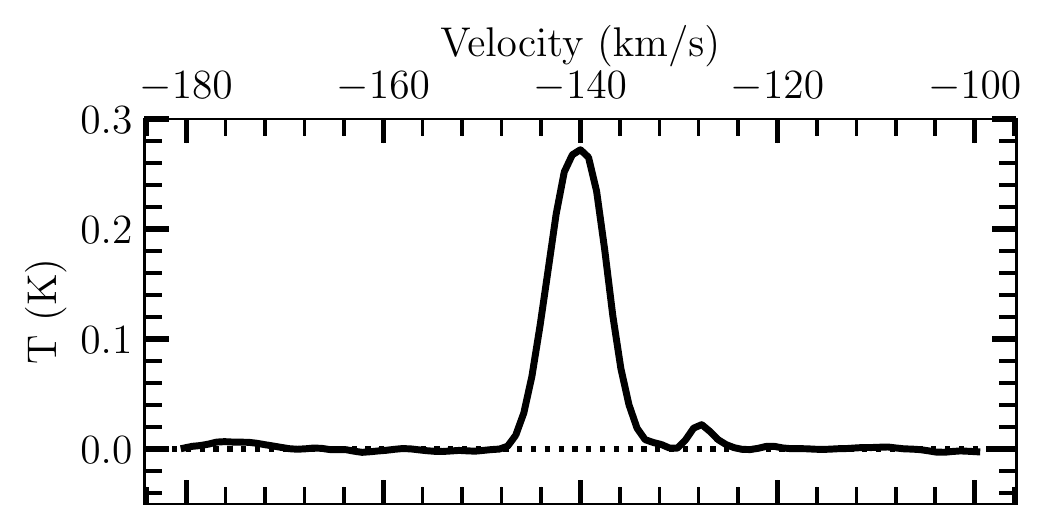}
  \includegraphics[width=0.48\textwidth]{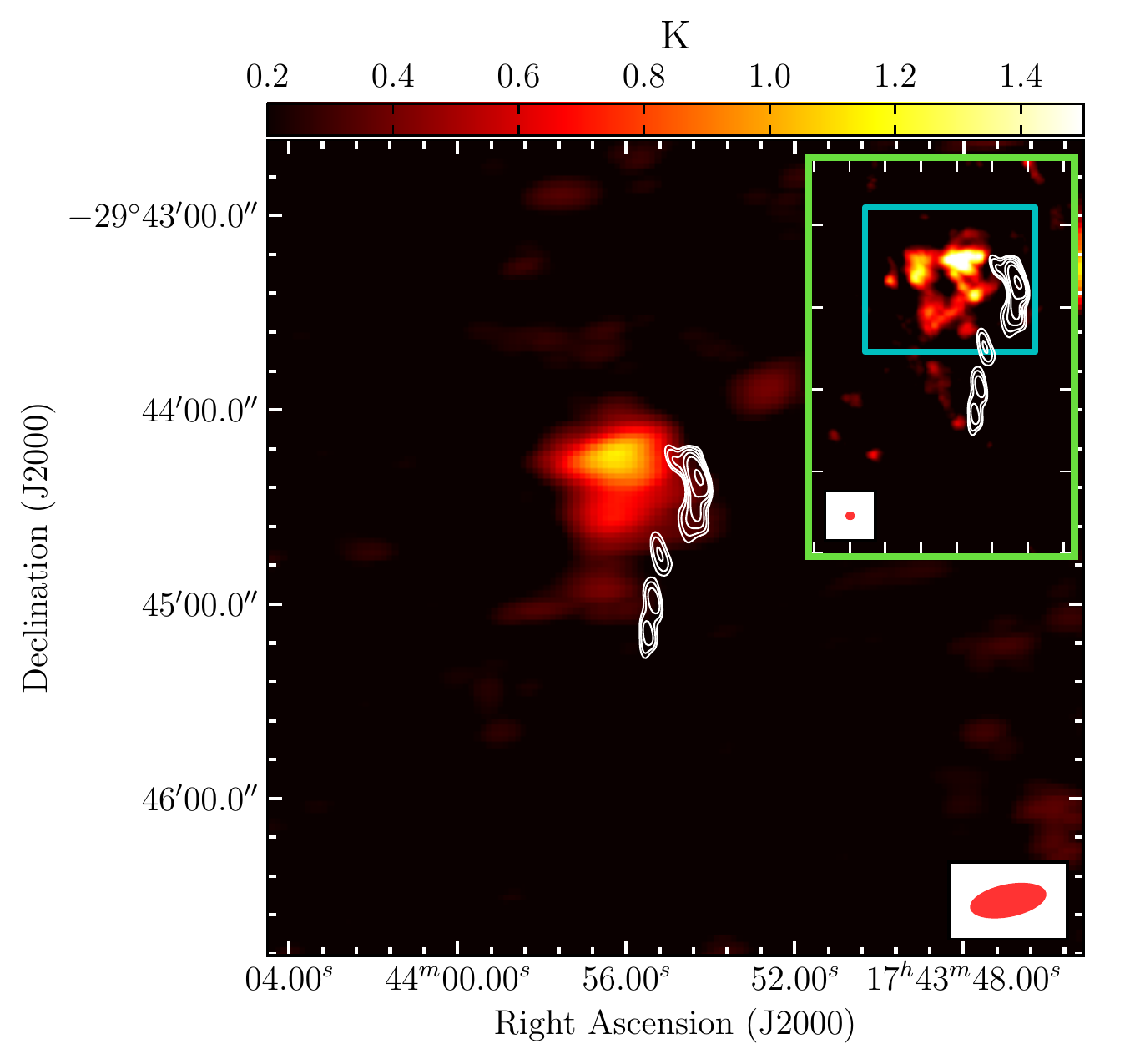}\\
 \caption{\label{fig:spec_alt}HCO$+$ emission (density tracer) detected near the BHXB 1E 1740.7$-$2942 in the velocity range $-180$ to $-100\,{\rm km\,s}^{-1}$ (in units of Kelvin). The \textit{bottom} panel displays maximum intensity maps, where the main panel displays ACA data alone, and the inset panel displays the combination of ACA + 12m array data. The colour scale represents the intensity of the molecular emission (the colour bar range for the inset panel has the same lower limits as the main panel, but an upper limit of 2.0 K), while the white contours represent continuum radio emission (masked to only show the radio lobes; contour levels are $2^{n}\times$ the rms noise of $8\mu{\rm Jy\,bm}^{-1}$, where $n=2.0, 2.5, 3.0, 3.5, 4.0, 5.0, 6.0$; see \S\ref{sec:vlarad}). The red ellipses indicate the ALMA beams. We detect an isolated molecular cloud in this velocity range, which is also seen in other density tracing HCN, HNCO, and CS molecules, to the north-east of the radio lobe structure. The \textit{top} panel displays the HCO$+$ spectrum (ACA + 12m) of this cloud, where the cyan square in the \textit{bottom} inset panel represents the spectral extraction region.}
\end{center}
 \end{figure}

In this Section, we show maps and spectra of other molecular emission detected in the field surrounding the BHXBs GRS 1758$-$258 and 1E 1740.7$-$2942, not shown in the main text. Figure~\ref{fig:specgrs_SiO} displays the emission from the SiO molecule in the velocity range $45$ to $105\,{\rm km\,s}^{-1}$ for GRS 1758$-$258, Figures~\ref{fig:spec1E_CS} -- \ref{fig:spec_CH3OH} display emission from the CS, HNCO, CO, and CH$_3$OH molecules in the velocity range $-30$ to $60\,{\rm km\,s}^{-1}$ for 1E 1740.7$-$2942, and Figure~\ref{fig:spec_alt} displays emission from the isolated cloud detected in the velocity range $-180$ to $-100\,{\rm km\,s}^{-1}$ for 1E 1740.7$-$2942.

\bsp	
\label{lastpage}
\end{document}